\documentclass[a4paper,nofootinbib,tightenlines,superscriptaddress]{revtex4}

\usepackage{amsfonts,amssymb,amsthm,bbm,amsmath}
\usepackage{hyperref}

\usepackage{color,psfrag}
\usepackage[dvips]{graphicx}

\usepackage{tikz}
\usetikzlibrary{calc}
\usetikzlibrary{decorations.pathmorphing}

\usepackage{subcaption}
\newcommand{\C}{{\mathbb C}}

\newcommand{\R}{{\mathbb R}}

\newcommand{\cG}{{\mathcal G}}

\newcommand{\cS}{{\mathcal S}}

\newcommand{\SU}{\mathrm{SU}}
\newcommand{\ISU}{\mathrm{ISU}}

\newcommand{\SL}{\mathrm{SL}}

\newcommand{\SB}{\mathrm{SB}}

\newcommand{\be}{\begin{equation}}
\newcommand{\ee}{\end{equation}}
\newcommand{\beq}{\begin{eqnarray}}
\newcommand{\eeq}{\end{eqnarray}}
\newcommand{\bes}{\begin{eqnarray}}
\newcommand{\ees}{\end{eqnarray}}

\newcommand{\mat} [2] {\left ( \begin{array}{#1}#2\end{array} \right ) }

\newcommand{\su}{{\mathfrak{su}}}
\newcommand{\isu}{{\mathfrak{isu}}}

\renewcommand{\sl}{{\mathfrak{sl}}}
\renewcommand{\sb}{{\mathfrak{sb}}}

\newcommand{\tr}{{\mathrm{Tr}}}
\newcommand{\f}{\frac}

\def\nn{\nonumber}
\def\pp{\partial}

\def\ka{\kappa}

\def\eps{\epsilon}

\newcommand{\id}{\mathbb{I}}
\def\ka{\kappa}

\begin{document}

\title{The closure constraint for the hyperbolic tetrahedron as a Bianchi identity}

\author{{\bf Christoph Charles}}\email{christoph.charles@ens-lyon.fr}

\author{{\bf Etera R. Livine}}\email{etera.livine@ens-lyon.fr}
\affiliation{Univ Lyon, ENS de Lyon, Univ Claude Bernard, CNRS, Laboratoire de	Physique,	
  		F-69342 Lyon, France}

\date{\today}

\begin{abstract}

The closure constraint is a central piece of the mathematics of loop quantum gravity. It encodes the gauge invariance of the spin network states of quantum geometry and provides them with a geometrical interpretation: each decorated vertex of a spin network is dual to a quantized polyhedron in $\R^{3}$. For instance, a 4-valent vertex is interpreted as a tetrahedron determined by the four normal vectors of its faces.
We develop a framework where the closure constraint is re-interpreted as a Bianchi identity, with the normals defined as holonomies around the polyhedron faces of a connection (constructed from the spinning geometry interpretation of twisted geometries).
This allows us to define closure constraints for hyperbolic tetrahedra (living in the 3-hyperboloid of unit future-oriented spacelike vectors in $\R^{3,1}$) in terms of normals living all in $\SU(2)$ or in $\SB(2,\C)$. The latter fits perfectly with the classical phase space developed for $q$-deformed loop quantum gravity supposed to account for a non-vanishing cosmological constant $\Lambda>0$.
This is the first step towards interpreting $q$-deformed twisted geometries as actual discrete hyperbolic triangulations.


\end{abstract}

\maketitle

\section*{Introduction}

Loop quantum gravity (for monographs see \cite{rovelli_quantum_2007, thiemann_modern_2007, vidotto_covariant_2014}) is based on a reformulation of general relativity as an $\mathrm{SU}(2)$ gauge theory of the Ashtekar-Barbero connection \cite{ashtekar_new_1986, barbero_g._real_1995, immirzi_real_1997, holst_barberos_1996} together with its canonically conjugated field, the densitized triad\footnotemark{}. At the quantum level, the kinematics of the theory is well-understood: the space of quantum  states of geometry are wave-functions of the   Ashtekar-Barbero connection and admits a canonical basis labelled by spin networks \cite{rovelli_spin_1995}. Spin networks are (abstract) graphs (which might have knotting information in 3+1d) colored with spins on the edges (hence the name) and intertwiners, which are invariant representations of the gauge group, on the vertices. The spin network basis diagonalize the geometrical operators. The spins carried by the edges corresponding to quanta of surfaces \cite{rovelli_discreteness_1995}. The volume operator is a bit more involved but we still can find a basis of intertwiners that diagonalize the volume operator.
\footnotetext{This new set of variables corresponds to a canonical transformation at the classical level and depends on a new dimensionless parameter, the Immirzi parameter. Classically, this parameter does not play any role in the equation of motion, at least in the vacuum case \cite{rovelli_immirzi_1998, alexandrov_immirzi_2008}.}

The spin network states have a natural geometrical interpretation in the twisted geometry framework \cite{freidel_twisted_2010}. Each node of the graph is interpreted as a flat convex polyhedron. Each edge coming out of the vertex corresponds to a face of the polyhedron and the area of this face is given by the spin carried by the edge. The intertwiner should encode the remaining quantum degrees of freedom. Then, the polyhedra corresponding to the vertices are glued together on their faces. By construction, the area of two glued faces match. Still, the geometry is said to be twisted since the shapes of the faces do not have to match. In the continuum, this corresponds to a discontinuous metric \cite{haggard_spin_2013}. There is a notable second geometrical parametrization that was proposed: spinning geometries \cite{freidel_spinning_2014}. The construction is roughly the same but the faces and edges of the polyhedra are not assumed to be (extrinsically) flat. As a consequence, spinning geometries correspond to the same phase space but gives it an interpretation with a continuous metric but with a non-trivial torsion\footnotemark{}.
\footnotetext{The torsion exists generally in the context of twisted geometries but does not reflect to a genuine torsion of the spin-connection. It corresponds to the torsion of the Ashtekar-Barbero connection with respect to the triad 
and encodes extrinsic curvature data.}

The dynamics of the theory is still a much researched point but should be implemented by yet-to-discover quantum Hamiltonian constraints. Several proposals exist to date, such as Thiemann's original regularization \cite{thiemann_anomaly_1996} or the Master Constraint Program \cite{thiemann_phoenix_2006}. Other proposals from a more covariant perspective have been studied, as in spinfoams (for a review, see \cite{perez_spin_2013} and references therein), with the EPRL vertex \cite{engle_lqg_2008}, or in group field theory (see \cite{oriti_group_2014} for a review). One of the salient point of the dynamics, beside its definition, is its compatibility with the classical theory. This is particularly addressed with coarse-graining. The issue of the coarse-graining or of the continuum limit of the theory and of its dynamics are related, first because a consistent theory of quantum gravity should solve the problem of perturbative non-renormalizability and second because a good theory of quantum gravity should describe all scales from the Planck scale to the continuum world we now know and love. One of the main challenge in the coarse-graining of the theory is the absence of any background to compare the scale to. We must, therefore, devise a background independent way of coarse-graining. Such an endeavour has been started in several directions. Koslowski and Sahlmann have, for instance, defined new vacuum states peaked on classical configuration of the triad and the spin network describe excitation over this configuration \cite{Koslowski:2011vn}. Dittrich and Geiller define yet another vacuum which corresponds to flat space (rather than degenerate metric) which appears naturally in BF theory and should be more suited to account for curvature defect \cite{Bahr:2015bra,Dittrich:2014wpa}. We also tried a similar approach in a recent work, trying to capture the degrees of freedom of a varying graph into a vertex and therefore using a fixed graph as a vacuum around which we expand \cite{Charles:2016xwc}. But in this paper, we pursue another course: we try to investigate the possibility of capturing only the large scale degrees of freedom, by capturing homogeneous curvature. Indeed, in a coarse-graining perspective, curvature builds up on large scales \cite{livine_deformation_2014}. So, our goal is to adapt the kinematical space of loop quantum gravity, so that the natural geometrical interpretation of the vertices will not be flat polyhedron but curved ones, or at least, so that the natural interpretation will be in the realm of homogeneously curved manifolds.

In the context of gravity, homogeneous curvature naturally leads to the question of the cosmological constant. And so our question might be linked to the inclusion of cosmological constant in the theory. We should note here that the question is actually more subtle, compared for instance to the 3d case. Indeed, because of the choice of variables and of the time-gauge, the time dependence is hidden in loop quantum gravity. Also, the question is more about homogeneous spatial curvature (as is supposed in cosmology for example) and not spatio-temporal curvature (as for a cosmological constant). But thanks to the choice of variable, it might very well be that, at the kinematical level, we only control the spatial part of the curvature. Of course, all this discussion will be settled once the dynamics is set. So far, let's just hope that such an idea might work and see where it leads. In three dimensional gravity, the cosmological constant enters the theory in the group structure of the theory. This is the case whether we consider combinatorial quantification \cite{witten_2+1-dimensional_1988} or covariant models \cite{mizoguchi_three-dimensional_1992}. The cosmological constant enters the theory as a quantum deformation parameter of the gauge group. For instance, for the covariant approach, the Ponzano-Regge model is deformed into the Turaev-Viro model based on the $\mathrm{SU}_q(2)$. At the canonical level, several directions are pursued, either by putting the quantum group structure directly at the kinematical level \cite{dupuis_observables_2013, bonzom_deformed_2014, dupuis_deformed_2014, bonzom_towards_2014} or by trying to retrieve it through the dynamics \cite{pranzetti_turaev-viro_2014}. In this new canonical development, we expect a kind of quantum deformed twisted geometry to appear, and indeed, the variables associated to the classical limit of the quantum deformation corresponds to the construction of hyperbolic polygons (at least in the 3-valent case)  \cite{bonzom_deformed_2014} and the spin networks allow the gluing of such polygons to construct a whole space.

With regard to the previous discussion, this raises the following question: in the four dimensional case, to what corresponds the introduction of a cosmological constant and the definition of a model based on $\mathrm{SU}_q(2)$? The hope would be of course to have a geometrical interpretation in terms of homogeneously curved geometries. For instance, for a real $q$ parameter, we expect such a theory to describe hyperbolic geometry. However, the present situation is less clear. It was shown that a quite natural way to describe hyperbolic geometry was to use $\mathrm{SU}(2)$ elements as normals instead of vectors \cite{charles_closure_2015, haggard_encoding_2015}. This can lead to a kind of quantum deformed theory (though not the same quantum deformation) \cite{haggard_sl2c_2015}. Though interesting in the study of the cosmological constant, the original question of the q-deformed $\SU(2)$ theory is left unanswered. And though, it should be possible to define a $q$-deformed version of 4d loop quantum gravity, the geometrical interpretation is much less understood. So this is our goal in this paper: to give a geometrical interpretation of the structures of such a $q$-deformed loop quantum gravity and explore the link with curved geometries. We concentrate on hyperbolic geometries (real $q$ deformation) leaving spherical construction for further studies. Following our first paper \cite{charles_closure_2015}, we use a top-down approach rather than a bottom-up one: we start from hyperbolic polyhedra and then try and construct semi-classical structures from geometrical data that have a natural matching in the $q$-deformed framework. 

\smallskip

In loop quantum gravity, for a given graph, it is natural to define some geometrical operators that correspond to a discretization adapted to the graph of the continuum variables, that is, of the densitized triad and of the Ashtekar-Barbero spatial connection \cite{freidel_twisted_2010}. First, it is natural to defined a vector associated to each end of each edge of the graph. This corresponds to the densitized triad integrated over the corresponding face, seen as the face of the polyhedron dual either to the source or to the target vertex of the considered edge. Because of the matching areas of the faces, the two vectors of the two ends of the edge have the same norm (which is, incidentally, the area of the face). Second, we can define an $\mathrm{SU}(2)$ group element for each edge which is the integrated version of the connection along the edge, which can also be understood in the twisted geometry framework as the parallel transport when crossing the face linking two polyhedra. All these quantities are operators as they correspond to geometric measurements. They also obey some constraints as the one already mentioned. So beside the matching norm constraint, they obey two further constraints. First, the vectors at each end of an edge are image of each other, up to a sign, by the $\mathrm{SU}(2)$ group element (which actually implies the area matching constraint). Second, the vectors at one vertex sum up to zero. This last constraint is the closure constraint and corresponds to the gauge invariance condition. It is one of our guiding principles for the construction in this paper.

\begin{figure}[h!]
\centering
\begin{tikzpicture}[scale=2]
\coordinate (O1) at (0,0,0);

\coordinate (A1) at (0,1.061,0);
\coordinate (B1) at (0,-0.354,1);
\coordinate (C1) at (-0.866,-0.354,-0.5);
\coordinate (D1) at (0.866,-0.354,-0.5);

\draw[blue] (A1) -- (B1);
\draw[blue] (A1) -- (C1);
\draw[blue] (A1) -- (D1);
\draw[blue] (B1) -- (C1);
\draw[dashed,blue] (C1) -- (D1);
\draw[blue] (D1) -- (B1);

\draw[dotted] (O1) -- ++(0,-0.531,0);
\draw (0,-0.531,0) -- ++(0,-0.531,0);

\draw[dotted] (O1) -- ++(0,0.177,-0.5);
\draw[dashed] (0,0.177,-0.5) -- ++(0,0.177,-0.5);

\draw[dotted] (O1) -- ++(0.433,0.177,0.25);

\draw[dotted] (O1) -- ++(-0.433,0.177,0.25);
\draw (-0.433,0.177,0.25) -- ++(-0.433,0.177,0.25);

\draw (O1) node{$\bullet$};
\draw[blue] (A1) node{$\bullet$};
\draw[blue] (B1) node{$\bullet$};
\draw[blue] (C1) node{$\bullet$};
\draw[blue] (D1) node{$\bullet$};

\coordinate (O2) at (3,0,0);

\coordinate (A2) at (3,1.061,0);
\coordinate (B2) at (3,-0.354,1);
\coordinate (C2) at (2.134,-0.354,-0.5);
\coordinate (D2) at (3.866,-0.354,-0.5);

\draw[blue] (A2) -- (B2);
\draw[blue] (A2) -- (C2);
\draw[blue] (A2) -- (D2);
\draw[blue] (B2) -- (C2);
\draw[dashed,blue] (C2) -- (D2);
\draw[blue] (D2) -- (B2);

\draw[dotted] (O2) -- ++(0,-0.531,0);
\draw (3,-0.531,0) -- ++(0,-0.531,0);

\draw[dotted] (O2) -- ++(0,0.177,-0.5);
\draw[dashed] (3,0.177,-0.5) -- ++(0,0.177,-0.5);

\draw[dotted] (O2) -- ++(0.433,0.177,0.25);
\draw (3.433,0.177,0.25) -- ++(0.433,0.177,0.25);

\draw[dotted] (O2) -- ++(-0.433,0.2,0.25);

\draw (O2) node{$\bullet$};
\draw[blue] (A2) node{$\bullet$};
\draw[blue] (B2) node{$\bullet$};
\draw[blue] (C2) node{$\bullet$};
\draw[blue] (D2) node{$\bullet$};

\draw (0.433,0.177,0.25) to[bend left=12] node[midway,above]{$g \in \mathrm{SU}(2)$} node[midway]{$>$} (2.567,0.2,0.25);
\end{tikzpicture}

\caption{The twisted geometry interpretation of spin networks: each node of the spin network is matched with a corresponding dual polyhedron (constructed from Minkowski theorem). Each edge then carries an $\mathrm{SU}(2)$ connection which corresponds to the parallel transport across the face.}
\end{figure}
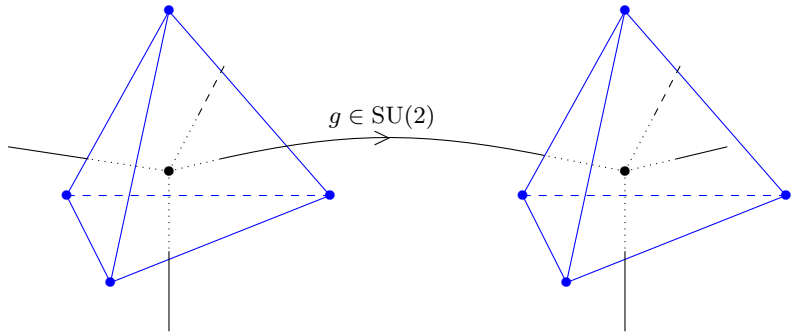

Of course, we can find a classical equivalent of this construction. In that case, the commutation relations become Poisson brackets in the classical limit. In the rest of this paper, we will work at the classical level. It was noted in \cite{bonzom_deformed_2014} that the Poisson structure constructed as above corresponds to the Heisenberg double of the $\mathrm{ISU}(2)$ group and can be understood as a limit $q \rightarrow 1$ of a similar construction on $\mathrm{SL}(2,\mathbb{C})$. The structure of the Heisenberg double is the classical equivalent of a quantum group. Therefore, looking at such construction at the classical level is the right framework for understanding the geometrical interpretation of a quantum deformed theory. From an algebraic perspective, everything is already set: the quantum deformation amounts, at the classical level, to a deformation of the vectors at each edge-end into $\mathrm{SB}(2,\mathbb{C})$ elements. The closure constraint is also deformed by using the non-commutative law on $\mathrm{SB}(2,\mathbb{C})$ (which is simply the matrix product). A braiding appears because of this non-commutativity. The non-deformed case had a nice geometrical interpretation: each vector corresponded to a normal to a face of the polyhedron. The closure constraint guaranteed the existence of a unique convex polyhedron with the given normals, thanks to Minkowksi's theorem and thus validated the twisted geometry interpretation. This sets our goal: writing down $\mathrm{SB}(2,C)$ elements that depends only on the geometrical data of faces of homogeneously curved polyhedron such that it satisfies the (generalized) closure condition and it has a natural interpretation as a normal. We will construct such structures in this paper.

\vspace{1em}

We actually offered such a proposal in \cite{charles_closure_2015}. But in the construction, we only used the closure constraint as a guiding line. This creates some problem when trying to put an actual Hamiltonian structure on top of this, as can be seen for example in the behaviour of the defined quantities under rotations (which is quite erratic). This problem will therefore be taken as one of our guide lines here: the $\mathrm{SB}(2,\mathbb{C})$ normals should transform correctly under rotation\footnote{That is, they should have the same behaviour under rotation as the one induced by the deformed closure constraint under Hamiltonian flow as defined in \cite{bonzom_deformed_2014}.} by construction. This means we want the elements to naturally  follow a closure relation by construction. If the previous construction showed nothing but one point, it is that it is quite easy to put group elements together so that they satisfy a closure condition. So, we want a more geometrical way of looking at it. For this, we will also follow an intuition based on the $\mathrm{SU}(2)$ normals construction of \cite{charles_closure_2015, haggard_encoding_2015}. In these cases, the $\mathrm{SU}(2)$ normals followed a closure condition, which could be interpreted quite naturally as a Bianchi identity. Building on this case, we will try and interpret any closure constraint as a discrete Bianchi identity. It turns out, this is pretty coherent and already warranted in the flat case by Freidel's construction of spinning geometry \cite{freidel_spinning_2014}. Indeed, we can define a connection on flat space such that the holonomy of this connection around flat polygons gives the normal. And the Bianchi identity of the connection does indeed gives the closure condition of the normals.

So, both known constructions of normals, that have some nice behaviour and geometrical interpretation, come from a connection. This is the case in flat space as pointed out in the previous paragraph but is also the case for the $\mathrm{SU}(2)$ normals for curved polyhedra which directly comes from the natural, triad compatible, spin connection on the corresponding curved manifold. Starting from these two cases, we are led to look for $\mathrm{SB}(2,\mathbb{C})$ connections on the three dimensional hyperboloid. Such a connection would, via a similar construction, lead to all the algebraic data we need for the classical limit of the quantum group structure. This task of finding well-behaved $\mathrm{SB}(2,\mathbb{C})$ connection is actually a bit difficult because of the particular transformation rules of the $\mathrm{SB}(2,\mathbb{C})$ elements under rotation. We will develop a similar technique and rather look for (non-trivial) $\mathrm{SL}(2,\mathbb{C})$ connections that we will then split into two parts according to the Iwasawa decomposition (splitting of $\mathrm{SL}(2,\mathbb{C})$ into $\mathrm{SB}(2,\mathbb{C})$ and $\SU(2)$). This will lead to two closure conditions, one in $\mathrm{SB}(2,\mathbb{C})$ and one in $SU(2)$. Both of these closures show a kind of braided structure which we expect for the $\mathrm{SB}(2,\mathbb{C})$ one but not for the $\mathrm{SU}(2)$ one. So, incidentally, we found a new $\mathrm{SU}(2)$ closure constraint for hyperbolic tetrahedra (and more generally for polyhedra). 

In order to look for such $\mathrm{SL}(2,\mathbb{C})$ connections, we will review the flat case and make a similar constructions there. In particular, we will generalize Freidel's connection to an $\mathrm{ISU}(2)$ to better match the deformed case. We will see that the natural connections to look at are the \textit{homogeneous} connections on the space. We will then mimic this construction of homogeneous connections on the three-dimensional hyperboloid. The resulting construction resembles the Immirzi-Barbero connection in a lot of ways, though it is expressed in intrinsic geometry terms only. Still, a lot of the analysis done on the Immirzi-Barbero connection can be transported to the new connection. In particular, we actually defined a one-(complex-)parameter family of connection. This leads to a one-parameter family of closure constraints that can be interpreted as a deformation of the previous $\mathrm{SU}(2)$ closure constraint.

\vspace{1em}

This paper is organized as follows. In the first section, we review and develop the flat case, interpreting the usual closure constraint for normals as a Bianchi identity for Freidel's spinning geometry connection (as defined in \cite{freidel_spinning_2014}). We then generalize this construction, still for flat tetrahedra and polyhedra, to non-abelian connections and closure relations. This procedure uses $\mathrm{SU}(2)$ connections and connections over the isometry group $\mathrm{ISU}(2)$ in order to get ready to its generalization to the curved case and hyperbolic tetrahedra. In the second section, we define a natural non-trivial $\mathrm{SL}(2,\mathbb{C})$ connection on the 3-hyperboloid and study its holonomy around hyperbolic triangles in order to decipher its geometrical meaning. It is found that the infinitesimal level behaves precisely as we expected, and the geometrical meaning of finite holonomies as normals to the triangles is discussed. In the final and third section, we apply this connection to the construction of an $\mathrm{SB}(2,\mathbb{C})$ closure constraint. We discuss the splitting of the natural $\mathrm{SL}(2,\mathbb{C})$ closure with respect to the Iwasawa decomposition and study the possibility of the reconstruction of the original hyperbolic tetrahedron.
Finally, we check that, in the framework of the $q$-deformed phase for loop quantum gravity \cite{dupuis_observables_2013,bonzom_deformed_2014, dupuis_deformed_2014}, the defined closure relation generates 3d rotations of the normals and of the hyperbolic tetrahedron as expected.


\section{Flat tetrahedron closure from holonomies}
\subsection{Closure as a Bianchi identity: a spinning geometry construction}

In this first section, we focus on the re-interpretation of the closure constraint of loop quantum gravity as a discrete Bianchi identity. Traditionally, we interpret the closure constraint as the geometric closure of convex polyhedra, or more specifically in the 4-valent case, as the geometric closure of tetrahedra. Indeed, given an arbitrary polyhedron, the sum of the outward oriented normal vectors of its faces (such that the norm of a vector gives the area of the corresponding face) vanishes. The reverse is warranted by Minkowski's theorem: given a set of vectors $\vec{N}_{i}\in\R^{3}$ summing to zero, there exists a unique convex polyhedron such that these are the normal vectors to its faces.
The goal of this section is to highlight the interpretation of the closure constraint as  a (discrete) Bianchi identity, as it was already put forward in \cite{freidel_spinning_2014} with the spinning geometry interpretation of loop quantum gravity's twisted geometry.

The key ingredient is the definition of the normals to the polyhedron faces as holonomies of a chosen connection around those faces. Once this first step is implemented, it becomes automatic that these ``normal holonomies'' satisfy a closure constraint (since the boundary of the polyhedron is a topologically trivial surface). Then we would like a homogeneous connection to ensure that the ``normal holonomies'' are related to the area of the face and that they record some notion of normal direction to the face.

In the flat case, we would like the ``normal holonomies'' to reproduce the standard notion of normal vectors to the polyhedron faces. Since the vectors are naturally valued in $\R^{3}$, we need to start with a  connection values in the abelian $\R^{3}$ Lie algebra and then define the normal vectors as the holonomies around each face.
This is exactly what was achieved by Freidel \textit{and al.} in their spinning geometry framework \cite{freidel_spinning_2014}.
The holonomy along each edge produces a vector. And the sum of those vectors along the boundary edges of each face gives the normal vector to that face. Finally, those normal vectors automatically satisfy a closure relation.
%

To this purpose, we define the following connection on flat space:
\begin{equation}
\label{eqn:Freidel}
A_\textrm{Fr}(\overrightarrow{x}) = \frac{1}{2} \epsilon^i_{~jk} T_i x^j \mathrm{e}^k
\end{equation}
Here $\mathrm{e}$ is the triad and the $T$'s are the generators of the $\mathbb{R}^3$ group. This connection lies in the Lie algebra of $\mathbb{R}^3$ and is commutative. There is some natural intuition to this connection: given a direction, we compute the cross product between the direction and a vector to some (arbitrarily chosen) origin. This is exactly the same construction as the usual construction of angular momentum. Of course, we will check shortly that the information of the connection does not depend on the choice of origin. This is quite intuitive when thinking in terms of angular momenta since the choice of origin should factor out and does not matter in the end. This will become obvious in our case once we will have checked that the holonomy around the closed loop bounding each face does indeed give the normal face which, of course, does not depend on the origin point.

So, how do we compute the holonomy around a close loop? Since the group $\R^{3}$ is commutative, the holonomy around a closed loop is equal to the integration of the curvature on (any of) the enclosed surface(s). In particular, if the loop is planar, there is a flat surface enclosed by it for which we can define the notion of a normal. Checking that the holonomy does give the normal amounts to checking that the curvature has an interpretation as an infinitesimal normal. We thus compute the curvature of the connection:
\begin{equation}
F[A_\textrm{Fr}] = \epsilon^i_{~jk} T_i \mathrm{e}^j \wedge \mathrm{e}^k
\end{equation}
At the infinitesimal level, the curvature of the connection, which corresponds to the holonomy around an infinitesimal surface, does indeed give the vector normal to the test surface, with its norm defining the area. And since the connection is commutative, this results  directly translates to the finite discrete case. Therefore, we found exactly what we were looking for: a connection such that its holonomies give the normal vector to the faces.

\begin{figure}[h!]

\centering

\begin{tikzpicture}[scale=2]
\coordinate (O) at (0,0,0);
\coordinate (A) at (1,1,-0.5);
\coordinate (B) at (0,2,0);

\fill[blue!75] (O) -- (A) -- (B) -- cycle;
\draw (O) node{$\bullet$} node[below left]{$O$};
\draw (A) node{$\bullet$} node[right]{$A$};
\draw (B) node{$\bullet$} node[above left]{$B$};

\draw (A) -- node[midway,rotate=135]{$>$}(B) -- (O) -- cycle;

\draw[->,>=stealth,very thick] (0.333, 1, -0.066) -- ++(-1,0,-1) node[left]{$\vec{M}$};
\end{tikzpicture}

\caption{The natural connection to construct the normal: the connection corresponds to the angular momentum of a particle moving along $AB$. Angular momentum naturally corresponds to the normal of $OAB$.}
\label{fig:AngularMomentum}
\end{figure}
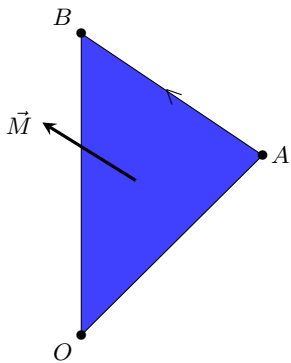

\begin{figure}[h!]

\centering

\begin{tikzpicture}[scale=1.5]
\coordinate (O) at (0,0,0);
\coordinate (B) at (-1,2,-1);
\coordinate (B2) at (-0.55,2,-0.8);
\coordinate (B1) at (-0.65,2,-0.7);
\coordinate (A) at (1,2,-1);
\coordinate (A2) at (0.6,2,-0.7);
\coordinate (A1) at (0.5,2,-0.8);
\coordinate (C) at (0,2,0.5);
\coordinate (C2) at (-0.075,2,0.2);
\coordinate (C1) at (0.0,2,0.2);

\fill[blue,opacity=0.25] (O) -- (A) -- (B) -- cycle;
\fill[blue,opacity=0.25] (O) -- (B) -- (C) -- cycle;
\fill[blue,opacity=0.25] (O) -- (C) -- (A) -- cycle;

\draw (A) -- (B) -- (C) -- cycle;
\draw (O) -- (A);
\draw (O) -- (B);
\draw (O) -- (C);

\draw[blue,rounded corners,very thick] (A1) -- (B2) -- (B1) -- (C2) -- (C1);
\draw[blue,very thick,->,>=stealth] (C1) -- (A2);

\draw (O) node{$\bullet$} node[below]{$O$};
\draw (A) node{$\bullet$} node[above]{$A$};
\draw (B) node{$\bullet$} node[above]{$B$};
\draw (C) node{$\bullet$} node[below right]{$C$};

\draw[->,>=stealth,very thick] (0.333, 1.333, -0.166) -- ++(0.75,-0.125,0.25);
\draw[->,>=stealth,very thick] (-0.333, 1.333, -0.166) -- ++(-0.75,-0.125,0.25);
\end{tikzpicture}

\caption{Thanks to the usual closure, the normal of the triangle $ABC$ is, up to a sign, the sum of the normals from $OAB$, $OBC$ and $OCA$. Therefore, the normal of $ABC$ can be computed from Freidel's spinning geometry connection \eqref{eqn:Freidel}.}
\label{fig:Normal}

\end{figure}
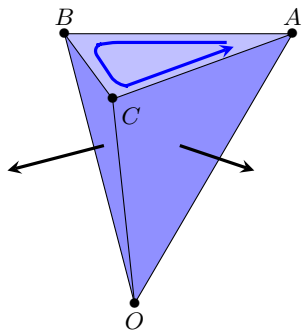

Let us work this out explicit for a polyhedron and see how the connection point of view fits with the usual  geometric construction.
%
We first evaluate the (open) holonomy along a line segment. The holonomy along oriented segment $\overrightarrow{AB}$ is:
\begin{equation}
\overrightarrow{v_{AB}}
= \frac{1}{2} \overrightarrow{OA}\times\overrightarrow{AB}
=-\overrightarrow{v_{BA}}\,,
\end{equation}
where $O$ is the space origin used in the definition of the connection. This holonomy is actually the normal vector to the triangle $OAB$ as illustrated on fig.\ref{fig:AngularMomentum}.
%
%
Adding up these contributions for all the boundary edges of a (planar) polygon then gives the normal vector to that polygon.
The closure condition of the cone whose base is the face and whose vertex is the origin $O$  implies that the normal to the face is, up to a sign, the sum of all the contributions from the segments as illustrated on fig.\ref{fig:Normal}. Of course, we can check this by an explicit computation. The holonomy for a triangle is:
\begin{equation}
\overrightarrow{N_{ABC}} = \overrightarrow{v_{AB}} + \overrightarrow{v_{BC}} + \overrightarrow{v_{CA}} = \frac{1}{2} \overrightarrow{AB}\times\overrightarrow{BC}\,,
\end{equation}
which is indeed the normal to the triangle.

Now we can check the closure condition for a polyhedron. For simplicity's sake, we focus on the tetrahedron. Of course, we already know that the normal vectors to a polyhedron faces sum up to zero, but we would like to illustrate how the connection point of view naturally leads to the closure constraint as a discrete Bianchi identity. It comes down to a very simple realization: the closed holonomies around the faces are built from the open holonomies along the edges; so that summing over all the faces forces to run through each edge twice in opposite direction thus leading to a cancellation of each edge contribution. So, for a tetrahedron $(ABCD)$, this goes as:
\beq
\overrightarrow{N_{ABC}}+\overrightarrow{N_{BDC}}+\overrightarrow{N_{CDA}}+\overrightarrow{N_{ADB}}
&=&
(\overrightarrow{v_{AB}} + \overrightarrow{v_{BC}} + \overrightarrow{v_{CA}})+
(\overrightarrow{v_{BD}} + \overrightarrow{v_{DC}} + \overrightarrow{v_{CB}}) \nn\\
&&+
(\overrightarrow{v_{CD}} + \overrightarrow{v_{DA}} + \overrightarrow{v_{AC}})+
(\overrightarrow{v_{AD}} + \overrightarrow{v_{DB}} + \overrightarrow{v_{BA}}) \nn\\
&=&0\,.
\eeq
Taken to the infinitesimal limit, this closure relation gives the usual Bianchi identity. In fact, this relation is of course well-known in discrete geometry. Given a surface, which is homeomorphic to the sphere, and a graph on it, which is therefore a planar graph, if we define a vector quantity for each face of the graph such that the total sums up to zero, we can decompose the quantities on the edges, up to an addition at the graph vertices. Their is no cocycle contribution, since the graph is planar. This is just the discrete equivalent of the usual:
\begin{equation}
\mathrm{d}F=0 \Leftrightarrow F=\mathrm{d}A
\end{equation}
as soon as the embedding space is simply connected (which is of course the case for a polyhedron which is homeomorphic to the sphere).

\medskip

Introducing such a connection (or its equivalent in discrete terms) makes some relations trivial (like the closure condition) but raises new questions. This is similar to going from Maxwell's equation, governing the electric and magnetic field, to a more modern and covariant way of writing electrodynamics using the gauge connection of $\mathrm{U}(1)$. In particular, the gauge transformation does not affect the electromagnetic fields, since the gauge group is abelian, but they do affect the connection (and the matter fields) and even have an interpretation in them, namely as transformation of the phase of the charged particles. Now that we have a connection to construct normals, it is natural to wonder what the gauge transformations correspond to. We should notice that, as for electromagnetism, the gauge group is abelian, and therefore the holonomies do not transform under gauge transformation. This is reflected in the fact that there is also no parallel transport needed in any of our constructs so far. This implies that the gauge transformation will not correspond to some rotation of space (otherwise they would transform the normals) or, except for a few possible - but implausible - exceptions (translations), to any geometrical transformation of the normals.

Let's dive a bit more into the precise transformations. Let's gauge transform our connection by a vector shift $\overrightarrow{\phi_{P}}$ at each point of space $P$. The fields transform as follows:
\begin{equation}
\left\{\begin{array}{rcl}
A_\textrm{Fr} &\rightarrow& A_\textrm{Fr} + d\phi^i T_i \\
\overrightarrow{v_{AB}} &\rightarrow& \overrightarrow{\phi_B} + \overrightarrow{v_{AB}} - \overrightarrow{\phi_A}
\end{array}\right.
\end{equation}
%
Do these transformations have a simple geometrical interpretation? Although they seem at first sight to be related to translation, this is not entirely the case. Indeed, let us consider a change of the origin point and distinguish the holonomies $\overrightarrow{v_{AB,O}}$ and $\overrightarrow{v_{AB,O'}}$ defined as the integrated connection along the oriented segment $AB$ but with different reference points $O$ and $O'$. We have:
\begin{equation}
\overrightarrow{v_{AB,O}} = \frac{1}{2} \overrightarrow{OA}\times\overrightarrow{AB} = \frac{1}{2} \overrightarrow{OO'}\times\overrightarrow{OB} + \overrightarrow{v_{AB,O'}} - \frac{1}{2}\overrightarrow{OO'}\times\overrightarrow{OA}
\end{equation}
So a change of origin from $O$ to $O'$ does correspond to a gauge transform with $\overrightarrow{\phi_P} = \frac{1}{2}\overrightarrow{OO'}\times\overrightarrow{OP}$. But this does not cover all the possible gauge transformations: only terms orthogonal to $\overrightarrow{OP}$ appear.


\subsection{Non-abelian normals}

Though our goal is to study the hyperbolic case, we  continue developing a few more new technics for the flat case. In particular, the spinning geometry construction of \cite{freidel_spinning_2014}, though elegant, is abelian and a number of subtleties, that will show up in the curved case, are absent. So, in this section, we will investigate a non-abelian connection, still on the flat space, in order to define a non-abelian extension of spinning geometries. The goal is to have a kind of toy-model to sharpen our intuition and understanding of the geometry linked to what we might call non-abelian normals but  still in a well-known controlled environment.

A non-abelian connection, in flat $\mathbb{R}^3$, would give us a definition of non-abelian normals. Such a proposal seems counter-intuitive at first. Indeed, the non-commutative nature of a connection is naturally connected with non-trivial  parallel transports and non-vanishing curvature. When investigating  flat space geometry, we usually expect to avoid such complications.
%
%
But we expect such a construction to be a first step towards naturally building non-abelian normals in the hyperbolic case.

So we want the normals to be invariant under translation and well-behaved under 3d rotations, which means looking for a homogeneous and isotropic connection transforming properly under the action of the rotation group.
%
%
This is a priori a non-trivial task.
%
%
But we will use the specificity of working in three dimensions and 
use exactly the same trick as in the construction of the Ashtekar-Barbero connection.
There is a canonical map between the rotation group generators and the directions in tangent space, and this map is defined by the triad. Explicitly, we define the following $\su(2)$-valued connection:
\begin{equation}
A_\textrm{nc} = \Gamma + a J_i \mathrm{e}^i = a J_i \mathrm{e}^i\,,
\label{Anc}
\end{equation}
where the $J$'s are the rotation generators and $a$ is an arbitrary real coefficient and the subscript ``nc'' stands for ``non-commutative''. The term $\Gamma$ is the usual spin connection and is required to ensure the correct behavior under gauge transformations. Nevertheless, 
as we fix the gauge to $e^{i}_{\mu}=\delta^{i}_{\mu}$,  $\Gamma$ is sent to $0$.

Geometrically, this connection gives a twist in the direction of the progression. This is the most intuitive notion of torsion.
Actually computing the torsion gives:
\begin{equation}
\mathrm{d}_{A_\textrm{nc}} \mathrm{e}^i = \mathrm{d} \mathrm{e}^i + a \epsilon^i_{~jk} \mathrm{e}^j \wedge \mathrm{e}^k = a \epsilon^i_{~jk} \mathrm{e}^j \wedge \mathrm{e}^k\,.
\end{equation}
The first term vanishes in the gauge we chose, but the second does not since $\epsilon^i_{~jk} A^j_\mu \mathrm{e}^k_\nu \neq 0$, so that 
the connection is no longer compatible with the triad.
%

\medskip

This new connection $A_\textrm{nc}$ is still somewhat related to the original abelian connection $A_\textrm{Fr}$, so that the non-abelian normal it defines are not completely disconnected to the usual normal vectors. Let us start by computing its curvature:
\begin{equation}
F[A_\textrm{nc}] = a^2 \epsilon^i_{~jk} J_i \mathrm{e}^j \wedge \mathrm{e}^k\,.
\label{curvatureAnc}
\end{equation}
We should remember here that the curvature of a connection gives the first order of its holonomy around infinitesimal surfaces. Getting out the $a^2$ factor and forgetting that $J_i \neq T_i$, we see here that the curvature of $A_\textrm{nc}$ is the same as the curvature of $A_\textrm{Fr}$. In more precise terms, it means that at the infinitesimal level, the holonomy precisely encodes the normal as a rotation around the normal axis. The angle is proportional to the area (we do not worry about compactness at the infinitesimal level). In a sense, we see here that $A_\textrm{Fr}$ encodes the first order of $A_\textrm{nc}$. But maybe, that's not that surprising when we consider that the cross product (crucial in the definition of $A_\textrm{Fr}$) naturally gives the infinitesimal transformation under rotation.

Let us go further and study the deviation from the usual normal. The holonomy of $A_\textrm{nc}$ can actually be computed exactly. Considering a triangle $ABC$, the closed holonomy around the triangle is the composition of three open holonomies corresponding to each edge. For an (oriented) edge $\overrightarrow{AB}$, the holonomy $g_{AB}$ is computed exactly as:
\begin{equation}
g_{AB} = \exp\left(\frac{\mathrm{i}a}{2} \overrightarrow{AB} \cdot \overrightarrow{\sigma}\right)
\end{equation}
where the $\sigma$'s are the Pauli matrices. The holonomy around the full triangle $ABC$ is then simply\footnotemark{}~:
\footnotetext{
Note here that the edge holonomies and even the closed holonomies depend on the particular chosen gauge, since the gauge group is non-abelian. Only the trace of closed holonomies is fully gauge-invariant.
}
\begin{equation}
h_{ABC} = g_{CA} g_{BC} g_{AB} = \exp\left(\frac{\mathrm{i}a}{2} \overrightarrow{CA} \cdot \overrightarrow{\sigma}\right) \exp\left(\frac{\mathrm{i}a}{2} \overrightarrow{BC} \cdot \overrightarrow{\sigma}\right) \exp\left(\frac{\mathrm{i}a}{2} \overrightarrow{AB} \cdot \overrightarrow{\sigma}\right)
\label{eq:prod}
\end{equation}
So one can consider $h_{ABC}$ as our new notion of non-abelian normal to the triangle. But in order to compare it to the usual normal vector, we can actually extract a notion of deformed normal from it.
A natural definition is to take the $\su(2)$ element generating it, which can be identified to a vector given by the rotation axis times the rotation angle. This gives  the vector $\vec{n}^a$ defined as:
\begin{equation}
h_{ABC} = \exp\left(\frac{\mathrm{i}a^2}{2} \vec{n}^a \cdot \overrightarrow{\sigma}\right)\,.
\end{equation}
Note here that we introduced a squared factor $a^2$, rather than just a linear factor in $a$.
%
%
Indeed the leading order in $a$ in the expansion of $h_{ABC}$ vanishes, as expected from the curvature formula above \eqref{curvatureAnc}. Second, we do want $\vec{n}^{a}$ to be like a normal vector whose norm gives the area of the surface. Since $a$ is an inverse length scale, it is natural to define  $\vec{n}^{a}$ with the appropriate dimension of a squared length.

We do not provide the full exact formula for $\vec{n}^a$, which is straightforwardly computable but cumbersome to interpret geometrically.
 Let's rather see the first deviation from the usual normal in order of $a$.
 Expanding the exponentials in terms of the deformation parameter $a$, the first order linear in $a$ involves the sum of the edges and vanishes (as a closure condition for the triangle). The second order gives back the exact normal vector and then we get higher order corrections:
%
%
\begin{equation}
h_{ABC} = \id + \frac{\mathrm{i}a^2}{4}\overrightarrow{n}\cdot \overrightarrow{\sigma} + \frac{\mathrm{i} a^3}{12}\left( \overrightarrow{CA}^2 \overrightarrow{CA} + \overrightarrow{BC}^2 \overrightarrow{BC} + \overrightarrow{AB}^2 \overrightarrow{AB} \right)\cdot \overrightarrow{\sigma} + \mathcal{O}(a^4)\,.
\end{equation}
And we find a deformed normal vector, with a first geometrical correction:
\begin{equation}
\vec{n}^a \simeq \vec{n} + \frac{a}{3}\left({CA}^2 \overrightarrow{CA} +{BC}^2 \overrightarrow{BC} + {AB}^2 \overrightarrow{AB} \right)+ \mathcal{O}(a^2)\,.
\end{equation}
Though this added term is not a usual geometrical observable for a triangle , it shows that the shape of the triangle influences our deformed normal.
On top of this, we naturally expect effects due to the compactness and topology of $\mathrm{SU}(2)$. For instance, a triangle with lengths of integer multiples of $\frac{2\pi}{a}$ necessarily has a trivial holonomy. This is due to the periodic nature of $\mathrm{SU}(2)$ which, therefore, cannot distinguish all the triangles.
%

\subsection{Non-abelian closure constraint}

Let us go on with showing how, once the connection is given,  the closure constraint for the non-abelian normals is guaranteed by the Bianchi identity. Given any polyhedron, we associate to each of its face a holonomy $h_f$. We will have a closure condition resembling:
\begin{equation}
h_n ... h_2 h_1 = \id
\end{equation}
There is however a subtlety here and it is linked to parallel transport. Indeed, we must remember that this connection is not abelian and therefore the $h$'s need to be appropriately transported.

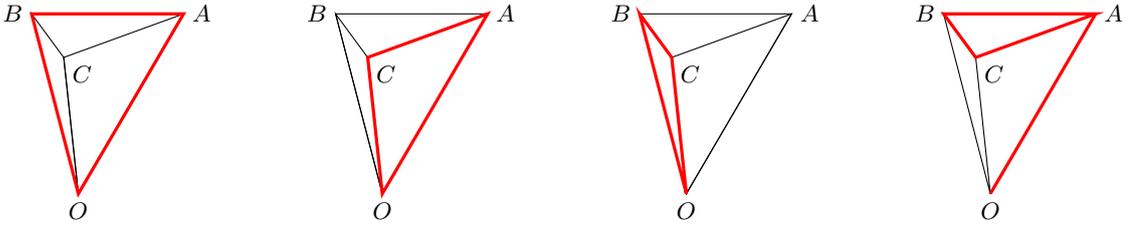
\begin{figure}[h!]

\centering

\begin{tikzpicture}

\coordinate (O1) at (0,0,0);
\coordinate (A1) at (1,2,-1);
\coordinate (B1) at (-1,2,-1);
\coordinate (C1) at (0,2,0.5);

\draw (O1) node[below]{$O$};
\draw (A1) node[right]{$A$};
\draw (B1) node[left]{$B$};
\draw (C1) node[below right]{$C$};

\draw (O1) -- (B1) -- (C1) -- cycle;
\draw (O1) -- (C1) -- (A1) -- cycle;
\draw[red,very thick] (O1) -- (A1) -- (B1) -- cycle;

\coordinate (O2) at (4,0,0);
\coordinate (A2) at (5,2,-1);
\coordinate (B2) at (3,2,-1);
\coordinate (C2) at (4,2,0.5);

\draw (O2) node[below]{$O$};
\draw (A2) node[right]{$A$};
\draw (B2) node[left]{$B$};
\draw (C2) node[below right]{$C$};

\draw (O2) -- (B2) -- (C2) -- cycle;
\draw (O2) -- (A2) -- (B2) -- cycle;
\draw[red,very thick] (O2) -- (C2) -- (A2) -- cycle;

\coordinate (O3) at (8,0,0);
\coordinate (A3) at (9,2,-1);
\coordinate (B3) at (7,2,-1);
\coordinate (C3) at (8,2,0.5);

\draw (O3) node[below]{$O$};
\draw (A3) node[right]{$A$};
\draw (B3) node[left]{$B$};
\draw (C3) node[below right]{$C$};

\draw (O3) -- (C3) -- (A3) -- cycle;
\draw (O3) -- (A3) -- (B3) -- cycle;
\draw[red,very thick] (O3) -- (B3) -- (C3) -- cycle;

\coordinate (O4) at (12,0,0);
\coordinate (A4) at (13,2,-1);
\coordinate (B4) at (11,2,-1);
\coordinate (C4) at (12,2,0.5);

\draw (O4) node[below]{$O$};
\draw (A4) node[right]{$A$};
\draw (B4) node[left]{$B$};
\draw (C4) node[below right]{$C$};

\draw (O4) -- (B4) -- (C4) -- cycle;
\draw[red,very thick] (O4) -- (A4);
\draw[red,very thick] (A4) -- (B4) -- (C4) -- cycle;

\end{tikzpicture}

\caption{When defining the holonomies for the closure constraint, we need to use the same reference point (root) for each holonomy. In the case of a tetrahedron for instance, this means that the last holonomy must have some parallel transport along an edge as shown on the forth figure.}
\label{fig:Transport}

\end{figure}

To start with, we define a reference point from which we will define all the holonomies.
For a tetrahedron, we choose an arbitrary point as origin. Three faces share it and so that origin point can be used as the root for the holonomies around those three faces. For the remaining face, we need to chose an edge for parallel transport (see fig.\ref{fig:Transport}). This edge must start at the origin and end at one of the vertices of the last face and therefore do not belong to it. We must also be careful to compose in the right order so that this parallel transport is cancelled. This is exactly the same procedure as the one  described in the previous work \cite{charles_closure_2015}. This procedure can be thought of as a gauge fixing on the polyhedron\footnotemark{}.
\footnotetext{
A gauge fixing needs a choice of origin. This is actually related to the necessary choice of origin in order to define the abelian spinning geometry connection $A_\textrm{Fr}$ given in \eqref{eqn:Freidel}. The non-abelian transport is  done through conjugation by holonomies. This is exactly the same operation as we would do for a gauge-transformation. In the previous case of $A_\textrm{Fr}$, gauge transformations was at least partially linked to a change of origin. Here, in the non-abelian context, it is implemented here in a much more concrete sense as an origin \textit{must} be selected to even define the holonomies. So even though the choice of an origin for space is not required to define the connection $A_\textrm{nc}$, it is still nevertheless needed to define the holonomies and all the non-abelian normals.
}

\medskip

The existence of such an $\mathrm{SU}(2)$ connection opens several interesting question. We might wonder, for instance, if we could develop a unifying framework encompassing any curvature for describing tetrahedron in a curved background. This would be done by deforming in a similar manner the usual $\mathrm{SU}(2)$ connection on hyperbolic and spherical manifolds and see if some condition, as the sign of the Gram matrix as defined in \cite{haggard_encoding_2015}, can allow the detection of the curvature and therefore allow some reconstruction. Contrary to the framework developed in the last reference, the non-triviality of the connection in the flat case would allow the reconstruction, at least partially since we have topological obstruction, even in the flat case. We will leave, however, this question for further investigation.

\medskip

A final thought concerns the free parameter $a$. Mathematically, we use it as the length scale entering the definition of the connection $A_\textrm{nc}$ and the deformation parameter for our non-abelian normals. It could also turn out useful when dealing with the coarse-graining of discrete geometries. We have a whole family of connections and consequently of geometrical closure relations, which could be relevant at different scales. For instance, \cite{Jacobson:2007uj} argued that the natural geometrical operators can evolve with the scale at which we probe the geometry. Here a set of four vectors which do not exactly sum to zero (i.e carrying a closure defect in the context of the coarse-graining of loop quantum gravity \cite{livine_deformation_2014,Charles:2016xwc}) may still be interpreted as defining the deformed normals of a closed flat tetrahedron for a specific fine-tuned choice of $a$. We will comment more on this in the hyperbolic case where the parameter $a$ becomes the Immirzi parameter.
%

\subsection{Duality and $\mathrm{ISU}(2)$ closure constraint}
\label{isu-closure}
\label{isu2}

Up to now, using connections and the re-interpretation of closure constraints as discrete Bianchi identities, we have recovered the usual flat abelian closure constraints for the tetrahedron (and polyhedron) in terms of standard normal vectors in $\R^{3}$  and a new non-abelian closure constraints in terms of new deformed normals valued in $\SU(2)$. We show below that we unify these two point of views by using a  $\mathrm{ISU}(2)$ connection. This will lead to a $\ISU(2)$ closure constraints, whose rotational part defines a non-abelian compact $\SU(2)$ closure constraint while its translational component produces a non-compact $\R^{3}$ closure constraint.

This is also very similar to our strategy to tackle the hyperbolic tetrahedron, and hyperbolic  polyhedra, in the next sections, where we will introduce a $\SL(2,\C)$ connection and use it to derive a $\SL(2,\C)$ closure constraints, which we will be able to split into two equivalent dual closure constraints with deformed normals living in the compact $\SU(2)$ Lie group or alternatively in the non-compact $\SB(2,\C)$ Lie group.

%

So, we use once more the triad to now define a $\isu(2)$ connection:
\begin{equation}
A_\textrm{ISU} = a J_i \mathrm{e}^i + b T_I \mathrm{e}^I
\end{equation}
where we mix both rotation and translation generators. $a$ and $b$ are two real parameters. The $a$ parameter is the same as in the $\SU(2)$-connection $A_\textrm{nc}$. So  our new connection is  a straightforward generalization of  $A_\textrm{nc}$ which we recover where $b$ is set to $0$.

The curvature of $A_\textrm{ISU}$ is still homogeneous:
\begin{equation}
F[A_\textrm{ISU}] = ab \epsilon^i_{~jk} T_i \mathrm{e}^j \wedge \mathrm{e}^k + b^2 \epsilon^I_{~JK} J_I \mathrm{e}^J \wedge \mathrm{e}^K
\end{equation}
Naively splitting into $\mathbb{R}^3$ and $\su(2)$ components, we see that at the infinitesimal level, the curvature is really the same as for $A_\textrm{Fr}$ and $A_\textrm{nc}$ apart from specific coefficients.
The curvature is still proportional to $e\wedge e$, so that the $\ISU(2)$-holonomies around triangles (and more generally polyhedron faces)will again encode the normal, at least, at the infinitesimal level, though we of course expect non-abelian deformations at the finite level. This curvature formula also highlights in which way $A_\textrm{ISU}$ is a generalization of $A_\textrm{Fr}$. If we send $b$ to $0$ while keeping $ab$ constant, the curvature tends to the curvature of $A_\textrm{Fr}$. So the two-parameter family of $\isu(2)$-connections in terms of $a$ and $b$ is actually an interpolation between the  previous two one-parameter family of connections valued in $\R^{3}$ and $\su(2)$.

As before, given a polyhedron, we choose a point as root and define the holonomies around every face of the polyhedron using that root as starting point. These holonomies are interpreted as new deformed normals to the faces. They naturally satisfy a discrete Bianchi identity, which defines a $\ISU(2)$-closure constraint between the normals:
\begin{equation}
g_n ... g_2 g_1 = \id
\end{equation}
As $\mathrm{ISU}(2)$ is the semi-direct product of $\mathbb{R}^3$ and $\mathrm{SU}(2)$, any $\mathrm{ISU}(2)$ group element has a unique decomposition, $g_i = (N_i, h_i)$, where $N_i \in \mathbb{R}^3$ and $h_i \in \mathrm{SU}(2)$.
We used the same notation as in the previous sections to highlight but those $N$'s and $h$'s do not match the holonomies of the $\R^{3}$ and $\su(2)$-connections. Since it is a semi-direct product, we indeed have a mixing between the rotational and translational components of the $\isu(2)$-connection when computing its holonomies.

We apply this decomposition to the closure constraint:
\begin{equation}
g_n ... g_2 g_1 = \id \Leftrightarrow \left\{\begin{array}{lcc}
N_n + h_n \triangleright N_{n-1} + ... + (h_n ... h_3) \triangleright N_2 + (h_n ... h_3 h_2) \triangleright N_1 = \overrightarrow{0} &&
\vspace*{2mm}\\
h_1 h_2 ... h_n =\id&&
\end{array}\right.
\end{equation}
where $\triangleright$ represents the natural action of $\mathrm{SU}(2)$ as 3d rotations  acting on vectors in  $\mathbb{R}^3$. So we obtain two dual closure constraints, one in terms of $\R^{3}$ vectors $\tilde{N}_{i}=(h_{n}..h_{i+1})\triangleright N_{i}$ and one in terms of $\SU(2)$ group elements $h_{i}$.
Some braiding  appears mixing  rotations and translations, and we have a similar behavior in the hyperbolic case in the next section.

Finally, we recover the previous pure $\R^{3}$ and $\SU(2)$-closure constraints in the two limit cases, respectively when $b$ goes to $0$ while keeping $ab$ constant  or $a$ is sent to 0.


\section{$\SL(2,\C)$ Holonomies and Hyperbolic Triangles}

\subsection{Hyperbolic space, Geometry and Lorentz connections}

In this section, we now turn to the hyperbolic case and focus on the definition of an $\sl(2,\mathbb{C})$ connection on the 3d hyperboloid $\mathcal{H}$. We will study of the holonomies of such a connection  around hyperbolic triangles\footnotemark{} in order to later use it to derive closure constraints for the hyperbolic tetrahedron.
\footnotetext{
Incidentally, holonomies of Lorentz connections on the 3d-spacelike hyperboloid around closed circular loops has already been studied and analyzed in e.g. \cite{charles_ashtekar-barbero_2015}.}
The upper sheet 3d hyperboloid of curvature radius $\kappa$ is the set of points $(t,x,y,z)$ in $\mathbb{R}^{3,1}$ such as:
\begin{equation}
t^2 - (x^2 + y^2 + z^2) = \kappa^2,\quad t>0
\end{equation}
This gives the natural embedding of the 3d hyperboloid into Minkowski space. Since this quadratic equation produces a two-sheet hyperboloid, we select the upper one with the condition $t>0$.
We use the identification of the Minkowski space $\mathbb{R}^{3,1}$ to the space of Hermitian two by two complex matrices, with the  Lorentzian pseudo-norm mapped to matrix determinant:
\begin{equation}
M = \begin{pmatrix}
t+z & x-\mathrm{i}y \\
x + \mathrm{i}y & t-z
\end{pmatrix}
\,,\qquad
\det M = t^2- (x^2 + y^2 + z^2)
\end{equation}
The hyperboloid $\mathcal{H}$ is then identified as a constant determinant subspace:
\begin{equation}
\mathcal{H} \simeq \{M \in \mathrm{H}_2(\mathbb{C}) / \det M = \kappa^2\,\, \& \,\,\tr M > 0\}\,.
\end{equation}
In this representation, the Lorentz group $\mathrm{SL}(2,\mathbb{C})$ simply acts by conjugation on the Hermitian matrices:
\begin{equation}
\forall \Lambda \in \mathrm{SL}(2,\mathbb{C}), \forall M \in \mathrm{H}_2(\mathbb{C}), \Lambda \triangleright M = \Lambda M \Lambda^\dagger
\end{equation}
This action preserves the determinant, as well as the sign of the trace and therefore acts naturally on the hyperboloid.

Let's now define an $\sl(2,\mathbb{C})$ connection on this hyperboloid.
This is the equivalent of the $\isu(2)$ connection that we defined in the flat case in the previous section.
Here we do not want a connection with trivial holonomies, so we will not use the most natural Lorentz connection on the hyperboloid, defined as the pull-back of the flat space-time connection on the four-dimensional Minkowski space.
Instead, we exploit the fact that $\mathrm{SL}(2,\mathbb{C})$ is the complexification of $\mathrm{SU}(2)$, so that a $\sl(2,\mathbb{C})$ connection can be thought of as a complex $\su(2)$ connection.
Let us write as before  $J_1$, $J_2$, $J_3$, for the $\su(2)$ generators (represented by the Pauli matrices up to a $\f12$ factor). Then the boost generators can be identified as $B_I$ are the boosts generators and we used $B_I = \mathrm{i} J_I$.
We define the following connection on the hyperboloid:
\begin{equation}
A_\textrm{SL} = \left(\Gamma^i + \frac{\beta}{\kappa} \mathrm{e}^i\right)J_i
= \left(\Gamma^i + \frac{\Re (\beta)}{\kappa} \mathrm{e}^i\right) J_i + \frac{\Im (\beta)}{\kappa} \mathrm{e}^I B_I\,.
\label{eq:connection}
\end{equation}
where  $\beta \in \mathbb{C}$ is a free parameter and $\Gamma^i$ is the unique spin connection on the hyperboloid compatible with the metric (and the triad) and without torsion. The $\kappa$ parameter is put here to keep $\beta$ dimensionless.

\medskip

This connection exactly matches the Ashtekar-Barbero connection on the 3d hyperboloid embedded in the flat Minkowski space-time, when $\beta$ is the Immirzi parameter. They are nevertheless subtly different in their definition. Indeed, the Ashtekar-Barbero connection $A_\textrm{A-B}$ crucially depends on the embedding of the spatial slice and its extrinsic curvature in space-time, while our connection $A_\textrm{SL}$ is entirely defined intrinsically in terms of the 3d triad field. For instance, if we were to change the embedding map of the 3-hyperboloid, to the Anti-de-Sitter space-time for example, the Ashtekar-Barbero connection would change while the new connection $A_\textrm{SL}$ would remain the same. However, for the specific embedding of the  hyperboloid in the flat Minkowski space, the extrinsic curvature equals the triad:
\begin{equation}
\frac{\mathrm{e}^i}{\kappa} = K^i\,,
\end{equation}
and $A_\textrm{SL}$ is equal to the Ashtekar-Barbero connection with a complex Immirzi parameter.
%
%
We compute the curvature of $A_\textrm{SL}$,
\begin{equation}
F[A_\textrm{SL}] = \frac{1+\beta^2}{\kappa^2} \epsilon^i_{~jk} J_i \mathrm{e}^j \wedge \mathrm{e}^k\,,
\qquad
F^{ij}[A_\textrm{SL}] = \Lambda \mathrm{e}^i \wedge \mathrm{e}^j\,,
\quad\textrm{with}\quad
\Lambda = \frac{1+\beta^2}{\kappa^2}
\end{equation}
Interpreted from the point of view of a $\su(2)$-connection in canonical loop quantum gravity, this can be interpreted as an effective complex cosmological constant $\Lambda =(1+\beta^2)/\kappa^2$. From a geometrical point of view, the fact that $F\propto e\wedge e$ ensures that the interpretation of the holonomies around triangles as normals, as in the flat case analyzed in the previous section. This allows us to interpret this connection $A_\textrm{SL}$ as defining the generalization of Freidel's spinning geometry to the hyperbolic case.

\smallskip

The precise behavior of $A_\textrm{SL}$ of course depends on the value of $\beta$ (or equivalently on the value of $\Lambda$). Here is a short list of interesting values for the Immirzi parameter and the corresponding properties of the connection:
\begin{itemize}
\item First, as for the Ashtekar-Barbero variables \textit{per se}, the values $\beta = \pm \mathrm{i}$ are very specific and induce very specific properties. In that case, the connection is the (anti-)self-dual $\mathrm{SL}(2,\mathbb{C})$ connection, which can be understood as the natural connection induced by the flat Minkowski connection. This connection is entirely flat, as can be seen from the value of $\Lambda = 0$. In particular, no information at all is preserved in the holonomies and all holonomies around closed loops (and triangles in particular) are the identity. So we obtain a trivial closure relation.


\item If $\beta \in \mathbb{R}$, the connection is pure $\mathrm{SU}(2)$. This is the standard choice in loop quantum gravity in order to avoid complex fields and the issues of reality conditions.
This gives us a generalization of $A_\textrm{nc}$ defined in \eqref{Anc}  for the flat case, which led to non-abelian closure constraints for the flat tetrahedron in terms of deformed normals defined as $\SU(2)$ holonomies.
Here, the special case $\beta = 0$ corresponds to the usual metric-compatible torsion-free spin-connection. And we have a whole family of  $\su(2)$ connections  which will lead to closure constraints for the hyperbolic tetrahedron in terms of normals living $\SU(2)$, similarly to what was developed in \cite{charles_closure_2015}. 

\item One particularly interesting choice is $\Lambda \in \mathrm{i}\mathbb{R}$, that is $(1 + \beta^2$) purely imaginary. 
The curvature and thus infinitesimal holonomies (around closed loops) are pure boosts. This can be interpreted in some sense as the orthogonal counterpart of a pure $\su(2)$-connection.
%
%
This would correspond to the reduction to the translational part of the connection $A_\textrm{ISU}$ in the flat case. 
%
%

\item Our last interesting values for the Immirzi parameter $\beta$  live on shifted real line  $\beta = \lambda\pm\mathrm{i}$ with  $\lambda \in \mathbb{R}$. This is a yet-to-be-explored sector for loop quantum gravity, but it has  a very natural geometrical interpretation in our context as we will see and we believe that it might offer new possibilities for loop quantum gravity.

So let us consider two points on the hyperboloid. The first point is mapped onto the second by a $\mathrm{SL}(2,\mathbb{C})$ transformation and the geodesic is generated by the corresponding $\sl(2,\C)$ vector. However such a Lorentz transformation is not unique, the stabilizer group of a point on the hyperboloid is the rotation subgroup $\SU(2)$ and we can indeed compose the map with an arbitrary rotation around the final point. The usual section is to choose pure boosts.

Nevertheless, another interesting construction is to set the rotation axis to be aligned with the boost axis. This corresponds to computing the holonomy along the geodesic of the  connection $A_\textrm{SL}$ with $\beta =  \lambda \pm {i} $. The parameter $\lambda\in\R$  is a helix parameter telling us how much we wind  around the boost axis. The geometrical resemblance with spinning geometry is cunning.
The equivalent in the flat case is to  set $b=\pm 1$ with $a$ left free in $A_\textrm{ISU}$.
\end{itemize}

\subsection{Holonomy around a finite hyperbolic triangle}

\begin{figure}[h!]

\centering

\begin{tikzpicture}[scale=1]

\coordinate (A) at (0,-1);
\coordinate (B) at (2,2);
\coordinate (C) at (-2,2);

\draw (A) to[bend left=15] node[midway,right]{$l_{AB}$} (B) to[bend left=15] node[midway,above]{$l_{BC}$} (C) to[bend left=15] node[midway,left]{$l_{AC}$} (A);
\draw (A) node[below]{$A$};
\draw (B) node[above right]{$B$};
\draw (C) node[above left]{$C$};

\draw ($(A)+(70:0.5)$) arc (70:110:0.5);
\draw ($(A)+(90:0.75)$) node{$\hat{a}$};
\draw ($(B)+(-167:0.5)$) arc (-167:-137:0.5);
\draw ($(B)+(-152:0.75)$) node{$\hat{b}$};
\draw ($(C)+(-43:0.5)$) arc (-43:-13:0.5);
\draw ($(C)+(-28:0.75)$) node{$\hat{c}$};

\end{tikzpicture}

\caption{Notations for a hyperbolic triangle: each angle is named after the corresponding point, the lengths are labeled by the end points of the edges. The lengths are given by the boost parameters, $l_{AB}=\ka\eta_{AB}$ for the edge $(AB)$.}
\label{fig:Notations}

\end{figure}
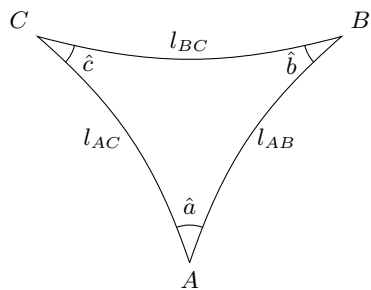

Let us now look at the holonomies of the $\sl(2,\C)$-connection $A_\textrm{SL}$ on the hyperboloid and compute the holonomies around closed hyperbolic triangles. The goal is to get an explicit formula for this holonomy and explain how it represents a notion of $\SL(2,\C)$-valued normal to the triangles. We start with a triangle $(ABC)$ as depicted in fig.\ref{fig:Notations}.
To compute explicitly the holonomies along the three edges of the triangle, we use the homogeneity of the hyperboloid. This allows to place, say, the point $A$ at the origin. Then we use the invariance of the triad and connection under rotation in order to place the whole triangle in the equatorial plane (at $z=0$). In the end, one can rotate back to an arbitrary triangle.

It is much simpler to use spherical coordinates parametrizing an arbitrary point on the hyperboloid in terms of the boost parameter $\eta\in\R_{+}$ and the angles $\theta\in\,[0,\pi]$ and $\phi\in\,[0,2\pi]$:
\be
\left\{
\begin{array}{lcl}
t&=&\ka\cosh\eta\\
x&=&\ka\sinh\eta\sin\theta\cos\phi \\
y&=&\ka\sinh\eta\sin\theta\sin\phi \\
z&=&\ka\sinh\eta\cos\theta \\
\end{array}
\right.
\ee
The metric on the hyperboloid induced by the flat 4d metric is the standard homogeneous hyperbolic metric $q_{ab}$:
\be
\mathrm{d}s^2=\ka^2\mathrm{d}\eta^2+\ka^2\sinh^2\eta\,(\mathrm{d}\theta^2+\sin^2\theta\mathrm{d}\phi^2)\,,
\ee
which gives a diagonal triad:
\be
q_{ab}=e^i_{a}e^i_{b}\,,
\quad
e^i_{a}=(\vec{e}_{\eta},\vec{e}_{\theta},\vec{e}_{\phi})
=\ka\,\mat{ccc}{1&&\\&\sinh\eta&\\&&\sinh\eta\sin\theta}\,,
\ee
where we use the vectorial notation for the coordinates on the tangent space. We compute the corresponding spin-connection $\Gamma^i_{a}$ compatible with the triad, which is given by the usual formula:
\be
\Gamma^i_{a}
=
\f12\eps^{ijk}e^b_{k}\,\big{(}
\pp_{[b}e^j_{a]}+e^c_{j}e^l_{a}\pp_{b}e^l_{c}
\big{)}\,,
\ee
where $e^a_{i}$ is the inverse triad, $e^a_{i}e^i_{b}=\delta^a_{b}$.
This gives the spin-connection and the Ashtekar-Barbero connection $A_{SL}$, for which we will drop the $SL$ subscript to simplify the notations:
\be
\Gamma^i_{a}=(\vec{\Gamma}_{\eta},\vec{\Gamma}_{\theta},\vec{\Gamma}_{\phi})
=\,
\mat{ccc}{0&0&\cos\theta\\0&0&-\sin\theta\cosh\eta\\0&-\cosh\eta&0}\,,
\ee
\be
A^i_{a}
=\Gamma^i_{a}+\beta\,\f{e^i_{a}}{\ka}
=(\vec{A}_{\eta},\vec{A}_{\theta},\vec{A}_{\phi})
=\,
\mat{ccc}{\beta&0&\cos\theta\\0&\beta\sinh\eta&-\sin\theta\cosh\eta\\0&-\cosh\eta&\beta\sinh\eta\sin\theta}\,.
\ee
In that setting, it is very simple to compute the holonomies along geodesic starting at the origin. In that case, the holonomy $g$ of the spin-connection is always trivial while the holonomy $h$ of the Ashtekar-Barbero connection is generated by a single Pauli matrix. For instance, for the hyperbolic edge $(AB)$ with $A$ at the origin and $B$ at the boost parameter $\eta_{AB}$ in an arbitrary direction, we have\footnotemark:
\be
h_{AB}
= e^{i\int_{0}^{\eta_{AB}} \mathrm{d}\eta\, \vec{A}_{\eta}\cdot\f{\vec{\sigma}}{2}}
=e^{i\beta\eta_{AB}\f{\sigma_{1}}{2}}
=\mat{cc}{\cos\f{\beta\eta_{AB}}{2}& i\sin\f{\beta\eta_{AB}}{2}\\i\sin\f{\beta\eta_{AB}}{2}&\cos\f{\beta\eta_{AB}}{2}}\,,
\quad
g_{AB}=h_{AB}^{(\beta=0)}=\id\,,
\ee
with a similar formula for the holonomy $h_{CA}=h_{AC}^{-1}$.
\footnotetext{In the Cartesian coordinates, as computed in  appendix \ref{app:cartesian}, the holonomy $h^{cartesian}_{AB}$ actually rotates with the direction of $B$. In the equatorial plane $\theta=\f{\pi}2$ to keep things at their simplest, we have:
$$
h^{cartesian}_{AB}=e^{\f i2\beta\eta_{AB}\,{\hat{u}_{B}\cdot\vec{\sigma}}}\,,
\quad \hat{u}_{B}=(\cos\phi,\sin\phi,0)\,,
$$
where $\phi$ is the angle of the direction of the point $B$. The holonomies $h^{cartesian}_{AB}$ are related to the holonomies $h_{AB}$ computed in spherical coordinate by a gauge transformation given by  rotations around the $z$-axis. This is reflected in the holonomies $h_{A'A''}$ computed in the spherical coordinate system around the origin $A$.
}
For an arbitrary complex  value of $\beta$, these holonomies are generically Lorentz transformations. For a real Immirzi parameter, they are $\SU(2)$ group elements, while they become pure boosts when $\beta$ is purely imaginary. 

All that's left is to compute the third holonomy along the edge $(BC)$. We could compute directly this holonomy by integrating the connections along that geodesic. We can actually sidestep this probably straightforward but lengthy computations by a trick. By placing the origin of the coordinate system at the point $C$ (or point $B$), the resulting holonomies $\tilde{g}_{BC}$ and $\tilde{h}_{BC}$ would be as given above. Using the fact that the hyperboloid and triad field are homogeneous, this means that the triad $e$, and thus the connection $A^i_{a}=\Gamma^i_{a}[e]+\beta e^i_{a}/\ka$, in the new coordinate system centered on $C$ are equal to the  initial fields defined in the initial coordinate system centered on $A$ up to a gauge transform. By determining this gauge transformation, one can easily deduce the holonomy $h_{BC}$ from $\tilde{h}_{BC}$.
The full explicit calculation can be found in appendix \ref{app:holoSL2C}.

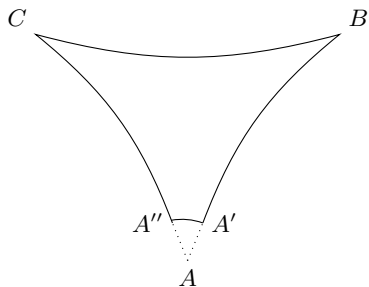
\begin{figure}[h!]

\centering

\begin{tikzpicture}[scale=1]

\coordinate (A) at (0,-1);
\coordinate (A1) at (0.2,-0.5);
\coordinate (A2) at (-0.2,-0.5);
\coordinate (B) at (2,2);
\coordinate (C) at (-2,2);

\draw (A1) to[bend left=15] (B) to[bend left=15] (C) to[bend left=15] (A2);
\draw[dotted] (A) -- (A1);
\draw[dotted] (A) -- (A2);
\draw (A) node[below]{$A$};
\draw (A1) node[right]{$A'$};
\draw (A2) node[left]{$A''$};
\draw (B) node[above right]{$B$};
\draw (C) node[above left]{$C$};

\draw (A1) arc (70:100:0.8);

\end{tikzpicture}

\caption{Regularization of the holonomy: because of the coordinate singularity at the origin $A$, we shift the point $A$ along the two triangle edges in order to define a regularized holonomy. In the limit where we send this shift to 0, we recover the holonomy around the triangle.}
\label{fig:Shift}

\end{figure}

The only tricky point with the spherical coordinate is that the origin of the coordinate system is a conical singularity. So we have to take special care of the holonomies going along arcs around the origin even in the infinitesimal limit. Indeed, we actually have to cut a little arc around $A$ between the edges $(AB)$ and $(AC)$ and introduce two points $A'$ and $A''$ as illustrated on fig.\ref{fig:Shift}.
The holonomy of the triangle is properly defined as $h^{(A)}_{\Delta}\equiv h_{A''A'}h_{CA''}h_{BC}h_{A'B}$ as the points $A'$ and $A''$ are sent back to the origin $A$. In that limit, the holonomy $h_{A''A'}$ remains non-trivial\footnotemark{}:
\be
h_{A'A''}=e^{i\int_{0}^{\hat{a}} \mathrm{d}\phi\,\vec{A_{\phi}}\cdot\f{\vec{\sigma}}{2}}
\,\,\underset{A',A''\rightarrow A}{=}\,
e^{i\,\hat{a}\f{\sigma_{3}}{2}}\,,
\ee
where we've place ourselves in the equatorial plane $\theta=\f{\pi}2$ as said earlier.
\footnotetext{
To remain as close as possible to the geometrical interpretation of the spatial  directions $(x,y,z)$ along the hyperboloid, we have associated the Pauli matrices to the tangent directions on the internal space as $\vec{\sigma}=(\sigma_{1},-\sigma_{3},\sigma_{2})$.
}
We could have avoided such considerations by working directly in the Cartesian coordinates (see appendix \ref{app:cartesian}), but calculations in spherical coordinates are actually much simpler since the triad remains diagonal.

\smallskip

The final result for an arbitrary triangle has a fairy simple structure at the end of the day:
\be
h = R B_{CA}^{\mathrm{i}\beta} B_{BC}^{\mathrm{i}\beta} B_{AB}^{\mathrm{i}\beta}\,,
\ee
where the exponentiation for a pure boost $B$ must be understood as a quick hand notation for:
\begin{equation}
B^{\alpha}=\left(\exp\left(\frac{\eta}{2} \hat{u} \cdot \overrightarrow{\sigma}\right)\right)^{\alpha} = \exp\left(\frac{\alpha \eta}{2} \hat{u} \cdot \overrightarrow{\sigma}\right)\,,\quad\forall \eta\in\R\,,\,\,\alpha\in\C
\,.
\end{equation}
Here we have introduced the boosts along each edge of the triangle, $B_{AB}$ that sends the point $A$ to the point $B$  (with $A$ at the origin, the 4d coordinates of $B$ are the projections $\tr\, B_{AB}B_{AB}^\dagger \sigma_{\mu}$  on the Pauli matrices) and so on. As pointed out in \cite{bonzom_deformed_2014,charles_closure_2015,haggard_encoding_2015}, these boosts satisfy an almost-closure relation around the triangle, in that their product is the identity up to now a possible rotation:
\be
B_{AC}B_{CB}B_{BA}=R^{-1}\,\in\SU(2)\,.
\label{defR}
\ee
As shown in \cite{charles_closure_2015,haggard_encoding_2015}, this rotation $R$ is actually the holonomy of the spin-connection around the hyperbolic triangle: its axis is orthogonal to the triangle plane at the root $A$ and its rotation angle measures the deficit angle $\theta=\pi-(\hat{a}+\hat{b}+\hat{c})$ which gives directly the area of the hyperbolic triangle.

\smallskip

When $\beta=0$ vanishes, the Ashtekar-Barbero connection $A$ reduces to the spin-connection and we do recover as expected $h=R$. This is actually the notion of $\SU(2)$-normal to a hyperbolic triangle developed in \cite{charles_closure_2015,haggard_encoding_2015}, which led to a $\SU(2)$ closure relation for hyperbolic tetrahedra. So we can naturally see our generic case with all the other possible connections $A_\textrm{SL}$ for arbitrary complex values of the Immirzi parameter $\beta$ as a generalization for those previous works. We do indeed derive a whole family of deformed normals valued in $\SL(2,\C)$. As we will see in the next section, this will provide us with a whole family of closure constraints for a same hyperbolic tetrahedron.

Finally, another special case is $\beta=i$, when $A_\textrm{SL}$ becomes the self-dual Lorentz connection. As expected, we get a trivial holonomy around the triangle, $h=\id$, directly from the  definition \eqref{defR} of $R$. This provides a good consistency check. Similarly the conjugate case $\beta=-i$ follows from taking the adjoint:
\be
h_{\beta=-i}=R B_{CA} B_{BC} B_{AB}=(R^{-1})^{\dagger} B_{CA}^{\dagger} B_{BC}^{\dagger} B_{AB}^{\dagger}
=(h_{\beta=+i}^{\dagger})^{-1}\,,
\ee
since $R$ is unitary and the pure boosts are Hermitian matrices. So the holonomy of the anti-self-dual Lorentz connection, for $\beta=-i$, around the hyperbolic triangle is also trivial.

\section{Closure Constraints for the Hyperbolic Tetrahedron}

\subsection{The $\mathrm{SL}(2,\mathbb{C})$ closure constraint and its  $\SB(2,\C)\rtimes\SU(2)$ splitting}

Now that we have settled the construction of $\SL(2,\C)$-normals for hyperbolic triangles, we can apply this technique to the derivation of a $\SL(2,\C)$-closure constraint for hyperbolic tetrahedra.
We proceed as in the flat case, where we naturally derived closure constraints for the flat tetrahedron from using the $\isu(2)$-connection in section \ref{isu-closure}.

Let us start with a hyperbolic tetrahedron $ABCD$ and choose the point $A$ as root for the holonomies.
We write the $\mathrm{SL}(2,\mathbb{C})$ holonomy $\Lambda_i$ around the face opposite to the vertex $i$, e.g. the holonomy $\Lambda_D$ goes around the face $ABC$. Then, we have:
\begin{equation}
\left\{\begin{array}{rcl}
\Lambda_B &=& g_{AD}^{-1} g_{CD} g_{AC}  \vspace*{0.8mm}\\
\Lambda_C &=& g_{AB}^{-1} g_{BD}^{-1} g_{AD} \vspace*{0.8mm}\\
\Lambda_D &=& g_{AC}^{-1} g_{BC} g_{AB}
\end{array}\right.
\end{equation}
each holonomy being rooted in $A$. The $g$'s are the holonomies along every oriented edges. Only three face holonomies are given in the previous equation. For the last face $(BCD)$, the root $A$ does not belong to its boundary and we have to parallel transport along an edge, say the edge $AC$, in order to define the corresponding holonomy. This gives:
\begin{equation}
\Lambda_A = g_{AC}^{-1} \left(g_{CD}^{-1} g_{BD} g_{BC}^{-1}\right) g_{AC}
\end{equation}
Then it is direct to check that the following discrete Bianchi identity holds:
\begin{equation}
\Lambda_D \Lambda_C \Lambda_B \Lambda_A = \id
\end{equation}
This is the $\mathrm{SL}(2,\mathbb{C})$ closure constraints derived from the connection. Each $\Lambda_{i}$ holonomy defines our new concept of non-abelian normal vector for each hyperbolic triangle of the hyperbolic tetrahedron.
This is the direct parallel of the $\mathrm{ISU}(2)$ closure constraint developed on flat space in the first section.

\medskip

Now, $\SL(2,\C)$ is 6-dimensional and we expect to work with 3-dimensional normal vectors. Roughly, $\SL(2,\C)$ is the complexification of $\SU(2)$ and we would like to split it into real and imaginary parts. In the flat case, this corresponded to the splitting of the $\isu(2)$-connection into its translation and rotation components, realized by the direct product decomposition of group elements $\mathrm{ISU}(2) \simeq \mathbb{R}^3 \rtimes \mathrm{SU}(2)$. Back to the hyperbolic case, we use the Iwasawa decomposition of Lorentz group elements:
\begin{equation}
\forall \Lambda \in \mathrm{SL}(2,\mathbb{C}),\ \exists! (L,H) \in \mathrm{SB}(2,\mathbb{C})\times\mathrm{SU}(2), \textrm{ such that }\Lambda = L H\,.
\end{equation}
This realizes the semi-direct product decomposition $\mathrm{SL}(2,\mathbb{C}) \simeq \mathrm{SB}(2,\mathbb{C}) \rtimes \mathrm{SU}(2)$. The subgroup $\SU(2)$ still consists in the 3d rotations, as in the flat case, and the group $\SB(2,\C)$ is to be interpreted as the non-abelian translations on the 3-hyperboloid. Applying the Iwasawa decomposition to the face holonomies, we split the $\mathrm{SL}(2,\mathbb{C})$ closure relation into two closure relations, the first one in terms of $\SB(2,\C)$ group elements and the other with $\SU(2)$ group elements.
Indeed, we write:
\begin{equation}
\label{LHdecomposition}
\left\{\begin{array}{rcl}
\Lambda_D &=& L_D H_D \\
\left(H_D\right) \Lambda_C \left(H_D\right)^{-1} &=& L_C H_C \\
\left(H_C H_D\right) \Lambda_B \left(H_C H_D\right)^{-1} &=& L_B H_B \\
\left(H_B H_C H_D\right) \Lambda_A \left(H_B H_C H_D\right)^{-1} &=& L_A H_A
\end{array}\right.
\end{equation}
The $\SL(2,\C)$ closure relation then reads:
\begin{equation}
\Lambda_D \Lambda_C \Lambda_B \Lambda_A = L_D L_C L_B L_A H_A H_B H_C H_D = \id
\end{equation}
Since the Iwasawa decomposition of a $\SL(2,\C)$ group element (here the identity $\id$) is unique, this implies that:
\begin{equation}
\left\{\begin{array}{rcl}
L_D L_C L_B L_A &=& \id \vspace*{.8mm}\\
H_A H_B H_C H_D &=& \id
\end{array}\right.
\end{equation}
So the original Lorentz closure constraint  implies two closure relations, one for the Borel subgroup $\SB(2,\C)$ and one for the rotational subgroup $\SU(2)$. These two closure relations can of course be assembled back to the original Lorentz closure.

Note that in order to get simple $\SU(2)$ and $\SB(2,\C)$ closure relations, we introduced non-trivial parallel transports through the rotational part of the Lorentz elements in \eqref{LHdecomposition}  when defining the $\SU(2)$ face normals $H_{i}$ and the $\SB(2,\C)$ normals $L_{i}$. This twist, coming from the semi-direct product structure, leads to a non-trivial behavior of the $L$'s and $H$'s under 3d rotations. This is expected for the Borel group elements $L_{i}$ since $\SB(2,\C)$ is not invariant under conjugation by $\SU(2)$ group elements, as underlined  in the previous work on the $\SL(2,\C)$ phase space for ($q$-deformed) loop quantum gravity \cite{bonzom_deformed_2014,dupuis_deformed_2014}. And this twist extends here to the $\SU(2)$-normals and closure relation between the $H$'s.
But all this behavior  is exactly what's expected from a phase space perspective and using the closure constraint as the generator for 3d-rotations as we discuss below.

\smallskip

So the $\SB(2,C)$ closure constraint is the one that we have been looking for in the context of $q$-deformed loop quantum gravity. Indeed, in \cite{dupuis_observables_2013,bonzom_deformed_2014,bonzom_towards_2014,dupuis_deformed_2014}, one imposes a $\SB(2,\C)$ closure constraint at every vertex of the graph but its geometrical meaning was not clear. The wish was that, similarly to the $\R^{3}$ closure constraint was interpreted as defining convex polyhedra dual to each vertex in standard loop quantum gravity, the new deformed $\SB(2,\C)$ closure constraint could be interpreted as defining some fundamental blocks of hyperbolic geometry. This is exactly what we achieved here: hyperbolic tetrahedra satisfy a $\SB(2,\C)$ closure relation on Borel group elements $L_{i}$ which can be interpreted as non-abelian normal vectors to the tetrahedron faces.

All this was made possible by the use of a complexified $\su(2)_{\C}$-connection $A_{SL}$ using complex values for the Immirzi parameter $\beta$.
Moreover, we get a whole one-parameter family of closure constraints. This means that one has to first decide the value of the parameter $\beta$ before reconstructing the hyperbolic tetrahedron from the non-abelian normals $L_{i}$.

As for the $\SU(2)$ closure constraint, it proposes a generalization of the $\SU(2)$ closure constraints for hyperbolic tetrahedra derived in \cite{charles_closure_2015, haggard_encoding_2015}. Indeed these recent studies relied on using the holonomies of the spin-connection to define $\SU(2)$ group elements normal to each hyperbolic triangle of the tetrahedron. This corresponds to the special case $\beta=0$ in ou new framework. 

\smallskip

It is interesting that we get both a non-compact closure constraint (in terms of $\SB(2,\C)$ group elements) and a compact closure constraint (in terms of $\SU(2)$ group elements) for hyperbolic tetrahedra. One could choose to use either one of these two dual closure constraints, following the spin-connection approach of \cite{charles_closure_2015, haggard_encoding_2015} or the $\SL(2,\C)$ phase space approach of \cite{bonzom_deformed_2014,dupuis_deformed_2014}. Here we have showed that both types of closure relations come from a unified framework of $\SL(2,\C)$ connections and $\SL(2,\C)$ closure constraints. The natural issue is then if these two dual closure constraints are equivalent and contain the same geometrical information. This question of the ``reconstruction of the hyperbolic tetrahedron'' will be discussed later in sections \ref{counting} and \ref{reconstruct}.

\subsection{Rotations, $\SL(2,\C)$ phase space and Poisson-Lie structure}

Let us see how these closure constraints for hyperbolic tetrahedra fit in the $\SL(2,\C)$ phase space structure developed for $q$-deformed loop quantum gravity and in particular focus on the behavior of the non-abelian normals and closure relations under 3d rotations.

A classical phase space for $q$-deformed loop quantum gravity was developed in \cite{bonzom_deformed_2014,dupuis_deformed_2014}. It is supposed to represent a non-vanishing cosmological $\Lambda$. More precisely, it deforms the basic $T^{*}\SU(2)$ phase space structure  of spin network states into a more complicated $\SL(2,\C)$ phase space based on the classical $r$-matrix for $\su(2)$. This leads to $q$-deformed spin network for a real deformation parameter $q\in\R$ \cite{dupuis_observables_2013}. This allowed to account for a negative cosmological constant $\Lambda<0$ in Euclidean 3d quantum gravity \cite{bonzom_deformed_2014,bonzom_towards_2014}, while the hope is to use that framework to deal with a positive cosmological constant $\Lambda>0$  in $3+1$-d Lorentzian quantum gravity. Here we will see that the $\SB(2,C)$ closure constraint generates 3d  rotations (realized as $\SU(2)$ transformations) in the $\SL(2,\C)$ phase space, the same way that the usual $\R^{3}$ closure constraint generate the $\SU(2)$ gauge invariance in standard loop quantum gravity.

\smallskip

So let us start with analyzing the behavior of the holonomies $\Lambda_{i}$ and group elements $L_{i}$ and $H_{i}$ under 3d rotations of the 3-hyperboloid. All the holonomies of the $\sl(2,\C)$-connection $A_{SL}$ behave covariantly under 3d rotations. For a rotation $k\in\SU(2)$, they transform simply under conjugation by $k$, so that the $\SL(2,\C)$ closure relation is trivially preserved:
\be
\Lambda_{i}\,\rightarrow\,k\Lambda_{i}k^{-1}
\,,
\qquad
\Lambda_D \Lambda_C \Lambda_B \Lambda_A =\id
\,\rightarrow\,
k
\,(\Lambda_D \Lambda_C \Lambda_B \Lambda_A )\,
k^{-1}=\id\,.
\ee
The Iwasawa decomposition leads to more complicated behaviors for the $\SB(2,\C)$ and $\SU(2)$ group elements:
\be
\Lambda=LH
\,\rightarrow\,
k\Lambda k^{-1} =(kL\tilde{k}^{-1})\,(\tilde{k}Hk^{-1})
\quad\Rightarrow\quad
\left\{
\begin{array}{l}
L
\,\rightarrow\,
kL\tilde{k}^{-1}\in\SB(2,\C)
\vspace*{.8mm}\\
H
\,\rightarrow\,
\tilde{k}Hk^{-1}\in\SU(2)
\end{array}\right.
\ee
where $\tilde{k}\in\SU(2)$ is the only group element such that $kL\tilde{k}^{-1}$ lies in the Borel subgroup $\SB(2,\C)$. This $\SU(2)$ group element $\tilde{k}$ depends non-linearly on both $k$ and $L$. Due to this twisted transformation law, the $L_{i}$ and $H_{i}$ do not all transform the same way:
\be
\Lambda_{D}=L_{D}H_{D}
\,\rightarrow\,
k\Lambda_{D} k^{-1} =(kL_{D}k_{D}^{-1})\,(k_{D}H_{D}k^{-1})\,,
\ee
\be
H_D\Lambda_C H_D^{-1} = L_C H_C
\,\rightarrow\,
\big{(}k_{D}H_Dk^{-1}\big{)}\big{(}k\Lambda_Ck^{-1}\big{)}\big{(}H_D^{-1} k_{D}^{-1}\big{)}
= k_{D}L_C H_Ck_{D}^{-1}
=(k_{D}L_Ck_{C}^{-1})(k_{C} H_C k_{D}^{-1})\,,
\ee
and so on, so that we get braided transformation laws:
\be
\label{LHrot}
\left|
\begin{array}{lcl}
L_{D}&\rightarrow&k L_{D}k_{D}^{-1}\\
L_{C}&\rightarrow&k_{D} L_{C}k_{C}^{-1}\\
L_{B}&\rightarrow&k_{C} L_{B}k_{B}^{-1}\\
L_{A}&\rightarrow&k_{B} L_{A}k_{A}^{-1}\\
\end{array}
\right.
\qquad
\left|
\begin{array}{lcl}
H_{D}&\rightarrow&k_{D} H_{D}k^{-1}\\
H_{C}&\rightarrow&k_{C} H_{C}k_{D}^{-1}\\
H_{B}&\rightarrow&k_{B} H_{B}k_{C}^{-1}\\
H_{A}&\rightarrow&k_{A} H_{A}k_{B}^{-1}\\
\end{array}
\right.
\ee
When the closure relations hold, the last $\SU(2)$ group element matches the original transformation, $k_{A}=k$ ensuring that the closure relations are consistently invariant under $\SU(2)$ rotations as expected.

\smallskip

It was shown in \cite{bonzom_deformed_2014} that this twisted $\SU(2)$ actions on $L$'s and $H$'s is exactly generated by the Poisson-Lie flow of the $\SB(2,\C)$ closure constraint $L_{D}L_{C}L_{B}L_{A}$ in the $\SL(2,\C)$ phase space. Let us review how this works.
We endow $\SL(2,\C)$ with a Poisson bracket defined with the classical $r$-matrix:
\be
\{\Lambda_1,\Lambda_2\} = -r \Lambda_1 \Lambda_2 - \Lambda_1 \Lambda_2 r^\dagger
\,,\qquad
r = \frac{\kappa}{4}\sum_i \tau_i \otimes \sigma_i = \frac{\mathrm{i}\kappa}{4}\begin{pmatrix}
1 & 0 & 0 & 0 \\
0 & -1 & 0 & 0 \\
0 & 4 & -1 & 0 \\
0 & 0 & 0 & 1
\end{pmatrix}\,,
\ee
where we use the standard tensor product convention for $\Lambda_1 = \Lambda\otimes \id$ and $\Lambda_2 = \id \otimes \Lambda$ and we defined modified Pauli matrices $\tau_i = \mathrm{i}(\sigma_i - \frac{1}{2} [\sigma_3,\sigma_i] ) = (\mathrm{i} \sigma_i + \epsilon^k_{~3i} \sigma_k)$.
This Poisson bracket translates to the Iwasawa decomposition $\Lambda=LH$:
\be
\{L_{1},L_{2}\}=-[r,L_{1}L_{2}],\quad
\{H_{1},H_{2}\}=[r^\dagger,H_{1}H_{2}],\quad
\{L_{1},H_{2}\}=-L_{1}rH_{2},\quad
\{H_{1},L_{2}\}=-L_{2}r^\dagger H_{1}\,.
\ee
For the $\SB(2,\C)$ component, it is convenient to parametrize them explicitly as:
\begin{equation*}
\begin{pmatrix}
\lambda & 0 \\
z & \lambda^{-1}
\end{pmatrix},\quad\textrm{with}\,\, \lambda > 0,\, z \in \mathbb{C}
\end{equation*}
The Poisson brackets then read:
\begin{equation}
\{\lambda, z\} = \frac{\mathrm{i} \kappa}{2} \lambda z, \quad \{\lambda, \overline{z}\} = -\frac{\mathrm{i} \kappa}{2} \lambda \overline{z}, \quad \{z,\overline{z}\} = \mathrm{i} \kappa \left(\lambda^2 - \lambda^{-2}\right)\,.
\end{equation}
We equip all four pairs of variables, $(L_{A},H_{A})$, $(L_{B},H_{B})$, $(L_{C},H_{C})$ and $(L_{D},H_{D})$, with this $\SL(2,\C)$ phase space structure.
Then a big result from \cite{bonzom_deformed_2014} is that the  Poisson-Lie flow of the Gauss constraint $\cG=L_{D}L_{C}L_{B}L_{A}$ generates the action of 3d rotations as given above in \eqref{LHrot}:
\be
\exp \left(\prod_k \lambda_k^{-2} \{\mathop{Tr} \,V_{\vec{\eps}}\, \mathcal{G} \mathcal{G}^\dagger, \cdot\}\right)
\,,
\quad\textrm{with}\quad
V_{\vec{\eps}}=
\left(
\begin{array}{cc}
2\eps_{z} & \eps_{-}\\ \eps_{+} & 0
\end{array}
\right)
\ee
where $V_{\vec{\eps}}$ parametrizes the $\SU(2)$ rotation.
This shows that we can use once more, as in the standard loop quantum gravity framework, the $\SB(2,\C)$ closure constraints for hyperbolic tetrahedra and polyhedra as generating the local $\SU(2)$ gauge invariance.

\smallskip

One might wonder if we could switch the roles of the $\SB(2,\C)$ closure constraint and $\SU(2)$ closure constraint, and in so use this latter as generating $\SU(2)$ rotations of the hyperbolic tetrahedron and holonomies. This seems to be possible using quasi-Poisson structures \cite{haggard_encoding_2015}. This suggests switching all together the roles of the $\SB(2,\C)$ and $\SU(2)$ holonomies in deformed loop quantum gravity. This could potentially lead to a duality between $q$-deformed hyperbolic spin networks (for $q$ real) and $q$-deformed spherical spin networks (for $q$ root of unity). 


\section{Reconstructing the Hyperbolic Tetrahedron}

\subsection{The non-abelian normal: counting of degrees of freedom}
\label{counting}

We have defined the $\SL(2,\C)$ holonomy of the Ashtekar-Barbero connection (for a chosen value of the Immirzi parameter $\beta\in\C$) around a hyperbolic triangle as our new notion of non-abelian normal to the triangle. The natural question is how much geometrical information about the triangle is recorded in that holonomy.

Let us look back at flat triangles. The usual normal vector $\vec{N}\in\R^{3}$ to a flat triangle has three components, but it has only one component invariant under 3d rotations: the only rotation-invariant data recorded by the normal vector is its norm $|\vec{N}|$, which gives the area of the triangle. Since a triangle is uniquely defined up to a rotation by three parameters, say its three edge lengths, we see clearly that the triangle area is not enough to reconstruct the initial triangle (up to 3d rotations). And we have a 2-parameter family of deformations of the triangle which do not change the triangle area\footnotemark.
\footnotetext{It would be interesting to repeat this analysis and counting of degrees of freedom with the Poincar\'e normal to flat triangles, as defined by the holonomy of the $\isu(2)$-connection introduced earlier in section \ref{isu2}. Here, we focus instead directly in the case of hyperbolic triangles.}

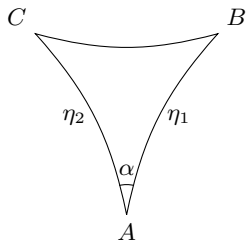
\begin{figure}[h!]

\centering

\begin{tikzpicture}[scale=0.8]

\coordinate (A) at (0,-1);
\coordinate (B) at (1.5,2);
\coordinate (C) at (-1.5,2);

\draw (A) to[bend left=15] node[midway,right]{$\eta_{1}$} (B) to[bend left=15] (C) to[bend left=15] node[midway,left]{$\eta_{2}$} (A);
\draw (A) node[below]{$A$};
\draw (B) node[above right]{$B$};
\draw (C) node[above left]{$C$};

\draw ($(A)+(76:0.5)$) arc (76:104:0.5);
\draw ($(A)+(90:0.75)$) node{$\alpha$};

\end{tikzpicture}

\caption{The geometry (up to 3d rotations) of a hyperbolic triangle is entirely determined by 3 parameters. We can consider for example its three edge lengths or its three angles. Here, for our purpose, we prefer a mixed parametrization using two edge lengths $\eta_{1}$ and $\eta_{2}$ and the angle $\alpha$ between them.
To compute explicitly the $\SL(2,\C)$ normals, we will choose the  hyperbolic triangles living in the equatorial hyperplane at $z=0$ with the first point $A$ at the origin and the second point $B$ defined by a boost in the $x$-direction.
}
\label{fig:hypertriangle}

\end{figure}

Let us now move on to hyperbolic triangles and their $\SL(2,\C)$ normals.
On the one hand, a hyperbolic triangle is defined up to 3d rotations by 3 parameters, say two edge lengths and the angle between as depicted on fig.\ref{fig:hypertriangle}, similarly to flat triangles.
On the other hand, $\SL(2,\C)$ group element is defined by 6 (real) parameters, but contains only three rotational invariants.  Indeed, under a 3d rotation  of the triangle defined by the group element $R\in\SU(2)$, the holonomy transforms as $\Lambda\rightarrow R\Lambda R^{-1}$. The only independent $\SU(2)$-invariants are $\tr\, \Lambda\in\C$ and $\tr \,\Lambda\Lambda^{\dagger}\in\R$.
So the natural question is how much of the geometrical information is recorded by the $\SU(2)$-invariant components of the $\SL(2,\C)$-holonomy? This means investigating the rank of the map  $(\eta_{1},\eta_{2},\alpha)\in\R^{3}\,\mapsto\,(\tr\, \Lambda,\tr \,\Lambda\Lambda^{\dagger}) \in\C\times\R$.

We analyze  this map locally. The first derivatives define a 3$\times$3 Jacobian matrix, which we study numerically. Its determinant always vanishes, so that we know the $\SL(2,\C)$ normal (for fixed $\beta$) has at most 2 independent components (up to rotations). Extensive numerics have shown that the rank of that matrix is actually 2, for a fixed complex value of the Immirzi parameter $\beta\notin\R$ and $\beta\ne \pm i$.
For the degenerate cases, the rank of the matrix is 1 when $\beta\in\R$ (as the $\SL(2,\C)$ holonomy reduces to a $\SU(2)$ group element) while it simply vanishes when $\beta=\pm i$ (since the holonomy is trivial).
Plots of the minors of the Jacobian matrix are given below in fig.\ref{ref:Inversion}.
\begin{figure}[h!]


\begin{subfigure}[t]{.45\linewidth}
\includegraphics[scale=0.6]{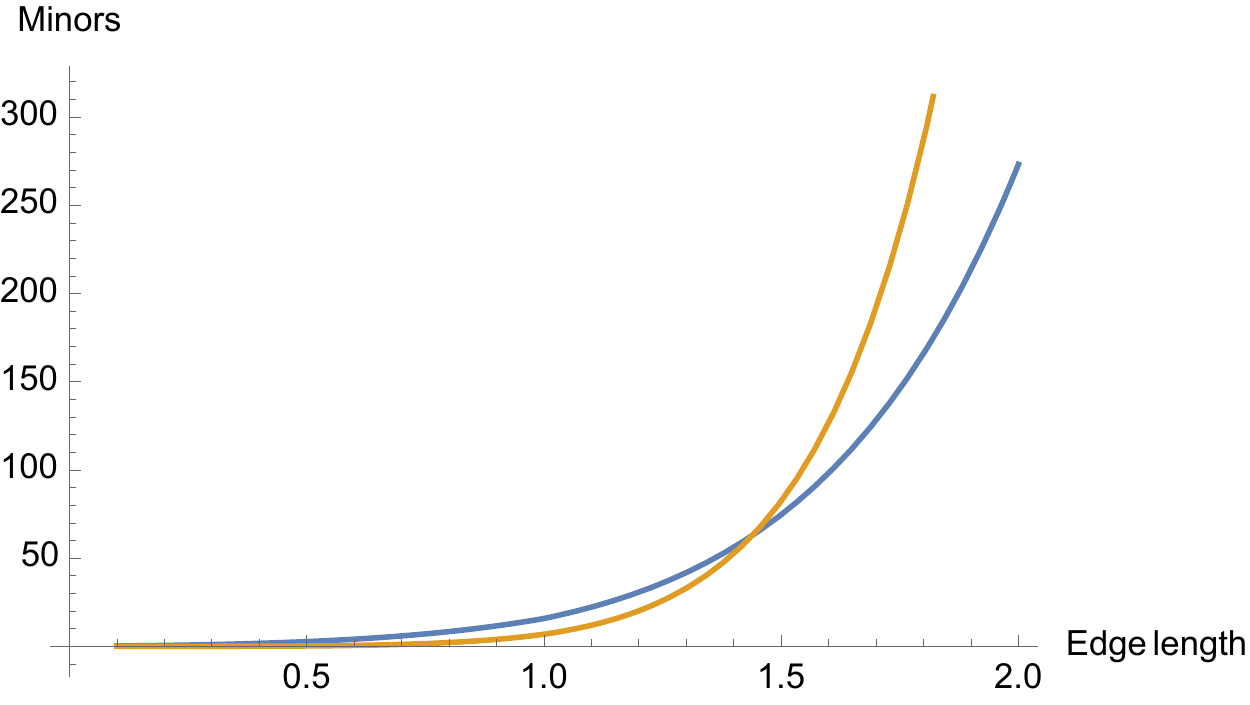}
\caption{Plot of the minors as one of the edge lengths, $\eta_1$, varies. Here $\ka=1$, $\beta = -2\mathrm{i}$, $\eta_2 = 1$ and $\alpha=1$. The minors never vanish.}
\end{subfigure}
\hspace{2mm}
\begin{subfigure}[t]{.45\linewidth}
\includegraphics[scale=0.6]{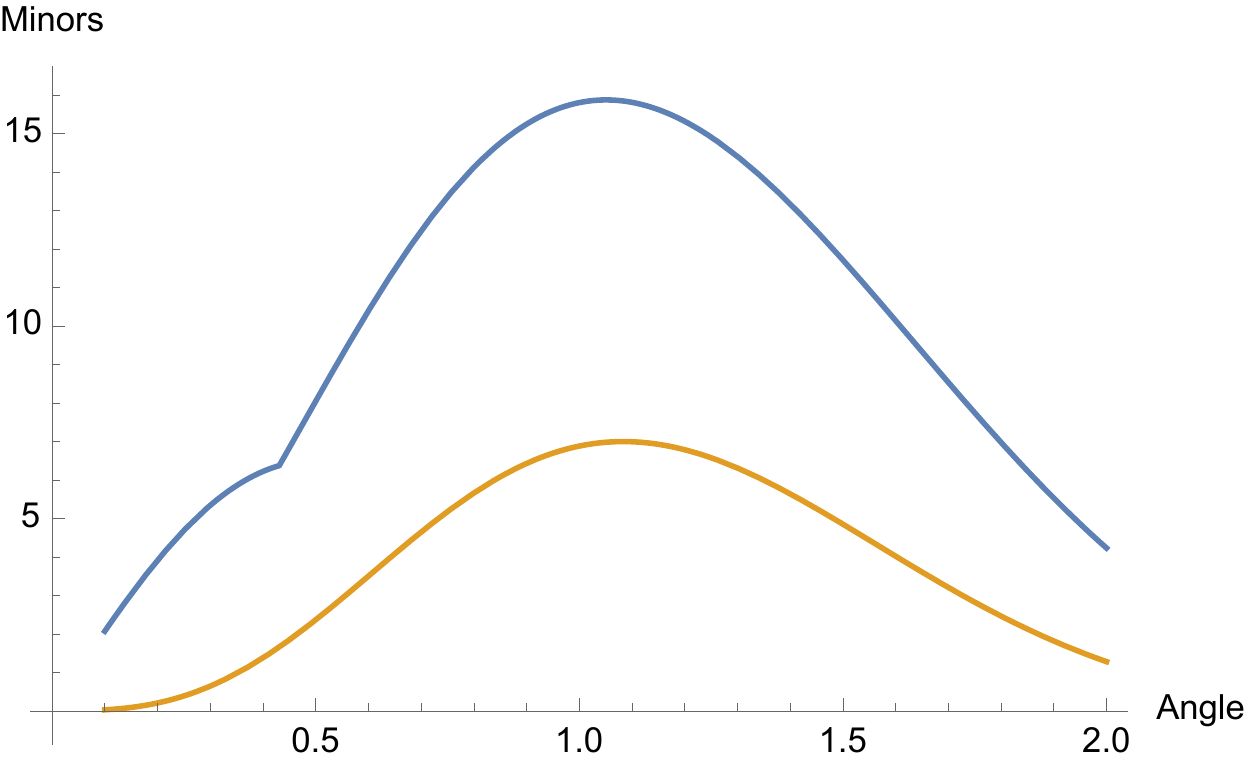}
\caption{Plot of the minors as a function of the angle $\alpha$, for fixed values $\beta = -2\mathrm{i}$, $\eta_1 = \eta_2 = 1$. The minors never vanish.}
\end{subfigure} \\
\begin{subfigure}[t]{.45\linewidth}
\includegraphics[scale=0.6]{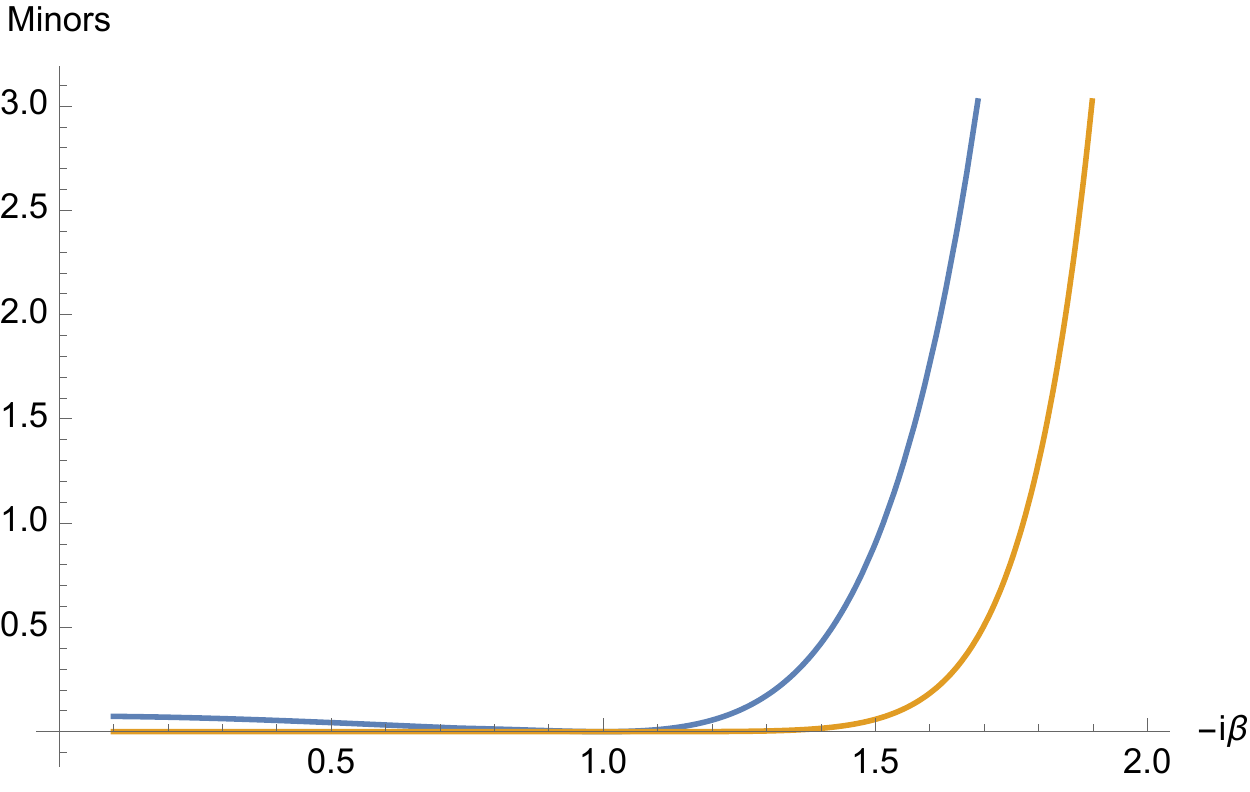}
\caption{Plot of the minors as a function of  purely imaginary values of the Immirzi parameter $\beta\in\,i\R$ and fixed triangle geometry $\alpha = \eta_1 = \eta_2 = 1$: the minors don't vanish except for the trivial case of $\beta = \mathrm{i}$ when the whole Jacobian matrix actually vanishes.}
\end{subfigure}
\hspace{2mm}
\begin{subfigure}[t]{.45\linewidth}
\includegraphics[scale=0.6]{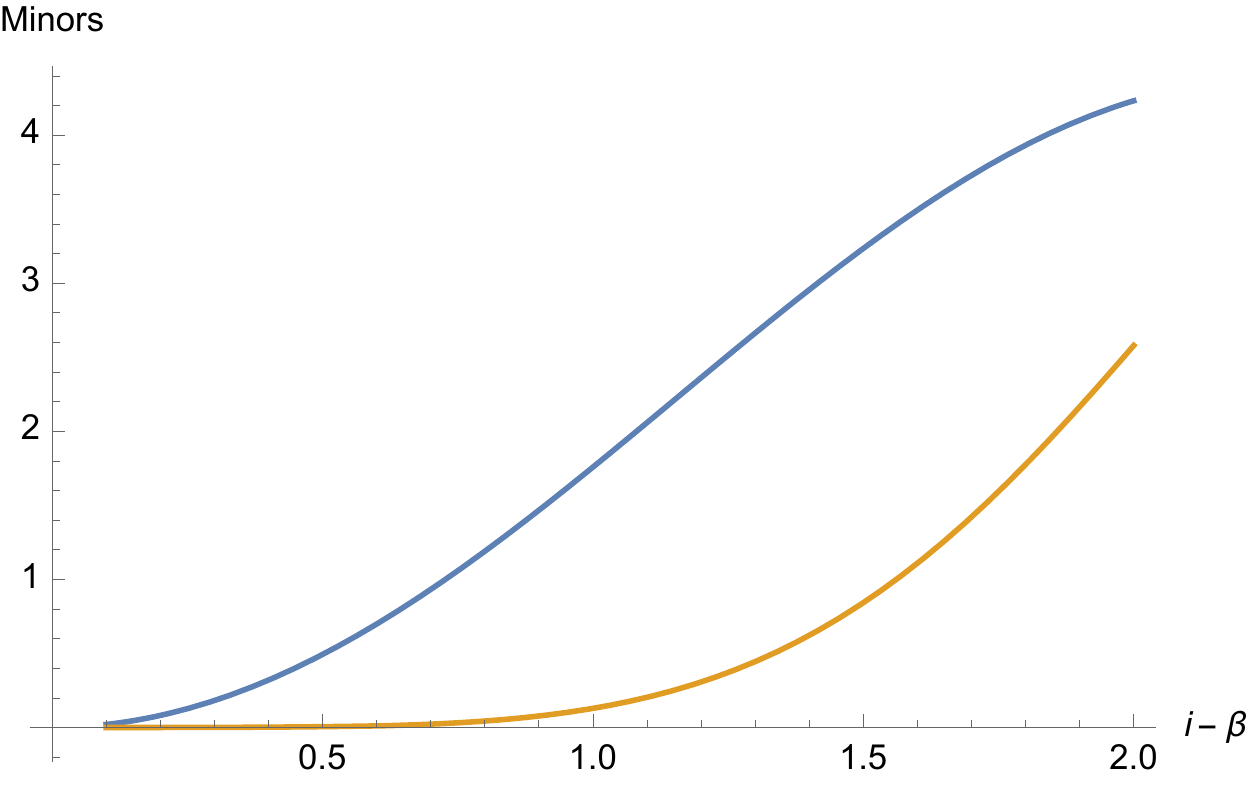}
\caption{Plot of the minors as a function of a shifted real Immirzi parameter $\beta\in\,i+\R$ and fixed triangle geometry $\alpha = \eta_1 = \eta_2 = 1$: the minors don't vanish except for the trivial case of $\beta = \mathrm{i}$ when the whole Jacobian matrix actually vanishes.}
\end{subfigure}

\caption{Numerical analysis of the information contained in the $\mathrm{SL}(2,\mathbb{C})$ holonomies: we compute the Jacobian matrix of the map $(\eta_{1},\eta_{2},\alpha)\in\R^{3}\,\mapsto\,(\tr\, \Lambda,\tr \,\Lambda\Lambda^{\dagger}) \in\C\times\R$ giving the rotational-invariant components of the $\SL(2,\C)$ holonomy as a function of the geometry of the hyperbolic triangle. The determinant vanishes, so we plot the maxima of the absolute value of the minors (the $1\times 1$ and $2\times 2$ minors in blue and orange respectively on each curve) of the Jacobian matrix in order to determine its rank.}

\label{ref:Inversion}
\end{figure}

Beyond the $\SL(2,\C)$ holonomy, what interest us are the $\SB(2,\C)$ and $\SU(2)$ normals: do they reflect the same geometrical data about the hyperbolic triangle and do they contain the same information as the initial $\SL(2,\C)$ holonomy? Writing $\Lambda=LH$ following the Iwasawa decomposition, with the resulting action of 3d rotations:
\be
\Lambda=LH \quad\longrightarrow\quad
R\Lambda R^{-1}=\,(RL\tilde{R}^{-1})\,(\tilde{R}H R^{-1})\,,
\ee
where $\tilde{R}$ depends on both $R$ ad $L$,
we realize that  the $\SB(2,\C)$-normal $L$  contains a single real degree of freedom invariant under rotation,  $\tr\, L L^{\dagger}=\tr\,\Lambda\Lambda^{\dagger}$. So this shows that, for hyperbolic triangles up to 3d rotations, the $\SB(2,\C)$-normal contains only one degree of freedom of the triangle geometry, compared to the $\SL(2,\C)$-normal which encodes two degrees of freedom. This means that the $\SU(2)$-normal $H$ also records an independent degree of freedom. 

At the end of the day, apart from the degenerate cases $\beta\in\R$ or $\beta=\pm i$, the $\SB(2,\C)$-normal contains the same amount of geometrical information about the hyperbolic triangle as the standard normal 3-vector about the flat triangle. And one needs both the $\SB(2,\C)$-normal and the $\SU(2)$-normal to reconstruct the whole $\SL(2,\C)$ holonomy around the triangle.

Finally, it would be very interesting to understand the geometrical meaning of the rotational invariant $\tr\, L L^{\dagger}=\tr\,\Lambda\Lambda^{\dagger}$ defined in terms of the $\SB(2,\C)$-normal and see how much it is related (or not) to the hyperbolic area of the triangle. For instance, at $\beta=0$, we know that the $\SU(2)$-normal $H$ encodes the hyperbolic area as its rotation angle \cite{charles_closure_2015,haggard_encoding_2015} (while the $\SB(2,\C)$-normal is trivial) and it would be enlightening to find a similar statement for the $\SB(2,\C)$-normal when $\beta$ has an arbitrary complex value.

\subsection{Hyperbolic tetrahedron reconstruction from the closure relation}
\label{reconstruct}

We have seen in the previous section that a hyperbolic tetrahedron leads to a closure constraint satisfied by the four $\SB(2,\C)$-normals to its triangle, $L_{4}L_{3}L_{2}L_{1}=\id$. The goal is now to understand if this relation is invertible and whether or not one can reconstruct the initial hyperbolic tetrahedron once given four $\SB(2,\C)$ group elements satisfying such a closure constraint.

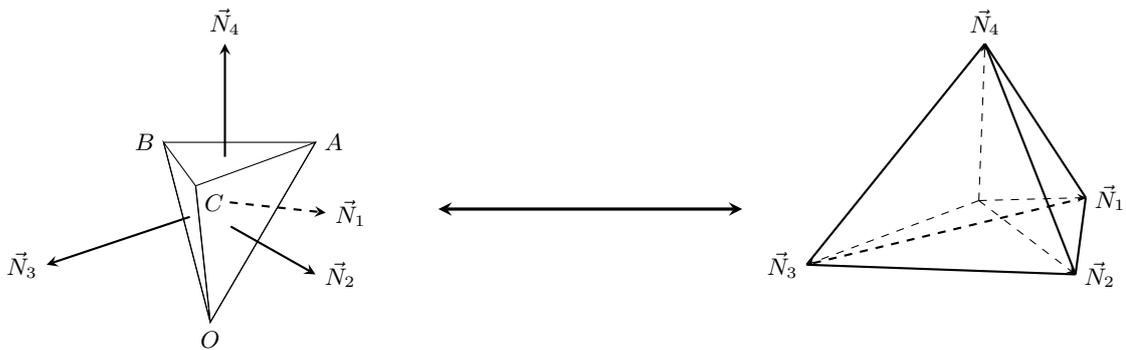
\begin{figure}[h!]

\centering

\begin{tikzpicture}
\coordinate (O) at (0,0,0);
\coordinate (A) at (1,2,-1);
\coordinate (B) at (-1,2,-1);
\coordinate (C) at (0,2,0.5);

\draw (O) node[below]{$O$};
\draw (A) node[right]{$A$};
\draw (B) node[left]{$B$};
\draw (C) node[below right]{$C$};

\draw (O) -- (B) -- (C) -- cycle;
\draw (O) -- (C) -- (A) -- cycle;
\draw (O) -- (A) -- (B) -- cycle;

\draw[->,>=stealth,thick,dashed] (0,1.333,-0.666) -- ++ (0.3,-1.1,-2.5) node[right]{$\vec{N}_1$};
\draw[->,>=stealth,thick] (0.333,1.333,0.166) -- ++ (1.5,-0.25,1) node[right]{$\vec{N}_2$};
\draw[->,>=stealth,thick] (-0.333,1.333,-0.166) -- ++ (-1.5,-0.25,1) node[left]{$\vec{N}_3$};
\draw[->,>=stealth,thick] (0,2,-0.5) -- ++ (0,1.5,0) node[above]{$\vec{N}_4$};

\draw[<->,>=stealth,very thick] (3,1.5) -- (7,1.5);

%

\coordinate (O2) at (10+0+.3,0.333+.1,-2.666-.5);
\coordinate (A2) at (10+1.833,1.333-0.25,1.166);
\coordinate (B2) at (10-1.833,1.333-0.25,1-.166);
\coordinate (C2) at (10+0,3.5,-0.5);
\draw (O2) node[right]{$\vec{N}_1$};
\draw (A2) node[right]{$\vec{N}_2$};
\draw (B2) node[left]{$\vec{N}_3$};
\draw (C2) node[above]{$\vec{N}_4$};
\draw[thick,dashed] (O2) -- (B2) ;
\draw[thick] (O2) -- (C2);
\draw[thick] (O2) -- (A2) ;
\draw[thick] (A2) -- (B2) ;
\draw[thick] (B2) -- (C2);
\draw[thick] (C2) -- (A2) ;
\coordinate (D) at (10,1.5,-.292);
\draw[->,>=stealth,dashed] (D) -- (O2) ;
\draw[->,>=stealth,dashed] (D) -- (A2) ;
\draw[->,>=stealth,dashed] (D) -- (B2) ;
\draw[->,>=stealth,dashed] (D) -- (C2) ;

\end{tikzpicture}

\caption{Construction of a dual tetrahedron from the normals in the flat case. There is some arbitrariness in the choice of the order of the normals. It turns out that the bidual is the original tetrahedron.}
\label{fig:Dual}

\end{figure}

For a flat tetrahedron, one usually defines the four normal 3-vectors to its triangle, $\vec{N}_{i=1..4}\in\R^{3}$, which satisfy the flat abelian closure constraint, $\vec{N}_{1}+\vec{N}_{2}+\vec{N}_{3}+\vec{N}_{4}=0$, as illustrated on the left hand side of fig.\ref{fig:Dual}. The reverse is also true. Starting from four vectors in $\R^{3}$ whose sum vanishes, there exists a unique flat tetrahedron such that these are its normal vectors. Each normal vector $\vec{N}_{i}$ can not determine by itself the corresponding triangle. For instance, it is possible to deform a triangle without changing its normal vector (any triangle in the same plane and with the same area has the same normal vector, for instance by moving one point parallel to the opposite side). And one needs the information of the normals to the other triangles in order to reconstruct each of the triangles. As reviewed in \cite{charles_closure_2015}, an edge of the tetrahedron will be recovered from the cross product of the normals to the two triangles sharing it. This can be recast as constructing the bidual to the tetrahedron: one defines the dual tetrahedron whose four vertices are given by the four normals, then one looks at the dual of the dual tetrahedron, i.e. the bidual of the initial tetrahedron. One finds that up to a volume factor of the tetrahedron (which can be obtained as the triple product of three of the normals), this bidual construction gives back exactly the initial tetrahedron, as drawn on fig.\ref{fig:Dual}.

Let us proceed similarly for a hyperbolic tetrahedron. We start with four $\SB(2,\C)$ group elements $L_{i}$. As we have seen above in the previous section \ref{counting}, each individual $\SB(2,\C)$-normal $L_{i}$ is not enough to reconstruct by itself the corresponding hyperbolic triangle. Actually, each $L_{i}$ does not contain enough information to recover the corresponding $\SL(2,\C)$ holonomy $\Lambda_{i}$ and this $\SL(2,\C)$ group element would still not be enough to recover the triangle. Like in the flat case, we truly need all four normals to build the corresponding hyperbolic tetrahedron.

So let us set out clearly the problem at hand. The initial hyperbolic tetrahedron is entirely determined, up to translation, by a triplet of edges attached to one of its vertices: we place one vertex at the hyperboloid origin and the tetrahedron is defined by the position of its other three vertices. These positions are each defined by a (pure) boost, i.e. a Hermitian 2$\times$2 unit-determinant matrix. So we start with three such matrices $(M_{1},M_{2},M_{3})\,\in\mathrm{SH}_{2}(\C)^{\times 3}\sim\,(\R^{3})^{\times 3}$. From these vertices, we define the four hyperbolic triangular faces of the tetrahedron and get the four $\SB(2,\C)$-normals $L_{i}$. Since they satisfy the closure constraints, the fourth normals is entirely determined by the other three, $L_{4}=(L_{3}L_{2}L_{1})^{-1}$. The goal is thus to study the invertibility of the map:
\be
\label{eqn:mapMtoL}
(M_{1},M_{2},M_{3})\,\in\mathrm{SH}_{2}(\C)^{\times 3}\quad\longmapsto\quad
(L_{1},L_{2},L_{3})\,\in\SB(2,\C)^{\times 3}\,.
\ee
We parametrize $\mathrm{SH}_{2}(\C)$ pure boost matrices by their $\R^{3}$ projection on the Pauli matrices:
$$
M\in\mathrm{SH}_{2}(\C)\,\mapsto\,\vec{M}=\f12\tr M\vec{\sigma}\,,
\qquad M=\sqrt{1+|\vec{M}|^{2}}\,\id+\vec{M}\cdot\vec{\sigma}\,,
$$
and we parametrize $\SB(2,\C)$ group elements as lower triangular matrices:
$$
L=\mat{cc}{\lambda & 0 \\ z & \bar{\lambda}}\,,\quad
\lambda \in\R_{+}\,,\quad z\in\C\,.
$$
\begin{figure}[h!]


\begin{subfigure}[t]{.45\linewidth}
\includegraphics[scale=0.6]{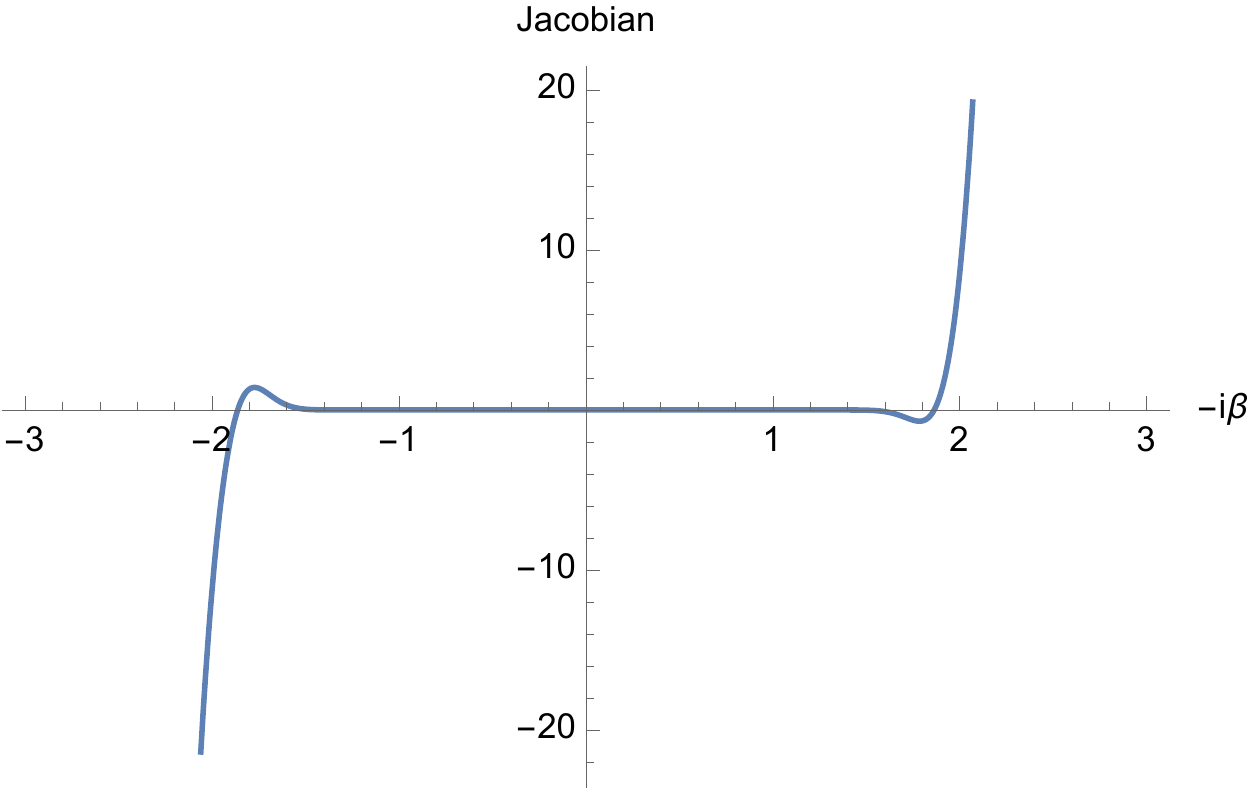}
\caption{ We plot the Jacobian of the map \eqref{eqn:mapMtoL}, defining the $\SB(2,\C)$-normals to the hyperbolic triangles from the position of the tetrahedron vertices,  as a function of the Immirzi parameter $\beta\in\, i\R$. We evaluate numerically the Jacobian for a rectangular tetrahedron, with the root vertex  at the origin of the hyperboloid, $O=\ka\,(1,0,0,0)$,  and each of the other three vertices along a different space axis, $A=\ka\,(\cosh 1,\sinh 1,0,0)$, $B=\ka\,(\cosh 1,0,\sinh 1,0)$, $C=\ka\,(\cosh 1,0,0,\sinh 1)$. The Jacobian determinant does not vanish except for the special values $\beta=\pm \mathrm{i}$ and $\beta=0$.}
\end{subfigure}
\hspace{2mm}
\begin{subfigure}[t]{.45\linewidth}
\includegraphics[scale=0.6]{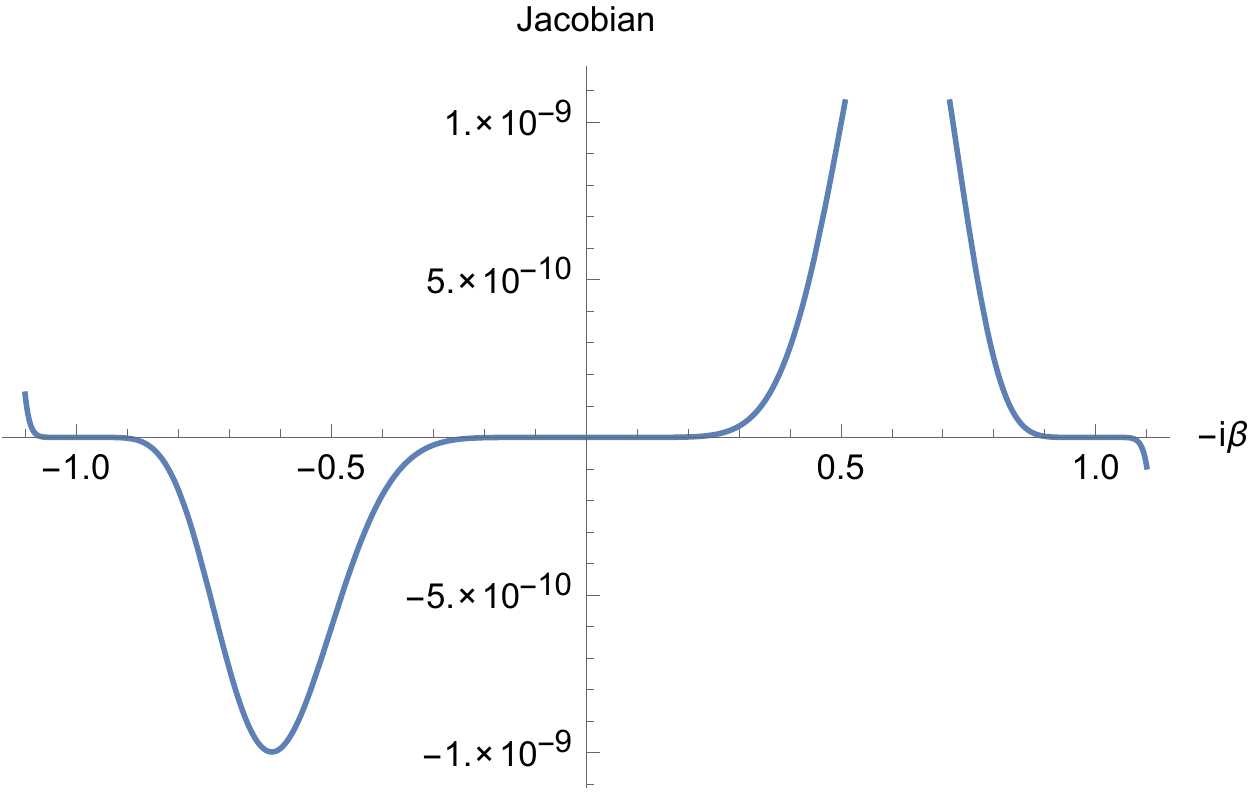}
\caption{We zoom on the plot of the Jacobian in terms of a purely imaginary Immirzi parameter $\beta\in\,i\R$. It vanishes at only three points. $\beta=0$ corresponds to using the spin connection, which has no $\sb(2,\C)$-component and thus leads to trivial $\SB(2,\C)$-normals. The two other special values, $\beta=\pm i$, corresponds to the self-dual and anti-self-dual Ashtekar-Barbero connections, which are flat and leads to trivial $\SL(2,\C)$ holonomies and thus in particular trivial $\SB(2,\C)$-normals.}
\end{subfigure}

\caption{Numerical analysis of the map from hyperbolic tetrahedra to the $\SB(2,\C)$-normals. The Jacobian of this map does not vanish except in degenerate cases, which  illustrates the local invertibility of the map and promises the existence of a reconstruction procedure of the hyperbolic tetrahedron from its $\SB(2,C)$-normals.}
\label{fig:Reconstruction1}

\end{figure}
\begin{figure}[h!]


\begin{subfigure}[t]{.6\linewidth}
\includegraphics[scale=0.6]{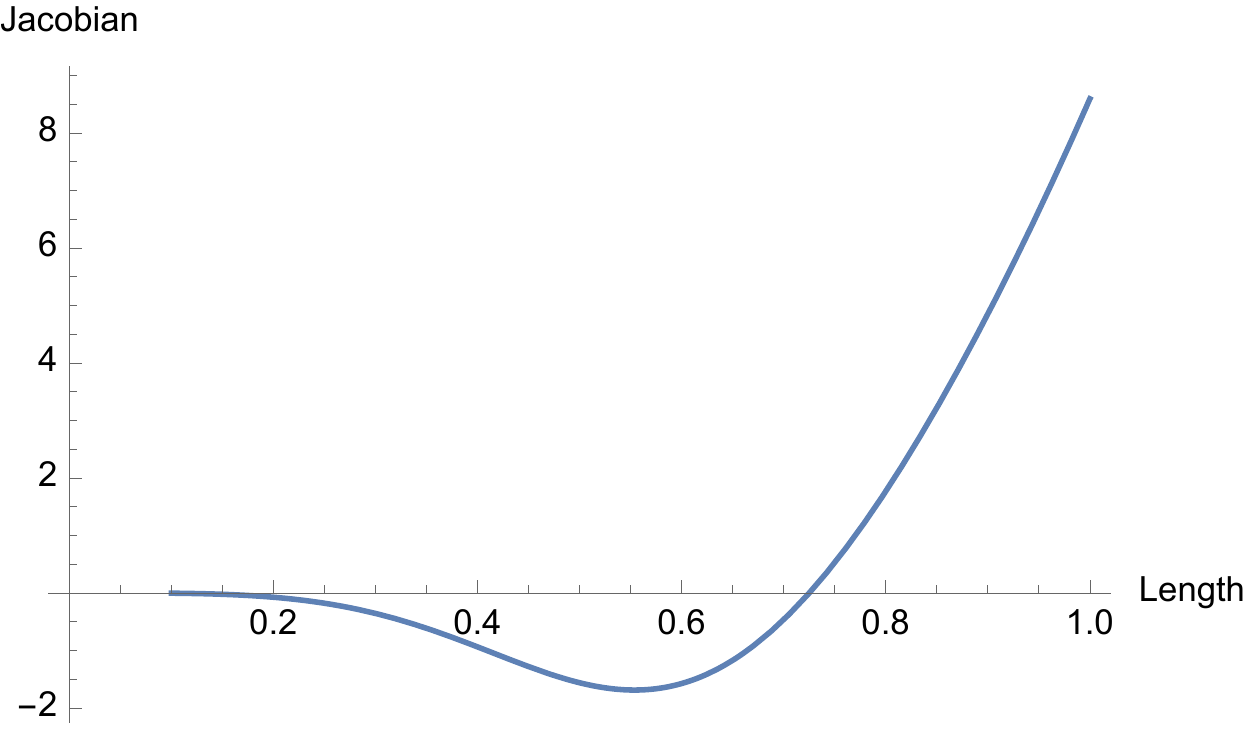}
\caption{We plot the Jacobian of the map \eqref{eqn:mapMtoL} from hyperbolic tetrahedra to its $\SB(2,\C)$-normals as a function of one of its edge length. The tetrahedron is a square tetrahedron and all the other lengths are fixed to $1$.
We see that the Jacobian generically does not vanish and that the map is locally invertible except at a special point where the plot crosses the axis for a length of about $0.72511$. We do not know what special about the tetrahedron geometry for this value and this should be investigated later in more detail.}
\end{subfigure}
\hspace{2mm}
\begin{subfigure}[t]{.35\linewidth}
\includegraphics[scale=0.5]{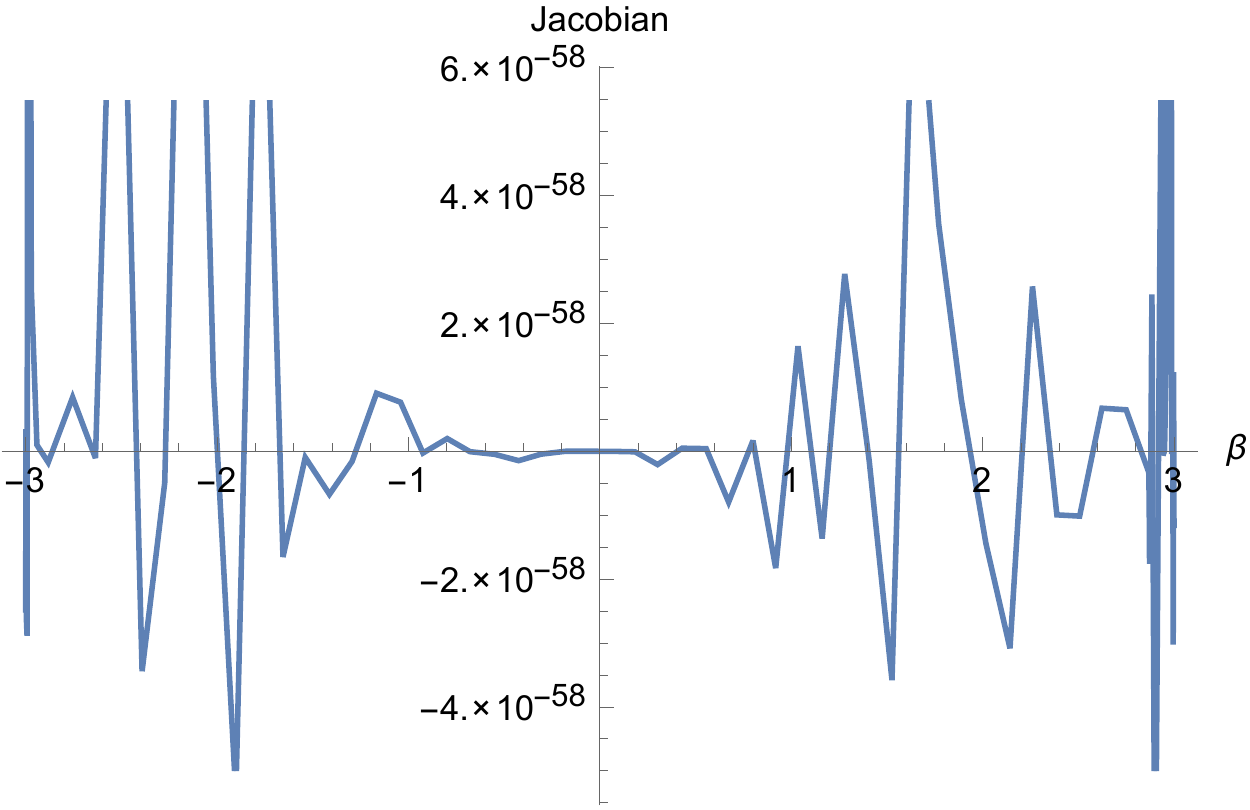}
\caption{We plot the Jacobian determinant as a function of a real Immirzi parameter $\beta\in\R$. In theory, the Ashtekar-Barbero connection is a pure $\su(2)$-connection and the $\SB(2,\C)$ should all be trivial. This is a consistency check of the stability of our numerical algorithm: we see only numerical noise around zero as expected.}
\end{subfigure}

\caption{Numerical analysis of the Jacobian of the map from hyperbolic tetrahedra to the $\SB(2,\C)$-normals.}
\label{fig:Reconstruction2}

\end{figure}
The invertibility of this map is actually a tough problem due to the non-linearities induced by the non-trivial parallel transport involved in the definition of the $\SB(2,\C)$-normals (both the natural parallel transport on the hyperboloid due to its curvature and the non-trivial conjugation by the holonomies of the Ashtekar-Barbero itself coming in the definition of the $\SB(2,\C)$-normals by the Iwasawa decomposition). We have not been able to show analytically that this map is indeed invertible. Nevertheless, we have analyzed the problem numerically by plotting the determinant of the corresponding 9$\times$9 Jacobian matrix and showed, as illustrated on fig.\ref{fig:Reconstruction1} and fig.\ref{fig:Reconstruction2}, that as long as $\beta\notin\R$ and $\beta\ne \pm i$, the determinant does not vanish and the map is locally invertible. This clearly indicates that closure constraints for $\SB(2,\C)$-normals should determine a unique hyperbolic tetrahedron and that a reconstruction procedure should exist.

%
%
%
%
%
%

We have tried to develop a bidual tetrahedron procedure, similarly to the flat case, by defining the dual hyperbolic tetrahedron with its four vertices defined by the $\SB(2,\C)$-normals to the initial triangles and then studying the dual of the dual of the initial tetrahedron. We further allowed a different Immirzi parameter $\tilde{\beta}$ for taking the dual the second time and searched for a value of $\tilde{\beta}$ depending on the original parameter $\beta$. Unfortunately, this idea has not been entirely conclusive. We nevertheless give more details  in  appendix \ref{app:reconstruct}. And we hope to be able to find a suitable reconstruction algorithm in the future or at least a definitive proof of the invertibility, which would provide an equivalent of Minkowski's theorem for convex polyhedra on the 3-hyperboloid.

\section{Conclusion \& Outlook}

In this papier, we provided a first step towards validating the geometrical interpretation of the spin network states of $q$-deformed loop quantum gravity in 3+1-dimensions \cite{dupuis_observables_2013,dupuis_deformed_2014,bonzom_towards_2014,bonzom_deformed_2014} as discrete hyperbolic geometry. Indeed we showed that the $\SB(2,\C)$ Gauss law or closure constraints imposed at the spin network vertices are related to hyperbolic tetrahedra, mimicking the correspondance given by the Minkowski theorem in the flat case between $\R^{3}$ closure constraints and (convex) polyhedra. This reinforces the intuition that the $q$-deformation, for $q\in\R$, is deeply related to taking into account a non-vanishing cosmological constant $\Lambda>0$.

\smallskip

We realize this by extending Freidel's spinning geometry framework \cite{freidel_spinning_2014} to the hyperbolic case. We reviewed the flat case, how to define the normal vectors $\vec{N}_{i}\in\R^{3}$ to a tetrahedron face as the holonomies of an abelian $\R^{3}$-connection and how this naturally leads to a $\R^{3}$ closure constraint for those normals, $\sum_{i}\vec{N}_{i}=0$. We then generalized this procedure to a one-parameter family of non-abelian $\isu(2)$-connection, defining a deformed notion of non-abelian normals valued in the 3d Poincar\'e group $\ISU(2)$. This led to a new non-abelian $\ISU(2)$ closure relation for a flat tetrahedron, from which we can derive two closure relations by splitting that relation into its  rotational part defining a $\SU(2)$-closure  and its translational part giving a $\R^{3}$-closure.

We followed the same route for hyperbolic tetrahedra, living in the space-like 3-hyperboloid $H_{3}\sim\SL(2,\C)/\SU(2)$. We introduced a one-parameter family of $\SL(2,\C)$ connections on the 3-hyperboloid, which actually match the Ashtekar-Barbero connections with complex Immirzi parameter $\beta\in\C$ as long as the 3-hyperboloid is embedded in the flat 3+1 Minkowski space-time. This naturally produce non-abelian normals to hyperbolic triangles valued in the Lorentz group $\SL(2,\C)$ and a corresponding $\SL(2,\C)$-closure constraints between the four normals to the faces of a hyperbolic tetrahedron. Acting the semi-direct splitting of the Lorentz group, $\SL(2,\C)=\SB(2,\C)\rtimes\SU(2)$, this Lorentz closure relation induces two dual closure relations, one between $\SU(2)$-normals and one between $\SL(2,\C)$-normals. We can use any of those two relations to characterize hyperbolic tetrahedra.
The $\SU(2)$ closure constraints we obtain are a generalization of the $\SU(2)$ closure constraint for hyperbolic tetrahedra derived in \cite{haggard_encoding_2015,charles_closure_2015} based on the holonomies of the spin-connection on the 3-hyperboloid. This actually correspond to a vanishing Immirzi parameter $\beta=0$ in our context.

We are more interested in the $\SB(2,\C)$ closure constraints. These are a great improvement on the previous proposal in \cite{charles_closure_2015} which did not behave properly under 3d rotations. Our framework shows that the $\SB(2,\C)$ Gauss law imposed at the vertices of spin networks in the $q$-deformed phase space proposal for loop quantum gravity \cite{dupuis_deformed_2014,bonzom_deformed_2014} can actually be interpreted geometrically as defining a hyperbolic tetrahedron dual to each vertex. This turns those $q$-deformed twisted geometries into actual discrete hyperbolic geometries made from gluing together hyperbolic tetrahedra. We would like to underline that this is only possible if choosing a non-real value for the Immirzi parameter $\beta\notin\R$ (else the $\sl(2,\C)$connection collapses to a real $\su(2)$-connection with a trivial boost component).

Moreover the $\SB(2,\C)$ closure constraint plays a double role as the usual $\R^{3}$ closure constraint in standard loop quantum gravity: it is a constraint between triangle normals ensuring the existence of a hyperbolic tetrahedron dual to each spin network vertex and its Poisson-Lie flow (under the Poisson bracket defined  by the $r$-mtrix on $\SL(2,\C)$) generates 3d rotations by the (braided) action of $\SU(2)$ on the $\SB(2,\C)$ normals.

Let us insist for we get a whole family of closure constraints labeled by a complex Immirzi parameter $\beta$. Imagine that we have four Borel group elements $L_{i}\in\SB(2,\C)$ satisfying the closure relation $L_{4}..L_{1}=\id$. Then we first need to choose a value for $\beta$, it will affect the holonomies of the Ashtekar-Barbero connection so that each different value of $\beta$ produces a different hyperbolic tetrahedron whose non-abelian normals are those $L_{i}$'s. This fits beautifully with the renormalization of the Immirzi parameter $\beta$ in loop quantum gravity (see e.g. \cite{Benedetti:2011nd}) and even with the recent proposal of seeing $\beta$ as the cut-off parameter of the renormalization scheme \cite{charles_ashtekar-barbero_2015}. Indeed, in the context of the coarse-graining of the quantum states of geometry in loop quantum gravity, we would have the Immirzi parameter run with the length at which we probe the geometry. And depending on the value of $\beta$ and thus on the length scale, we would get a different reconstruction scheme for the hyperbolic tetrahedra and thus for the discrete space manifold.

\smallskip

There are some possible technical improvements on our result.
\begin{itemize}
\item We should investigate more precisely the reconstruction process of the hyperbolic tetrahedron from non-abelian normals satisfying ether the $\SU(2)$ closure constraint or the $\SB(2,\C)$ closure constraint, once the Immirzi parameter $\beta$ and the curvature radius $\kappa$ are fixed. This is made very complicated by the various $\SU(2)$ and $\SL(2,\C)$ parallel transport used to define the non-abelian normals from the original boosts defining the tetrahedron points on the hyperboloid. Even if an explicit reconstruction procedure might not be possible, we should strive to prove that there is a one-to-one correspondance between normals satisfying the closure constraints and hyperbolic tetrahedra.

\item We should understand better the geometrical meaning of the $\SU(2)$ and $\SB(2,\C)$ normals to the hyperbolic triangles. In the case $\beta=0$,  the $\SB(2,\C)$ normal is trivial and the  $\SU(2)$ normal is defined as a holonomy of the spin-connection around the triangle, with its rotation angle giving the deficit angle and thus the hyperbolic area of the triangle and its rotation axis giving the normal direction to the triangle plane (at the root chosen to define the holonomy). But this clear geometrical meaning is lost as we deformed the holonomies to arbitrary values for $\beta$. For instance, it seems essential to understand the geometrical meaning of the $\SU(2)$-invariant $\tr L L^{\dagger}$ for the $\SB(2,\C)$-normal $L$. Indeed it gives the matching condition for $q$-deformed twisted geometries: gluing two hyperbolic triangles along a spin network edge, we do not match the $L$ at the edge source with the $\tilde{L}$ at the edge target, but we simply match $\tr L L^{\dagger}$ and  $\tr \tilde{L} \tilde{L}^{\dagger}$.
This ``area-matching'' condition ensures that there exists a $\SU(2)$ holonomy $G$ mapping $L$ to $\tilde{L}=LG$.
In the flat case (or flat limit at $\kappa\rightarrow\infty$), it does imply that the flat triangles have the same area.  This interpretation is of course deformed as soon as $\kappa$ is non-trivial and  we need to understand what observable of the triangle geometry $\tr L L^{\dagger}$ measures.
We could also drop this area-matching condition and study the possibility of a generic $\SL(2,\C)$ holonomy between $L$ and $\tilde{L}$ (for instance coming the Ashtekar-Barbero connection for $\beta\in\C$).

\end{itemize}

\smallskip

Finally, we would like to draw two lessons for loop quantum gravity from this present work, beyond the $\SL(2,\C)$ closure constraints and hyperbolic tetrahedra. 
First we would like to stress the importance of using a complex Immirzi paramter $\beta\in\C$. For a real value of $\beta$, the $\SB(2,\C)$ part of the $\sl(2,\C)$-connection disappears and we lose the connection with the $q$-deformed phase space for loop quantum gravity. Coming back to the purely imaginary values $\beta=\pm i$ has already been pushed forward  in the recent litterrature \cite{Frodden:2012dq,Achour:2014eqa,Achour:2015xga}. But we would like to argue for a whole analytical continuation to the complex plane. For instance, values on the shifted real line $\beta\in i+\R$ appear to admit a very nice geometrical interpretation for the $\SL(2,\C)$ holonomies.

\begin{figure}[h!]
\includegraphics[scale=.7]{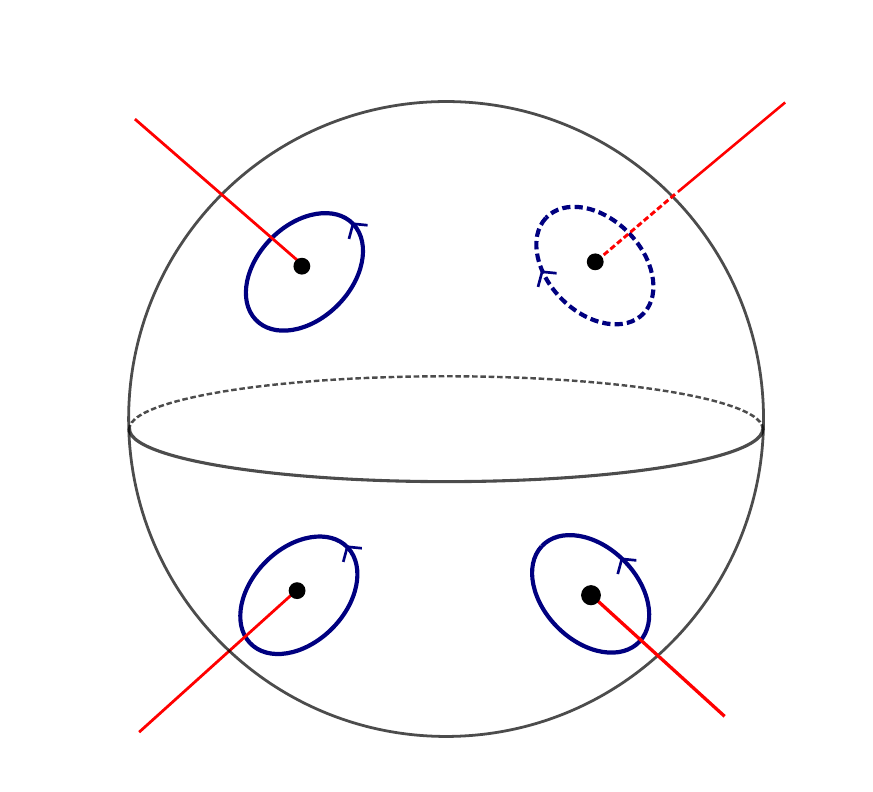}
\caption{In loop quantum gravity, spin network vertices are thought of as carrying volume excitations, of an abstract region of space bounded by a surface dual to the vertex.  Quantum states of geometry are then usually defined as excitations of the holonomies of the Ashtekar-Barbero connection along the (transversal) edges puncturing the surface. Our derivation of closure contraints as discrete Bianchi identities relies on interpreting the holonomies on the dual surface as defining (non-abelian) normals to the  surface. This strongly suggests using new dual spin network structure, as a graph dressed with the data of holonomies along the edges and also around those edges.}
\label{fig:SurfaceHolonomy}
\end{figure}
Then we would like to insist on the new double role of the Ashtekar-Barbero connection. Loop quantum gravity usually looks at its holonomies along the edges of the spin network graph and these are the basic degrees of freedom of the theory. Here, we use the holonomies around the faces of the tetrahedron (or more generally polyhedron) dual the spin network vertices, that holonomies looping around the spin network edges. Of course, we are actually considering holonomies of our intrinsically defined Lorentz connection, which only matches the Ashtekar-Barbero connection for a local embedding of the geometry around the vertex in a constant spatial curvature hyperboloid embedded itself the a flat space-time. Such an embedding is nevertheless natural from the point of view of the local equivalence principle. And the key point is that these new holonomies explore exactly directions orthogonal to the usual one considered in loop quantum gravity. As illustrated in fig.\ref{fig:SurfaceHolonomy}, this suggests introducing a double spin network structure, based on graphs with holonomies both along the spin network edges and also around those edges. Such double graph have already been hinted at in previous works, see e.g. \cite{Sahlmann:2011rv,Dittrich:2014wpa} (also see \cite{Dittrich:2016typ} for the applications of similar ideas in 3d quantum gravity in order to properly encode matter with both mass and spin). But they usually consider exponentiated flux variables (i.e. holonomies representing the triad), while we propose here to use again the holonomies of the Ashtekar-Barbero connection\footnotemark{} instead as the normal variables to the surfaces dual to the spin network vertices. 

\footnotetext{
From that perspective, one is then free to choose a different complex Immirzi parameter at each vertex to modify the local hyperbolic geometry reconstruction from the holonomies. The idea of using Ashtekar-Barbero connections with different Immirzi parameters to define quantum states of geometry has already appeared in work by Pranzetti and Perez in the context of black holes in \cite{Perez:2010pq}.
}

\section*{Acknowledgement}

We would like to thank Alexandre Feller for making the figure \ref{fig:SurfaceHolonomy} and Mait\'e Dupuis, Laurent Freidel, Florian Girelli, Aldo Riello, Simone Speziale and Christophe Goeller for the many discussions on the project. All the numerical simulations were realized with Mathematica 10.

\appendix

\section{Holonomy around a finite hyperbolic triangle}
\label{app:holoSL2C}

\subsection{Triad and holonomies in spherical coordinates}

We parametrize the hyperboloid, $t^2-x^2-y^2-z^2=\ka^2$, $t>0$ in terms of spherical coordinates:
\be
\left\{
\begin{array}{lcl}
t&=&\ka\cosh\eta\\
x&=&\ka\sinh\eta\sin\theta\cos\phi \\
y&=&\ka\sinh\eta\sin\theta\sin\phi \\
z&=&\ka\sinh\eta\cos\theta \\
\end{array}
\right.\,,
\ee
with  the boost parameter $\eta\in\R_{+}$ and the angles $\theta\in\,[0,\pi]$ and $\phi\in\,[0,2\pi]$
The metric on the hyperboloid induced by the flat 4d metric is the standard homogeneous hyperbolic metric
$\mathrm{d}s^2=\ka^2\mathrm{d}\eta^2+\ka^2\sinh^2\eta\,(\mathrm{d}\theta^2+\sin^2\theta\mathrm{d}\phi^2)$:
\be
q_{ab}=\ka^2\,\mat{ccc}{1&&\\&\sinh^2\eta&\\&&\sinh^2\eta\sin^2\theta}\,.
\,,
\ee
This gives a diagonal triad field $e^i_{a}$ such that $q_{ab}=e^i_{a}e^i_{b}$:
\be
e^i_{a}=(\vec{e}_{\eta},\vec{e}_{\theta},\vec{e}_{\phi})
=\ka\,\mat{ccc}{1&&\\&\sinh\eta&\\&&\sinh\eta\sin\theta}\,,
\ee
where we use the vectorial notation for the coordinates on the tangent space. We compute the corresponding spin-connection $\Gamma^i_{a}$ compatible with the triad, which is given by the usual formula:
\be
\pp_{[a}e^i_{b]}-\eps^i{}_{jk}\Gamma^j_{[a}e^k_{b]}=0\,,
\qquad
\Gamma^i_{a}
=
\f12\eps^{ijk}e^b_{k}\,\big{(}
\pp_{[b}e^j_{a]}+e^c_{j}e^l_{a}\pp_{b}e^l_{c}
\big{)}\,,
\ee
where $e^a_{i}$ is the inverse triad, $e^a_{i}e^i_{b}=\delta^a_{b}$.
This gives the spin-connection:
\be
\Gamma^i_{a}=(\vec{\Gamma}_{\eta},\vec{\Gamma}_{\theta},\vec{\Gamma}_{\phi})
=\,
\mat{ccc}{0&0&\cos\theta\\0&0&-\sin\theta\cosh\eta\\0&-\cosh\eta&0}\,,
\ee
Finally we get the  Ashtekar-Barbero connection $A$ by adding the triad to the spin-connection:
\be
A^i_{a}
=\Gamma^i_{a}+\beta\,\f{e^i_{a}}{\ka}
=(\vec{A}_{\eta},\vec{A}_{\theta},\vec{A}_{\phi})
=\,
\mat{ccc}{\beta&0&\cos\theta\\0&\beta\sinh\eta&-\sin\theta\cosh\eta\\0&-\cosh\eta&\beta\sinh\eta\sin\theta}\,.
\ee

\smallskip

Now we would like to compute the holonomy around a hyperbolic triangle. We first choose a first vertex, $A$, of the triangle as the origin of our coordinate system. And using the invariance of the triad and connection under rotation, we place the whole triangle in the equatorial plane at $z=0$, by fixing the angle $\theta=\f\pi 2$, and further put the point $B$ on the $x$-axis (at $\phi=0$).

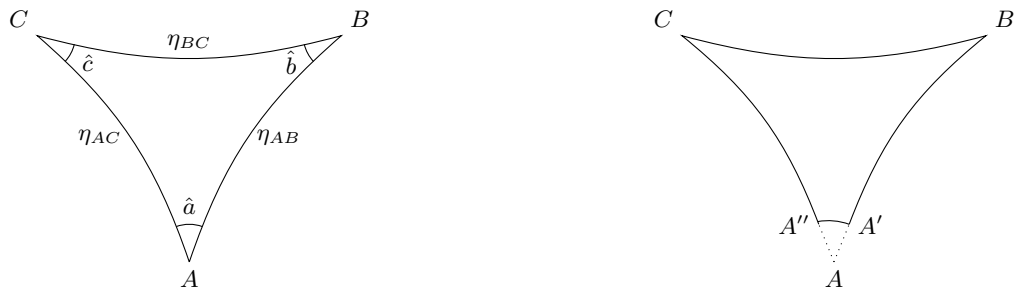
\begin{figure}[h!]


\begin{subfigure}[t]{.45\linewidth}
\begin{tikzpicture}[scale=1]

\coordinate (A) at (0,-1);
\coordinate (B) at (2,2);
\coordinate (C) at (-2,2);

\draw (A) to[bend left=15] node[midway,right]{$\eta_{AB}$} (B) to[bend left=15] node[midway,above]{$\eta_{BC}$} (C) to[bend left=15] node[midway,left]{$\eta_{AC}$} (A);
\draw (A) node[below]{$A$};
\draw (B) node[above right]{$B$};
\draw (C) node[above left]{$C$};

\draw ($(A)+(70:0.5)$) arc (70:110:0.5);
\draw ($(A)+(90:0.75)$) node{$\hat{a}$};
\draw ($(B)+(-167:0.5)$) arc (-167:-137:0.5);
\draw ($(B)+(-152:0.75)$) node{$\hat{b}$};
\draw ($(C)+(-43:0.5)$) arc (-43:-13:0.5);
\draw ($(C)+(-28:0.75)$) node{$\hat{c}$};

\end{tikzpicture}

\caption{Each angle of the hyperbolic triangle is noted with respect to the corresponding vertex, while the length along each edge is given by the boost parameter $\eta$.}
\end{subfigure}
\hspace*{2mm}
\begin{subfigure}[t]{.45\linewidth}
\begin{tikzpicture}[scale=1]

\coordinate (A) at (0,-1);
\coordinate (A1) at (0.2,-0.5);
\coordinate (A2) at (-0.2,-0.5);
\coordinate (B) at (2,2);
\coordinate (C) at (-2,2);

\draw (A1) to[bend left=15] (B) to[bend left=15] (C) to[bend left=15] (A2);
\draw[dotted] (A) -- (A1);
\draw[dotted] (A) -- (A2);
\draw (A) node[below]{$A$};
\draw (A1) node[right]{$A'$};
\draw (A2) node[left]{$A''$};
\draw (B) node[above right]{$B$};
\draw (C) node[above left]{$C$};

\draw (A1) arc (70:100:0.8);

\end{tikzpicture}
\caption{We regularize the holonomy around the triangle by introducing the shifted points $A'$ and $A''$ along the geodesics $(AB)$ and $(AC)$ in order to avoid the conical singularity at the origin $A$ of the spherical coordinate system.}

\end{subfigure}

\caption{The hyperbolic triangle}

\label{app:triangle}

\end{figure}

The subtlety about spherical coordinates is the conical-like defect at the origin. To address this issue, we need to regularize our holonomies.  As illustrated on fig.\ref{app:triangle}, we introduce two shifts of the point $A$ along the geodesics $(AB)$ and $(AC)$. The holonomy around the hyperbolic triangle is then defined as the holonomy along the loop $(A'BCA'')$ in the limit where we send the shifted points $A'$ and $A''$ back to $A$.

The holonomies $h$ of the Ashtekar-Barbero connection $A$ along the geodesics starting at the origin are simply computed by integration:
\be
h_{AB}
= e^{i\int_{0}^{\eta_{AB}} \mathrm{d}\eta\, \vec{A}_{\eta}\cdot\f{\vec{\sigma}}{2}}
=e^{i\beta\eta_{AB}\f{\sigma_{1}}{2}}
=\mat{cc}{\cos\f{\beta\eta_{AB}}{2}& i\sin\f{\beta\eta_{AB}}{2}\\i\sin\f{\beta\eta_{AB}}{2}&\cos\f{\beta\eta_{AB}}{2}}
\,\equiv L^x_{\beta\eta_{AB}}
\,.
\ee
where we introduce the notation $L^x$ for the Lorentz transformations generated by the Pauli matrix $\sigma_{1}$ for arbitrary complex exponentiation parameter.
The holonomies of the spin-connection $\Gamma$ corresponds to the value $\beta=0$ and are trivial along those geodesics as expected (since $\vec{\Gamma}_{\eta}=0$):
\be
g_{AB}=h_{AB}^{(\beta=0)}=\id\,.
\ee
Similarly it is easy to compute the holonomies along the infinitesimal arc $(A'A'')$ (in the equatorial plane):
\be
h_{A'A''}=e^{i\int_{0}^{\hat{a}} \mathrm{d}\phi\,\vec{A_{\phi}}\cdot\f{\vec{\sigma}}{2}}
\,\,\underset{A',A''\rightarrow A}{=}\,
e^{i\,\hat{a}\f{\sigma_{3}}{2}}
\,\equiv
R^z_{\hat{a}}\,,
\ee
where we introduce the notation $R^z$ for $\SU(2)$ transformations generated by $\sigma_{3}$. Here, we point out that we chose to associate the Pauli matrices to the tangent directions on the internal space as $\vec{\sigma}=(\sigma_{1},-\sigma_{3},\sigma_{2})$ in order to keep the intuition of the $z$-direction as normal to the triangle.

\smallskip

In order to compute the holonomy $h_{BC}$ along the edge $(BC)$, we change coordinate system and choose new spherical coordinates centered on the origin $C$. Since the hyperboloid is homogeneous, the new triad will be equal to the initial triad by a simple $\SU(2)$-gauge transformation. Similarly, the new holonomies of the spin-connection and of the Ashtekar-Barbero connections will be obtained from the initial holonomies by the same gauge transformation. Let us look at the  holonomies of the spin-connection, writing $g$'s for the initial holonomies in the coordinate system with $A$ as origin and $\tilde{g}$ for the new holonomies with $C$ as origin.

\begin{figure}[h!]

\centering

\begin{tikzpicture}[scale=1]

\coordinate (A) at (0,-1);
\coordinate (B) at (2,2);
\coordinate (C) at (-2,2);
\coordinate (C2) at (-5,4);
\coordinate (B2) at (-1.5,5);

\draw (A) to[bend left=15] (B) to[bend left=15] (C) to[bend left=15] (A);
\draw[dashed] (C) to[bend right=15] (C2);
\draw[dashed] (C) to[bend left=15] (B2);
\draw (A) node[below]{$A$};
\draw (B) node[above right]{$B$};
\draw (C) node[below left]{$C$};

\draw ($(A)+(70:0.5)$) arc (70:110:0.5);
\draw ($(A)+(90:0.75)$) node{$\hat{a}$};

\draw ($(C)+(93:0.5)$) arc (93:132:0.5);
\draw ($(C)+(111.5:0.75)$) node{$\hat{a}$};

\draw ($(C)+(-42:0.3)$) arc (-42:94:0.3);
\draw ($(C)+(-42:0.35)$) arc (-42:94:0.35);
\draw ($(C)+(26:0.8)$) node{$\pi - \hat{a}$};

\end{tikzpicture}

\caption{Illustration of the rotation of the coordinates when we choose $C$ as the reference point in our coordinates.}
\label{fig:Transform}

\end{figure}
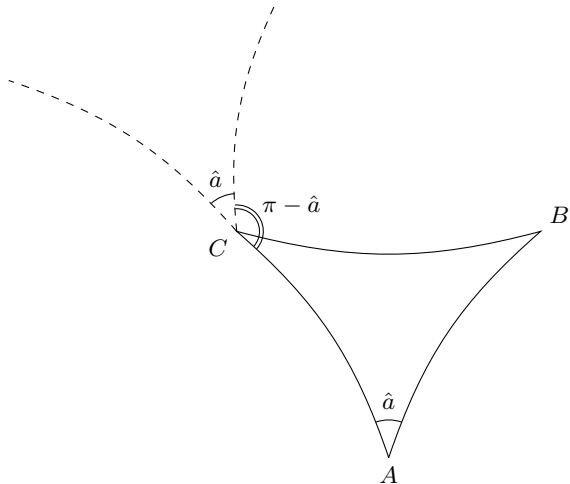

To make everything work without singularity, we need to introduce  shifts $C'$ and $C''$ of the vertex $C$ along the edges of the triangle.
The original holonomies are:
\be
\left\{
\begin{array}{ccl}
g_{A'B}&=&\id \\
g_{BC'}&=&R^z_{\pi-\hat{b}-\hat{c}}\\
g_{C'C''}&=&\id \\
g_{C''A''}&=&\id \\
g_{A''A'}&=&R^z_{-\hat{a}}
\end{array}
\right.\,,
\ee
so that the overall holonomy around the triangle $g_{ABC}=g_{A''A'}g_{C''A''}g_{C'C''}g_{BC'}g_{A'B}=R^z_{\varphi}$ reproduces the deficit angle $\varphi\equiv\pi-\hat{a}-\hat{b}-\hat{c}$ as already known \cite{charles_closure_2015}.
The new holonomies using $C$ as origin should result from those by a $\SU(2)$ gauge-transformation $k_{v}$ for all points $v$:
\be
\left\{
\begin{array}{ccccl}
\tilde{g}_{A'B}&=&k_{B}^{-1}g_{A'B}k_{A'}&=& R^z_{\pi-\hat{a}-\hat{b}}\\
\tilde{g}_{BC'}&=&k_{C'}^{-1}g_{BC'}k_{B}&=&\id \\
\tilde{g}_{C'C''}&=&k_{C''}^{-1}g_{C'C''}k_{C'}&=&R^z_{-\hat{c}}\\
\tilde{g}_{C''A''}&=&k_{A''}^{-1}g_{C''A''}k_{C''}&=&\id\\
\tilde{g}_{A''A'}&=&k_{A'}^{-1}g_{A''A'}k_{A''}&=&\id\\
\end{array}
\right.\,,
\ee
This first implies that all the $k$'s are also rotations around the $z$-axis. Setting $k_{A'}=R^z_{\psi}$, we easily solve the system to:
\be
k_{B}=R^z_{\psi+\hat{a}+\hat{b}-\pi}\,,\quad
k_{C'}=R^z_{\psi+\hat{a}-\hat{c}}\,,\quad
k_{C''}=k_{A''}=R^z_{\psi+\hat{a}}\,.
\ee
One way to determine the angle $\psi$ is to work out the explicit coordinate change. By a simple look at the triangle geometry, as shown on fig.\ref{fig:Transform}, we easily get $\psi=\pi-\hat{a}$, giving the gauge transformation:
\be
k_{B}=R^z_{\hat{b}}\,,\quad
k_{C'}=R^z_{\pi-\hat{c}}\,,\quad
k_{C''}=k_{A''}=R^z_{\pi}\,.
\ee
%
Applying this gauge transformation to the holonomies of the Ashtekar-Barbero connection, we get:
\be
\tilde{h}_{BC}
=k_{C'}^{-1} h _{BC} k_{B}
\quad
\Longrightarrow
\quad
h_{BC}=k_{C'} \tilde{h}_{BC} k_{B}^{-1}
= R^z_{\pi-\hat{c}}\,L^x_{-\beta\eta_{BC}}\,R^z_{-\hat{b}}\,.
\ee
Putting all the pieces together, we finally get the holonomy around the hyperbolic triangle:
\be
h_{ABC}=h_{A''A'}h_{CA''}h_{BC}h_{A'B}
=
R^z_{-\hat{a}}\,L^x_{-\beta\eta_{AC}}\,R^z_{\pi-\hat{c}}\,L^x_{-\beta\eta_{BC}}\,R^z_{-\hat{b}}\,L^x_{\beta\eta_{AB}}\,.
\ee
Using that $L^x_{-\eta}=R^z_{-\pi}L^x_{+\eta}R^z_{+\pi}$, it is convenient to re-organize this expression in a more symmetrical fashion to clarify its geometrical interpretation:
\be
h_{ABC}=
R^z_{\varphi}\,
\Big{(}\big{(}R^z_{\pi-\hat{c}}R^z_{\pi-\hat{b}}\big{)}^{-1}\,L^x_{\beta\eta_{AC}}\,\big{(}R^z_{\pi-\hat{c}}R^z_{\pi-\hat{b}}\big{)}\Big{)}
\Big{(}\big{(}R^z_{\pi-\hat{b}}\big{)}^{-1}\,L^x_{\beta\eta_{BC}}\,\big{(}R^z_{\pi-\hat{b}}\big{)}\Big{)}
\,L^x_{\beta\eta_{AB}}\,.
\ee
The first $\SU(2)$ group element $R^z_{\varphi}$ is the rotation around the normal direction to the triangle with angle given by the deficit angle $\varphi=\pi-\hat{a}-\hat{b}-\hat{c}$. This is exactly the holonomy $g_{ABC}$ of the spin-connection $\Gamma^{i}_{a}$ around the triangle. Then the three following terms are the pure boosts along the edges of the hyperbolic triangle, respectively $(CA)$, $(BC)$ and finally $(AB)$. Moving back to an arbitrary hyperbolic triangle (not necessarily in the equatorial plane), this structure directly generalizes to a rather simple formula:
\be
h_{ABC}\, =\, R B_{CA}^{\mathrm{i}\beta} B_{BC}^{\mathrm{i}\beta} B_{AB}^{\mathrm{i}\beta}\,.
\ee
The pure boost $B_{AB}\in\mathrm{SH}_{2}(\C)$ maps the vertex $A$ to the vertex $B$, and so on around the triangle. The rotation $R\in\SU(2)$ is the holonomy of the spin-connection around the triangle. It enters the ``almost-closure'' relation of boosts around the triangle:
\be
R=B_{AB}B_{BC}B_{CA}=B_{BA}B_{CB}B_{AC}\,,
\ee
where the second equality holds by taking the inverse and self-adjoint.
Finally the complex exponentiation for a pure boost $B$ must be understood as a quick hand notation for:
\begin{equation}
B^{\alpha}=\left(\exp\left(\frac{\eta}{2} \hat{u} \cdot \overrightarrow{\sigma}\right)\right)^{\alpha} = \exp\left(\frac{\alpha \eta}{2} \hat{u} \cdot \overrightarrow{\sigma}\right)\,,\quad\forall \eta\in\R\,,\,\,\alpha\in\C\,,\,\,\hat{u}\in\cS^{2}
\,,
\end{equation}
which is strictly defined only for boosts, though $\alpha$ can be complex.

\smallskip

The case $\beta=0$ clearly reduces the Ashtekar-Barbero holonomies to those of the original spin-connection. The cases $\beta=\pm i$ correspond to the self-dual and anti-self-dual Lorentz connection, which are flat, and give as expected $h_{ABC}^{(\beta=i)}=h_{ABC}^{(\beta=-i)}=\id$.

\subsection{Triad and holonomies in the Cartesian coordinate system}
\label{app:cartesian}

We could use the Cartesian coordinates $(x,y,z)$ instead of spherical coordinates to parametrize the hyperboloid. In that case, the induced 3d metric is not diagonal anymore:
\be
q=
\f{1}{\ka^{2}+\vec{x}^{2}}\,\mat{ccc}{\ka^{2}+y^{2}+z^{2} & -xy& -xz \\ -xy & \ka^{2}+x^{2}+z^{2} & -yz \\ -xz & -yz & \ka^{2}+x^{2}+y^{2}}
\,\,,\qquad
q_{ab}=\delta_{ab}-\f1{\ka^{2}+\vec{x}^{2}}x_{a}x_{b}\,.
\ee
The triad $e^{i}_{a}$ can be expressed as a cross-product:
\be
e^{i}_{a}=\f1{\sqrt{\ka^{2}+\vec{x}^{2}}}\,\big{(}\ka \delta^{i}_{a}-\eps^{iab}x_{b}\big{)}\,,\quad
q_{ab}=e^{i}_{a}e^{i}_{b}\,.
\ee
The inverse triad is similarly computed:
\be
e^{a}_{i}=\f\ka{\sqrt{\ka^{2}+\vec{x}^{2}}}\,\big{(}
\delta^{a}_{i}+\f{x_{a}x_{i}}{\ka^{2}}-\eps^{iab}\f{x_{b}}{\ka}
\big{)}\,.
\ee
This allows to compute the spin-connection:
\be
\Gamma^{i}_{a}=\f1\ka\Big{(}\delta^{i}_{a}-\f{x_{i}x_{a}}{\ka^{2}+\vec{x}^{2}}\Big{)}\,.
\ee
From here, we can easily compute holonomies of the Ashtekar-Barbero connection $A^{i}_{a}=\Gamma^{i}_{a}+\beta e^{i}_{a}/\ka$ along geodesics starting at the origin and see that they go along the direction of the geodesic itself.

\section{About the reconstruction of the tetrahedra: the hyperbolic bidual}
\label{app:reconstruct}

In order to reconstruct the hyperbolic tetrahedron from the $\SB(2,\C)$-normals to its triangles, we try to adapt the bi-dual tetrahedron construction of the flat case where a dual tetrahedron is constructed with the dual vertices defined as the normals of the initial tetrahedron. In that case, we know that taking the dual twice leads back to the initial tetrahedron up to a global re-scaling  (by the tetrahedron volume).

So let us start with a hyperbolic tetrahedron. We place one of its vertices at the hyperboloid origin. The other three vertices are defined by three pure boosts (in $\mathrm{SH}_{2}(\C)$), or equivalently by three group elements in $\SB(2,\C)\sim\SL(2,\C)\SU(2)$. Following our definitions, we choose a complex value of the Immirzi parameter $\beta$ and compute the $\SL(2,\C)$ holonomies $\Lambda_{i}$ of the Ashtekar-Barbero connection around the four hyperbolic triangles. These holonomies satisfy a $\SL(2,\C)$ closure constraint, $\Lambda_{D}\Lambda_{C}\Lambda_{B}\Lambda_{A}=\id$. Finally taking their Iwasawa decompositions leads to the $\SB(2,\C)$-normals to the hyperbolic triangles, which also satisfy a $\SB(2,\C)$ closure relation:
\begin{equation}
L_D L_C L_B L_A = \id\,.
\end{equation}
We can now use these four new  $\mathrm{SB}(2,\mathbb{C})$ group elements as the Lorentz boosts defining the vertices of a new tetrahedron, as $L_{A}L_{A}^{\dagger}$ and so on. Unfortunately, this definition would not behave properly under 3d rotations. Indeed, due to the non-abelian nature $\SB(2,\C)$ and the twisting induced by the Iwasawa decomposition, we know that the four $\SB(2,\C)$-normals don't transform with the same $\SU(2)$ transformation under rotations and the correct transformation law was earlier given in \eqref{LHrot}:
\be
\left|
\begin{array}{lcl}
L_{D}&\rightarrow&k L_{D}k_{D}^{-1}\\
L_{C}&\rightarrow&k_{D} L_{C}k_{C}^{-1}\\
L_{B}&\rightarrow&k_{C} L_{B}k_{B}^{-1}\\
L_{A}&\rightarrow&k_{B} L_{A}k^{-1}\\
\end{array}
\right.
\qquad
\left|
\begin{array}{lcl}
H_{D}&\rightarrow&k_{D} H_{D}k^{-1}\\
H_{C}&\rightarrow&k_{C} H_{C}k_{D}^{-1}\\
H_{B}&\rightarrow&k_{B} H_{B}k_{C}^{-1}\\
H_{A}&\rightarrow&k H_{A}k_{B}^{-1}\\
\end{array}
\right.\,,
\nn
\ee
where we recall that the $L$'s and $H$'s were both derived from the $\SL(2,\C)$-holonomies as:
\begin{equation}
\left\{\begin{array}{rcl}
\Lambda_D &=& L_D H_D \\
\left(H_D\right) \Lambda_C \left(H_D\right)^{-1} &=& L_C H_C \\
\left(H_C H_D\right) \Lambda_B \left(H_C H_D\right)^{-1} &=& L_B H_B \\
\left(H_B H_C H_D\right) \Lambda_A \left(H_B H_C H_D\right)^{-1} &=& L_A H_A
\end{array}\right.\nn
\end{equation}
So under a 3d rotation, the dual vertex  $L_{D}L_{D}^{\dagger}$ would get conjugated by the $\SU(2)$ group element $k$ while the dual vertex $L_{C}L_{C}^\dagger$ would get conjugated by $k_{D}$. The way out is to use the $H$'s to compensate the twist in the action of rotations. This actually amounts to coming back to the original $\SL(2,\C)$ holonomies. Indeed, these follow simple homogeneous law under rotation:
\be
\Lambda_{i}\,\longrightarrow\, k\Lambda_{i}k^{-1}\,.
\ee
In the end, our dual hyperbolic tetrahedron has its four dual vertices defined as:
\be
\left|
\begin{array}{l}
\Lambda_{D}\Lambda_{D}^{\dagger}=L_{D}L_{D}^{\dagger}\\
\Lambda_{C}\Lambda_{C}^{\dagger}=H_{D}^{-1}\,(L_{C}L_{C}^{\dagger})\,H_{D}\\
\Lambda_{B}\Lambda_{B}^{\dagger}=(H_{C}H_{D})^{-1}\,(L_{B}L_{B}^{\dagger})\,(H_{C}H_{D})\\
\Lambda_{A}\Lambda_{A}^{\dagger}=(H_{B}H_{C}H_{D})^{-1}\,(L_{A}L_{A}^{\dagger})\,(H_{B}H_{C}H_{D})\,.\\
\end{array}
\right.
\ee
Finally, we can always translate back the first vertex to the hyperboloid origin\footnotemark{}, considering the translated dual vertices defined as $\id$, $\Lambda_{D}^{-1}\Lambda_{C}\Lambda_{C}^{\dagger}\Lambda_{D}^{-1\dagger}$, $\Lambda_{D}^{-1}\Lambda_{B}\Lambda_{B}^{\dagger}\Lambda_{D}^{-1}$ and $\Lambda_{D}^{-1}\Lambda_{A}\Lambda_{A}^{\dagger}\Lambda_{D}^{-1}$. Let  us point out that we could also decide to translate  using the Borel group element $L_{D}$ instead of the full  Lorentz transformation $\Lambda_{D}$. This stills translates the first vertex back to the origin and is more consistent with the logic of distinguishing the 3d rotations from 3d translations on the hyperboloid by selecting a proper section of the coset $\SL(2,\C)/\SU(2)$ and choosing the Borel group $\SB(2,\C)$ as defining the hyperbolic translations. The resulting translated dual tetrahedra only differ by an overall rotation (by $H_{D}$), so we should keep in mind that we might recover the initial tetrahedron only up to a global rotation through this bi-dual procedure.
\footnotetext{
The hyperboloid is defined as the coset $\SL(2,\C)/\SU(2)$ as the set of Lorentz group elements $\Lambda$ up to $\SU(2)$ transformations $\Lambda \rightarrow \Lambda H$ for arbitrary $H\in\SU(2)$. The hyperbolic distance between two points on the hyperboloid defined by two Lorentz transformations $\Lambda$ and $\tilde{\Lambda}$ is $\tr\,(\tilde{\Lambda}^{-1}\Lambda)(\tilde{\Lambda}^{-1}\Lambda)^{\dagger}$ and is invariant under translations and rotations on the hyperboloid both realized as Lorentz transformations $\Omega\in\SL(2,\C)$ acting as $\Lambda \rightarrow \Omega \Lambda$.
}

\smallskip

We are now ready to take the dual of the dual tetrahedron. We give ourself the freedom to choose a different Immirzi parameter $\tilde{\beta}$ for computing the $\SL(2,\C)$-holonomies around the dual triangles and defining the $\SB(2,\C)$-normals to the dual tetrahedron. In this scheme, we can potentially adjust the new Immirzi parameter $\tilde{\beta}$ in function of the initial value $\beta$ in order to recover the initial hyperbolic tetrahedron.

We have worked out preliminary numerical simulations focusing on the simplified case of a square tetrahedron: the three edges meeting at the root vertex are orthogonal to each other and furthermore of equal length. Such a tetrahedron is defined entirely by the value of a single edge length. Indeed, once the length of the three edges meeting at the right angles is given, the length of the remaining three edges is fixed.

\begin{figure}[h!]


\begin{subfigure}[t]{.45\linewidth}
\includegraphics[scale=0.6]{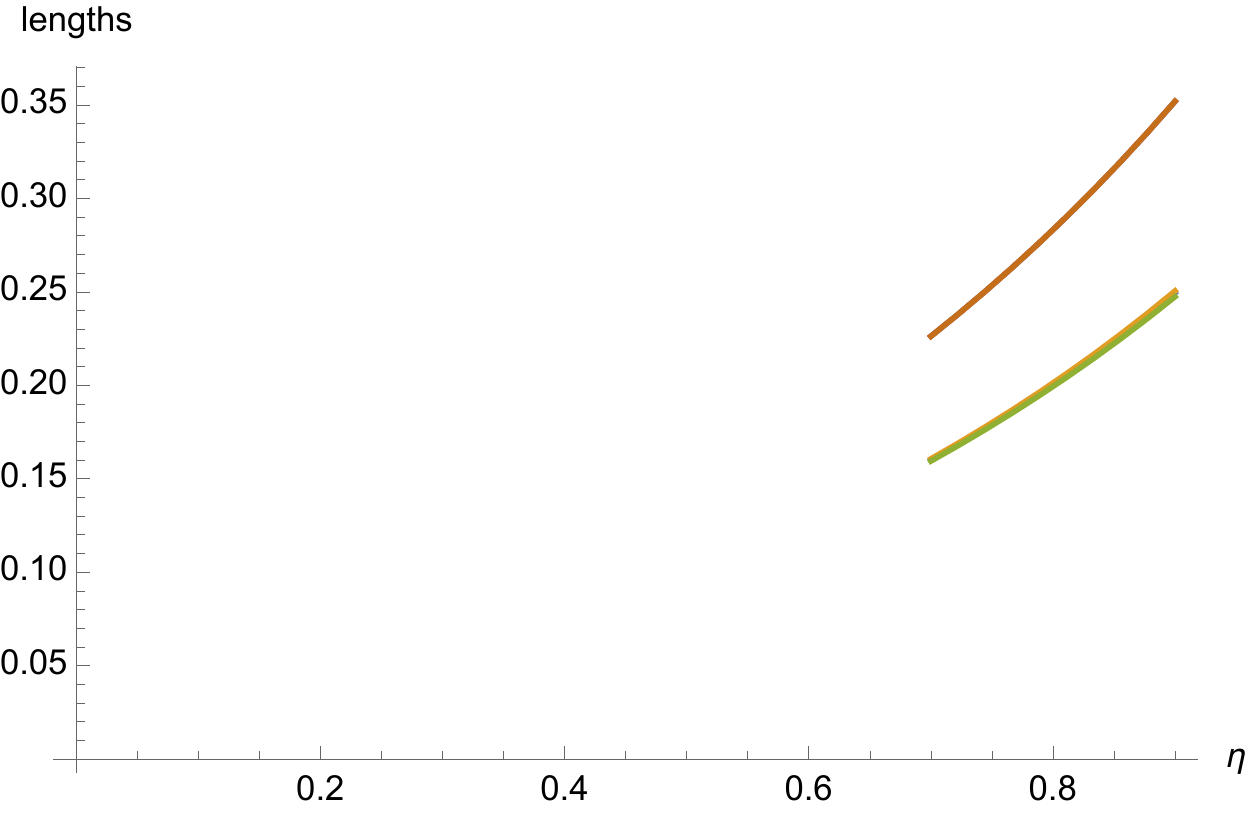}
\caption{We plot the lengths of the edges of the bidual of a square tetrahedron with edge length $0.2$ at the square. The original Immirzi parameter is taken to be $\cosh(1)+\mathrm{i}\sinh(1)$ for the construction of the dual and the bidual is constructed using an Immirzi parameter of the form $\cosh(\eta)+\mathrm{i}\sinh(\eta)$. The six lengths do indeed group into two different lengths: the length of the edges meeting at the square angle and the length of the opposite edges.}
\end{subfigure}
\hspace{2mm}
\begin{subfigure}[t]{.45\linewidth}
\includegraphics[scale=0.6]{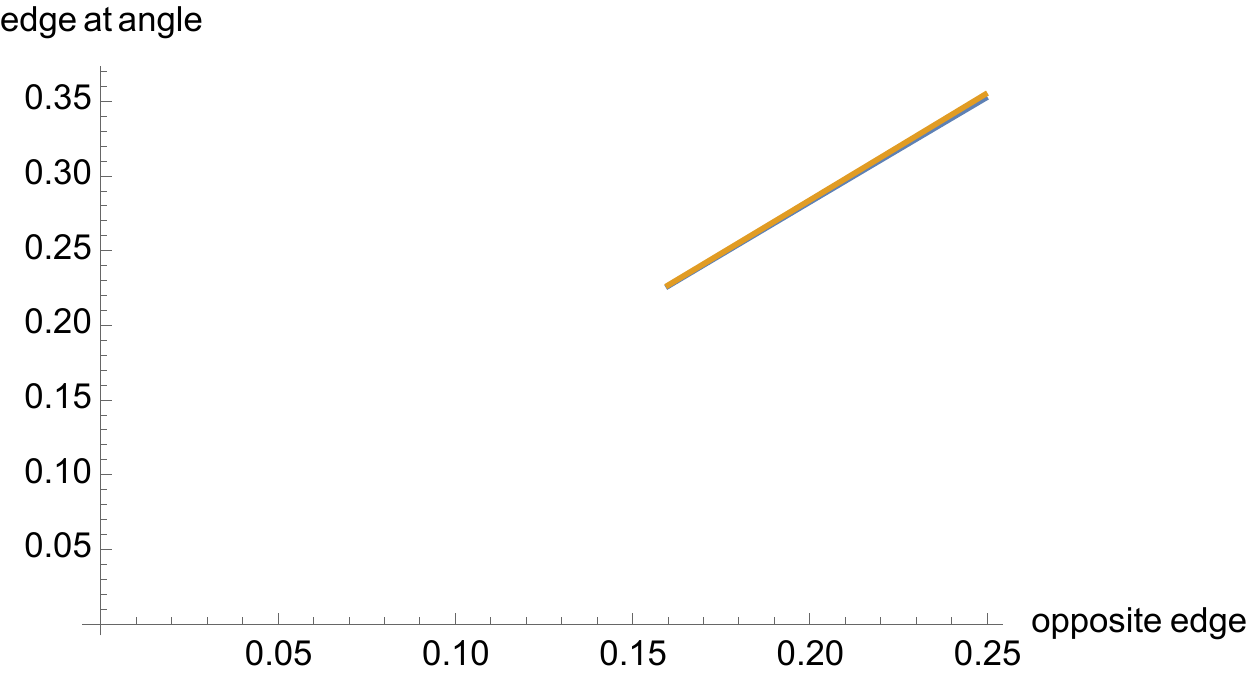}
\caption{We plot the two lengths of the edges of the bidual of a square tetrahedron with edge length $0.2$ at the square against each other. The original Immirzi parameter is taken to be $\cosh(1)+\mathrm{i}\sinh(1)$ for the construction of the dual and the bidual is constructed using an Immirzi parameter of the form $\cosh(\eta)+\mathrm{i}\sinh(\eta)$. We also plot the expected relation between the two for a square tetrahedron and find that the curves match perfectly. It seems therefore that the bidual is indeed square.}
\end{subfigure}

\caption{Numerical analysis of the bidual of the tetrahedron in order to check that it is indeed also square. The original tetrahedron is a square tetrahedron of edge length $0.2$. The dual is constructed with Immirzi parameter $\cosh(1)+\mathrm{i}\sinh(1)$ and the bidual with Immirzi parameter $\cosh(\eta)+\mathrm{i}\sinh(\eta)$. This illustrates the fact that the bidual is still square and therefore the reconstruction procedure, in this peculiar case, can be tested using only one edge.}
\label{fig:bidualCheck}

\end{figure}

As displayed on figure \ref{fig:bidualCheck}, we first checked that the bidual tetrahedron of such a square hyperbolic tetrahedron is again a square tetrahedron. We plotted the six lengths and checked that indeed they group into two lengths: the length of the edges meeting at the root vertex and the length of the opposite edges.
This by itself is not enough to ensure that the angles at the root vertex are right again. Therefore, we further plotted the two lengths against each other and compared to the curve expected for a square tetrahedron and the match seems perfect. This provides a numerical check that the bidual of a square tetrahedron is again a square tetrahedron.

We also checked the shape of the dual of a square tetrahedron.
In the flat space case, the dual tetrahedron would be square too, but this is actually not the case in the hyperbolic case. The six lengths are indeed still grouped into two groups, reflecting some preservation of the symmetry, but these two lengths do not have the correct relations between themselves and therefore the dual is not a square tetrahedron.

\begin{figure}[h!]


\begin{subfigure}[t]{.45\linewidth}
\includegraphics[scale=0.6]{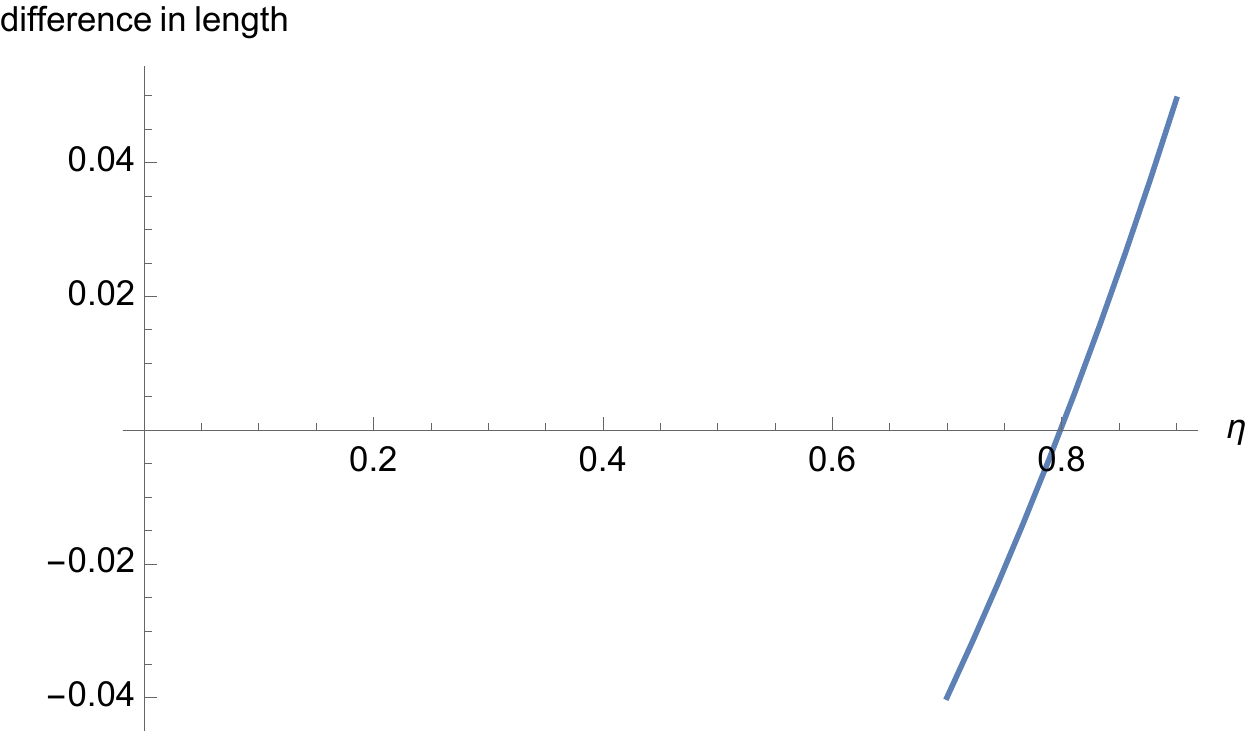}
\caption{We plot the difference between the length of one of the edge at the corner of the bidual of a square tetrahedron with edge length $0.2$ at the square and that of the original tetrahedron. The original Immirzi parameter is taken to be $\cosh(1)+\mathrm{i}\sinh(1)$ for the construction of the dual. The bidual is constructed using an Immirzi parameter of the form $\cosh(\eta)+\mathrm{i}\sinh(\eta)$. The difference does cross the zero-line, signaling that the reconstruction procedure works at least in the case of a square tetrahedron.}
\end{subfigure}
\hspace{2mm}
\begin{subfigure}[t]{.45\linewidth}
\includegraphics[scale=0.6]{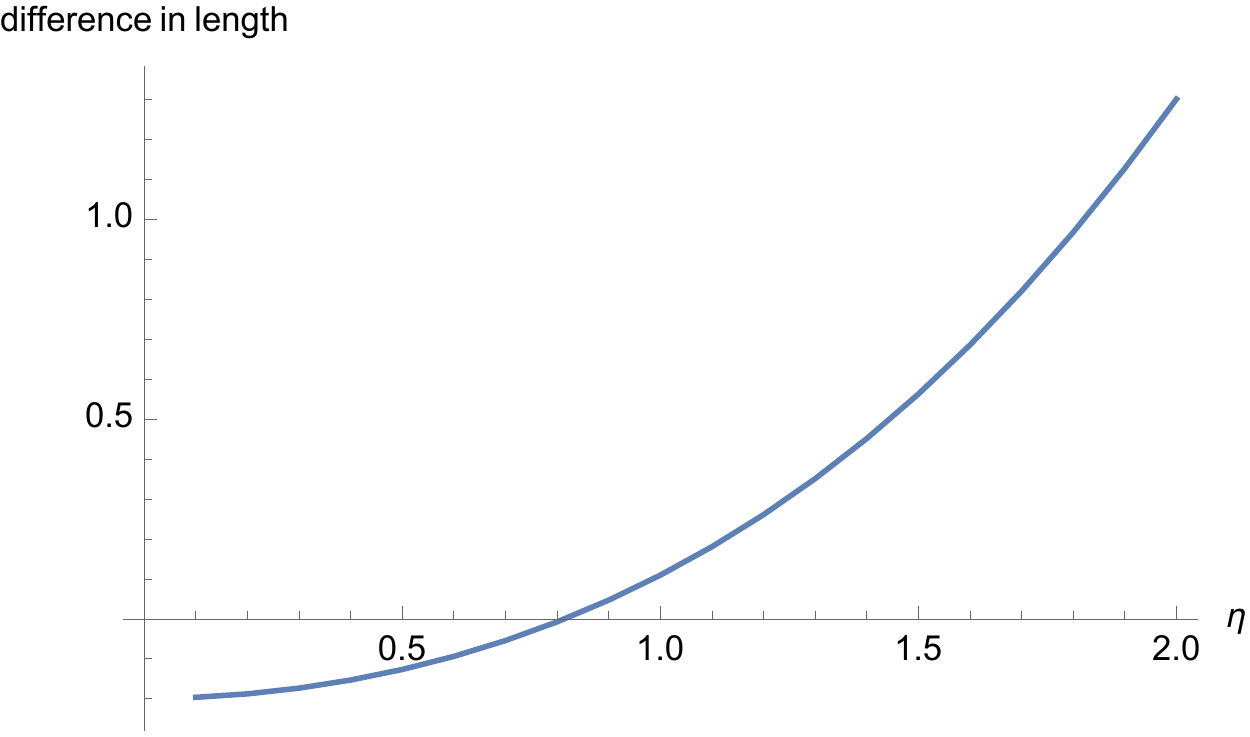}
\caption{We plot the difference between the length of one of the edge at the corner of the bidual of a square tetrahedron with edge length $0.2$ at the square and that of the original tetrahedron. The original Immirzi parameter is taken to be $\cosh(1)+\mathrm{i}\sinh(1)$ for the construction of the dual. The bidual is constructed using an Immirzi parameter of the form $\eta (\cosh(1)+\mathrm{i}\sinh(1))$. The difference does cross the zero-line, signaling that the reconstruction procedure works at least in the case of a square tetrahedron.}
\end{subfigure}
\caption{Numerical analysis of the reconstruction procedure of a tetrahedron via the bidual. The original tetrahedron is a square tetrahedron of edge length $0.2$. The dual is constructed with Immirzi parameter $\cosh(1)+\mathrm{i}\sinh(1)$. The bidual is constructed using two different schemes as explained in the captions. The plots represent the difference between the caracterizing length of the bidual and that of the original tetrahedron. In both cases, the curves cross the zero-line signaling the success of the procedure, at least in the square tetrahedron case.}
\label{fig:BidualReconstruct}

\end{figure}

In this context, it is easy to proceed to check whether the bidual tetrahedron will match or not with the initial square tetrahedron: we fix the initial edge length (here 0,2 in our numerical simulations) and we vary the Immirzi parameter until the edge length of the bidual matches that initial edge length.
We plotted the difference of the initial edge length with the bidual edge length in fig.\ref{fig:BidualReconstruct}, using two different ansatz for the Immirzi parameter: for an Immirzi parameter proportional to the initial one and for a (complex) cosmological constant proportional to the initial one.
Both schemes seem to be able to reconstruct the original tetrahedron. This was to be expected as we have an $\mathrm{SU}(2)$ freedom on each normal which does not change the reconstruction. Therefore we expect that a (one-real-parameter) family of Immirzi parameters should work and allow the reconstruction.

These preliminary numerical simulations  are very promising but they are not yet conclusive. They should be studied in the more general setting of an arbitrary tetrahedron and we believe this line of research worth investigating further.

\begin{figure}[h!]


\begin{subfigure}[t]{.45\linewidth}
\includegraphics[scale=0.6]{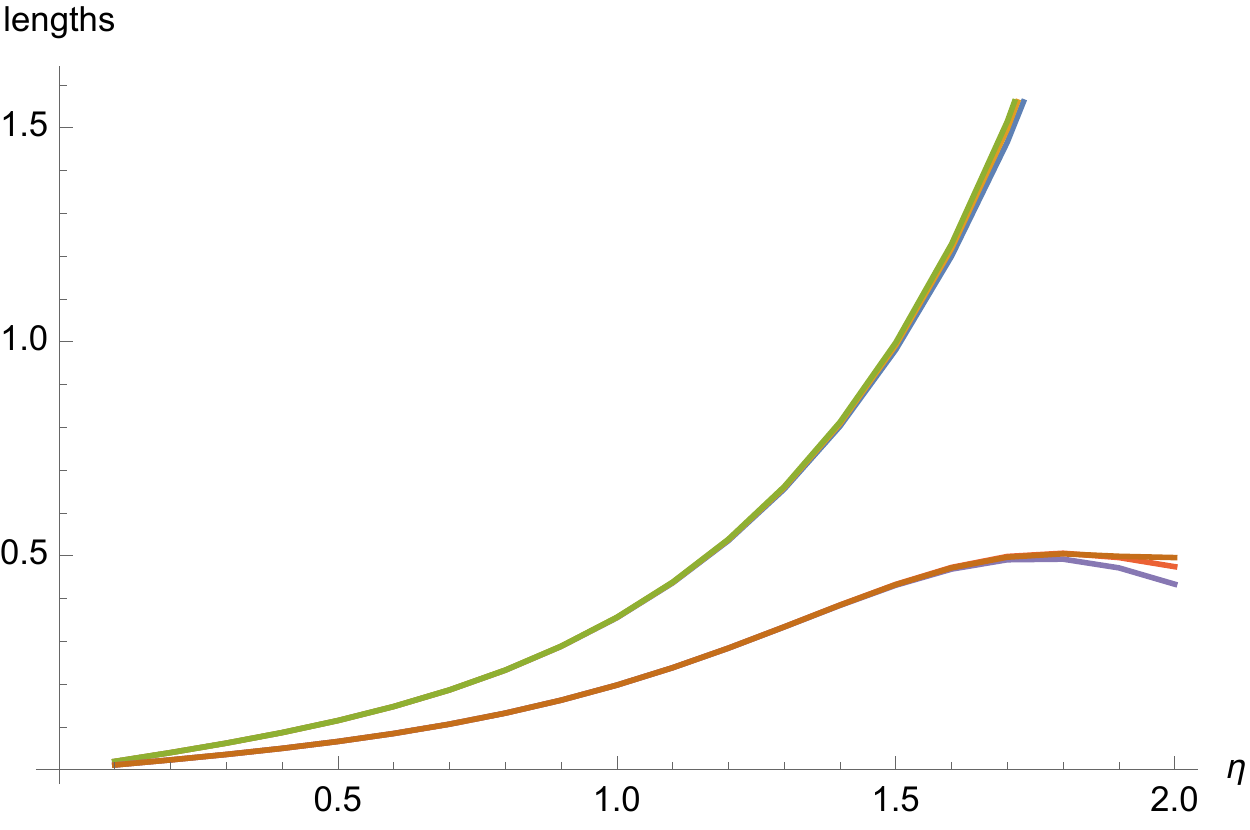}
\caption{We plot the lengths of the edges of the dual of a square tetrahedron with edge length $0.2$ at the square. The Immirzi parameter is taken to be $\cosh(\eta)+\mathrm{i}\sinh(\eta)$ for the construction. The six lengths do indeed group into two different lengths as for the bidual.}
\end{subfigure}
\hspace{2mm}
\begin{subfigure}[t]{.45\linewidth}
\includegraphics[scale=0.6]{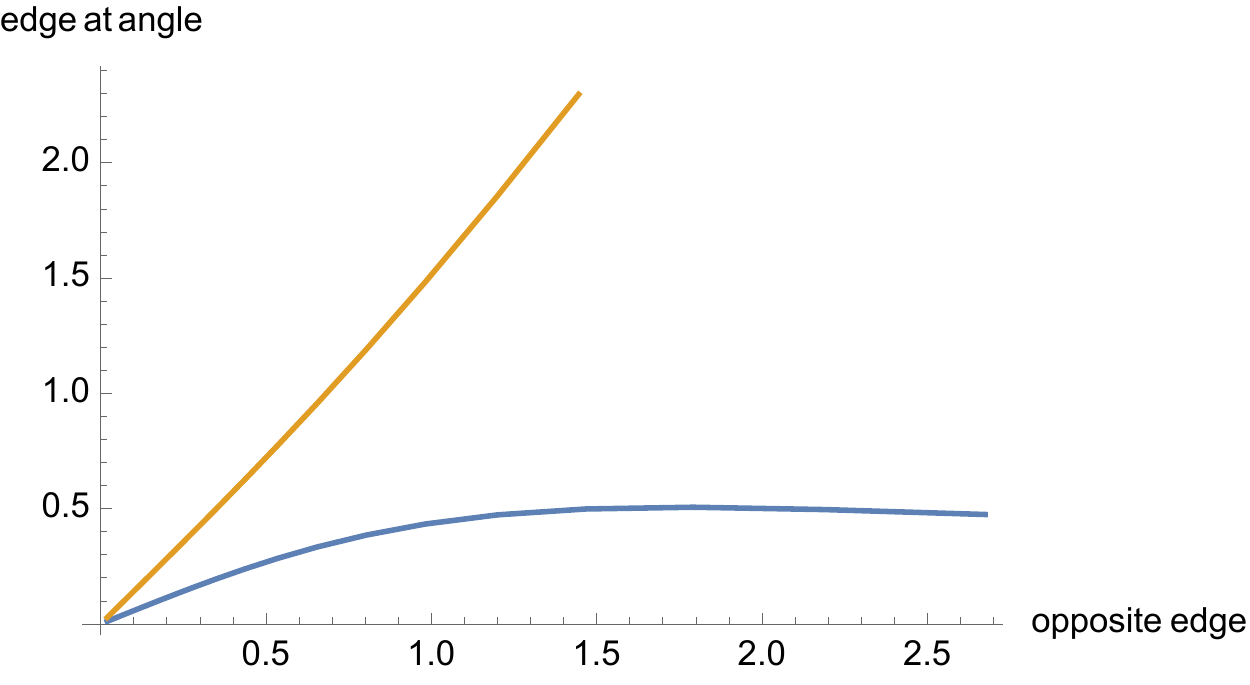}
\caption{We plot the two lengths of the edges of the dual of a square tetrahedron with edge length $0.2$ at the square against each other. The Immirzi parameter is taken to be $\cosh(\eta)+\mathrm{i}\sinh(\eta)$. We also plot the expected relation between the two for a square tetrahedron and find that the curves do not match at all. It seems therefore that the dual is not square.}
\end{subfigure}
\caption{Numerical analysis of the dual of the tetrahedron in order to understand its shape. The original tetrahedron is a square tetrahedron of edge length $0.2$. The dual is constructed with Immirzi parameter $\cosh(\eta)+\mathrm{i}\sinh(\eta)$. Though the dual is not square, it seems to retain some symmetry in the grouping of the edges lengths.}
\label{fig:dualCheck}

\end{figure}

\bibliographystyle{bib-style}
\bibliography{LQGquantumgroups}

\providecommand{\href}[2]{#2}\begingroup\raggedright\begin{thebibliography}{10}

\bibitem{rovelli_quantum_2007}
C.~Rovelli, {\em Quantum {Gravity}}.
\newblock Cambridge {Monographs} on {Mathematical} {Physics}. Cambridge
  University Press, 2007.

\bibitem{thiemann_modern_2007}
T.~Thiemann, {\em Modern {Canonical} {Quantum} {General} {Relativity}}.
\newblock Cambridge {Monographs} on {Mathematical} {Physics}. Cambridge
  University Press, 2007.

\bibitem{vidotto_covariant_2014}
F.~Vidotto and C.~Rovelli, {\em Covariant {Loop} {Quantum} {Gravity}}.
\newblock Cambridge {Monographs} on {Mathematical} {Physics}. Cambridge
  University Press, 2014.

\bibitem{ashtekar_new_1986}
A.~Ashtekar, ``New {Variables} for {Classical} and {Quantum} {Gravity},'' Phys.
  Rev. Lett. {\bf 57} (Nov., 1986) 2244--2247.

\bibitem{barbero_g._real_1995}
J.~F. Barbero~G., ``Real {Ashtekar} variables for {Lorentzian} signature space
  times,'' Phys.Rev. {\bf D51} (1995) 5507--5510.

\bibitem{immirzi_real_1997}
G.~Immirzi, ``Real and complex connections for canonical gravity,''
  Class.Quant.Grav. {\bf 14} (1997) L177--L181.

\bibitem{holst_barberos_1996}
S.~Holst, ``Barbero's {Hamiltonian} derived from a generalized
  {Hilbert}-{Palatini} action,'' Phys.Rev. {\bf D53} (1996) 5966--5969.

\bibitem{rovelli_spin_1995}
C.~Rovelli and L.~Smolin, ``Spin networks and quantum gravity,'' Phys.Rev. {\bf
  D52} (1995) 5743--5759.

\bibitem{rovelli_discreteness_1995}
C.~Rovelli and L.~Smolin, ``Discreteness of area and volume in quantum
  gravity,'' Nucl.Phys. {\bf B442} (1995) 593--622.

\bibitem{rovelli_immirzi_1998}
C.~Rovelli and T.~Thiemann, ``The {Immirzi} parameter in quantum general
  relativity,'' Phys.Rev. {\bf D57} (1998) 1009--1014.

\bibitem{alexandrov_immirzi_2008}
S.~Alexandrov, ``Immirzi parameter and fermions with non-minimal coupling,''
  Class. Quant. Grav. {\bf 25} (2008) 145012.

\bibitem{freidel_twisted_2010}
L.~Freidel and S.~Speziale, ``Twisted geometries: {A} geometric parametrisation
  of {SU}(2) phase space,'' Phys.Rev. {\bf D82} (2010) 084040.

\bibitem{haggard_spin_2013}
H.~M. Haggard, C.~Rovelli, W.~Wieland, and F.~Vidotto, ``Spin connection of
  twisted geometry,'' Phys.Rev. {\bf D87} (2013), no.~2, 024038.

\bibitem{freidel_spinning_2014}
L.~Freidel and J.~Ziprick, ``Spinning geometry = {Twisted} geometry,''
  Class.Quant.Grav. {\bf 31} (2014), no.~4, 045007.

\bibitem{thiemann_anomaly_1996}
T.~Thiemann, ``Anomaly - free formulation of nonperturbative, four-dimensional
  {Lorentzian} quantum gravity,'' Phys.Lett. {\bf B380} (1996) 257--264.

\bibitem{thiemann_phoenix_2006}
T.~Thiemann, ``The {Phoenix} project: {Master} constraint program for loop
  quantum gravity,'' Class.Quant.Grav. {\bf 23} (2006) 2211--2248.

\bibitem{perez_spin_2013}
A.~Perez, ``The {Spin} {Foam} {Approach} to {Quantum} {Gravity},'' Living
  Rev.Rel. {\bf 16} (2013) 3.

\bibitem{engle_lqg_2008}
J.~Engle, E.~Livine, R.~Pereira, and C.~Rovelli, ``{LQG} vertex with finite
  {Immirzi} parameter,'' Nucl.Phys. {\bf B799} (2008) 136--149.

\bibitem{oriti_group_2014}
D.~Oriti, ``Group {Field} {Theory} and {Loop} {Quantum} {Gravity},''.

\bibitem{Koslowski:2011vn}
T.~Koslowski and H.~Sahlmann, ``Loop quantum gravity vacuum with nondegenerate
  geometry,'' SIGMA {\bf 8} (2012) 026,
\href{http://arXiv.org/abs/1109.4688}{{\texttt{arXiv:1109.4688}}}.

\bibitem{Bahr:2015bra}
B.~Bahr, B.~Dittrich, and M.~Geiller, ``{A new realization of quantum
  geometry},''
\href{http://arXiv.org/abs/1506.08571}{{\texttt{arXiv:1506.08571}}}.

\bibitem{Dittrich:2014wpa}
B.~Dittrich and M.~Geiller, ``{A new vacuum for Loop Quantum Gravity},'' Class.
  Quant. Grav. {\bf 32} (2015), no.~11, 112001,
\href{http://arXiv.org/abs/1401.6441}{{\texttt{arXiv:1401.6441}}}.

\bibitem{Charles:2016xwc}
C.~Charles and E.~R. Livine, ``{The Fock Space of Loopy Spin Networks for
  Quantum Gravity},''
\href{http://arXiv.org/abs/1603.01117}{{\texttt{arXiv:1603.01117}}}.

\bibitem{livine_deformation_2014}
E.~R. Livine, ``Deformation {Operators} of {Spin} {Networks} and
  {Coarse}-{Graining},'' Class. Quant. Grav. {\bf 31} (2014) 075004.

\bibitem{witten_2+1-dimensional_1988}
E.~Witten, ``(2+1)-{Dimensional} {Gravity} as an {Exactly} {Soluble}
  {System},'' Nucl.Phys. {\bf B311} (1988) 46.

\bibitem{mizoguchi_three-dimensional_1992}
S.~Mizoguchi and T.~Tada, ``Three-dimensional gravity from the {Turaev}-{Viro}
  invariant,'' Phys.Rev.Lett. {\bf 68} (1992) 1795--1798.

\bibitem{dupuis_observables_2013}
M.~Dupuis and F.~Girelli, ``Observables in {Loop} {Quantum} {Gravity} with a
  cosmological constant,''.

\bibitem{bonzom_deformed_2014}
V.~Bonzom, M.~Dupuis, F.~Girelli, and E.~R. Livine, ``Deformed phase space for
  3d loop gravity and hyperbolic discrete geometries,''.

\bibitem{dupuis_deformed_2014}
M.~Dupuis, F.~Girelli, and E.~R. Livine, ``Deformed {Spinor} {Networks} for
  {Loop} {Gravity}: {Towards} {Hyperbolic} {Twisted} {Geometries},''
  Gen.Rel.Grav. {\bf 46} (2014), no.~11, 1802.

\bibitem{bonzom_towards_2014}
V.~Bonzom, M.~Dupuis, and F.~Girelli, ``Towards the {Turaev}-{Viro} amplitudes
  from a {Hamiltonian} constraint,''.

\bibitem{pranzetti_turaev-viro_2014}
D.~Pranzetti, ``Turaev-{Viro} amplitudes from 2+1 {Loop} {Quantum} {Gravity},''
  Phys. Rev. {\bf D89} (2014), no.~8, 084058.

\bibitem{charles_closure_2015}
C.~Charles and E.~R. Livine, ``Closure constraints for hyperbolic tetrahedra,''
  Class. Quant. Grav. {\bf 32} (2015), no.~13, 135003.

\bibitem{haggard_encoding_2015}
H.~M. Haggard, M.~Han, and A.~Riello, ``Encoding {Curved} {Tetrahedra} in
  {Face} {Holonomies}: a {Phase} {Space} of {Shapes} from {Group}-{Valued}
  {Moment} {Maps},''.

\bibitem{haggard_sl2c_2015}
H.~M. Haggard, M.~Han, W.~Kamiński, and A.~Riello, ``{SL}(2,{C})
  {Chern}-{Simons} {Theory}, a non-{Planar} {Graph} {Operator}, and 4D {Loop}
  {Quantum} {Gravity} with a {Cosmological} {Constant}: {Semiclassical}
  {Geometry},'' Nucl. Phys. {\bf B900} (2015) 1--79.

\bibitem{Jacobson:2007uj}
T.~Jacobson, ``{Renormalization and black hole entropy in Loop Quantum
  Gravity},'' Class. Quant. Grav. {\bf 24} (2007) 4875--4879,
\href{http://arXiv.org/abs/0707.4026}{{\texttt{arXiv:0707.4026}}}.

\bibitem{charles_ashtekar-barbero_2015}
C.~Charles and E.~R. Livine, ``Ashtekar-{Barbero} holonomy on the hyperboloid:
  {Immirzi} parameter as a cutoff for quantum gravity,'' Phys. Rev. {\bf D92}
  (2015), no.~12, 124031.

\bibitem{Benedetti:2011nd}
D.~Benedetti and S.~Speziale, ``{Perturbative quantum gravity with the Immirzi
  parameter},'' JHEP {\bf 06} (2011) 107,
\href{http://arXiv.org/abs/1104.4028}{{\texttt{arXiv:1104.4028}}}.

\bibitem{Frodden:2012dq}
E.~Frodden, M.~Geiller, K.~Noui, and A.~Perez, ``{Black Hole Entropy from
  complex Ashtekar variables},'' Europhys. Lett. {\bf 107} (2014) 10005,
\href{http://arXiv.org/abs/1212.4060}{{\texttt{arXiv:1212.4060}}}.

\bibitem{Achour:2014eqa}
J.~Ben~Achour, A.~Mouchet, and K.~Noui, ``{Analytic Continuation of Black Hole
  Entropy in Loop Quantum Gravity},'' JHEP {\bf 06} (2015) 145,
\href{http://arXiv.org/abs/1406.6021}{{\texttt{arXiv:1406.6021}}}.

\bibitem{Achour:2015xga}
J.~Ben~Achour and K.~Noui, ``{Analytic continuation of real Loop Quantum
  Gravity : Lessons from black hole thermodynamics},'' PoS {\bf FFP14} (2015)
  158,
\href{http://arXiv.org/abs/1501.05523}{{\texttt{arXiv:1501.05523}}}.

\bibitem{Sahlmann:2011rv}
H.~Sahlmann and T.~Thiemann, ``{Chern-Simons expectation values and quantum
  horizons from LQG and the Duflo map},'' Phys. Rev. Lett. {\bf 108} (2012)
  111303,
\href{http://arXiv.org/abs/1109.5793}{{\texttt{arXiv:1109.5793}}}.

\bibitem{Dittrich:2016typ}
B.~Dittrich and M.~Geiller, ``{Quantum gravity kinematics from extended
  TQFTs},''
\href{http://arXiv.org/abs/1604.05195}{{\texttt{arXiv:1604.05195}}}.

\bibitem{Perez:2010pq}
A.~Perez and D.~Pranzetti, ``{Static isolated horizons: SU(2) invariant phase
  space, quantization, and black hole entropy},'' Entropy {\bf 13} (2011)
  744--777,
\href{http://arXiv.org/abs/1011.2961}{{\texttt{arXiv:1011.2961}}}.

\end{thebibliography}\endgroup

\end{document}